\documentstyle[twoside,11pt,fleqn]{article}
\makeatletter
\def\@revmess#1#2{\typeout{CJPhys LaTeX #1: #2}}
\ifx\selectfont\undefined %
\@revmess{message}{NFSS not detected. Assuming OFSS.}
\let\reset@font\relax
\def\mathhexbox{\protect\mathhexbox@}
\def\mathhexbox@#1#2#3{\relax
\ifmmode\mathpalette{}{\m@th\mathchar"#1#2#3}%
\else\leavevmode\hbox{$\m@th\mathchar"#1#2#3$}\fi}
\def\text#1{%
\relax
\ifmmode %
\mathchoice
{\hbox{\everymath{\displaystyle}\rm #1}}%
{\hbox{\everymath{\textstyle}\rm #1}}%
{\hbox{\everymath{\scriptstyle}%
\def\prm{\fam\z@ \the\scriptfont\z@ \relax}%
\def\pit{\fam\itfam \the\scriptfont\itfam \relax}%
\rm #1}%
}%
{\hbox{\everymath{\scriptscriptstyle}%
\def\prm{\fam\z@ \the\scriptscriptfont\z@ \relax}%
\def\pit{\fam\itfam \the\scriptscriptfont\itfam \relax}%
\rm #1}%
}%
\else %
\leavevmode\hbox{#1}%
\fi
}
\def\bbox#1{%
\leavevmode\text{%
\textfont0 \the\textfont\bffam
\scriptfont0 \the\scriptfont\bffam
\scriptscriptfont0 \the\scriptscriptfont\bffam
\@temptokena\everymath \boldmath \everymath\@temptokena
{$\m@th\relax#1$}%
}%
}
\font\fivbf=cmbx5 \font\sixbf=cmbx6 \font\sevbf=cmbx7 \font\egtbf=cmbx8
\expandafter\def\expandafter\ixpt\expandafter{\ixpt
\scriptfont\bffam\sixbf \scriptscriptfont\bffam\fivbf}
\expandafter\def\expandafter\xpt\expandafter{\xpt
\scriptfont\bffam\sevbf \scriptscriptfont\bffam\fivbf}
\expandafter\def\expandafter\xipt\expandafter{\xipt
\scriptfont\bffam\egtbf \scriptscriptfont\bffam\sixbf}
\expandafter\def\expandafter\xiipt\expandafter{\xiipt
\scriptfont\bffam\egtbf \scriptscriptfont\bffam\sixbf}
\expandafter\def\expandafter\xivpt\expandafter{\xivpt
\scriptfont\bffam\tenbf \scriptscriptfont\bffam\sevbf}
\else %
\@revmess{message}{NFSS detected! Assuming NFSS.}
\def\text#1{%
\relax\ifmmode
\mathchoice
{\hbox{{\everymath{\displaystyle}#1}}}%
{\hbox{{\everymath{\textstyle}#1}}}%
{\hbox{{\everymath{\scriptstyle}\let\f@size\sf@size\selectfont#1}}}%
{\hbox{{\everymath{\scriptscriptstyle}\let\f@size\ssf@size\selectfont#1}}}%
\glb@settings
\else
\mbox{#1}%
\fi
}
\def\bbox#1{%
\relax\ifmmode
\mathchoice
{{\hbox{\boldmath$\displaystyle#1$}}}%
{{\hbox{\boldmath$\textstyle#1$}}}%
{{\hbox{\boldmath$\scriptstyle#1$}}}%
{{\hbox{\boldmath$\scriptscriptstyle#1$}}}%
\glb@settings
\else
\mbox{#1}%
\fi
}
\def\mathhexbox{\protect\mathhexbox@}
\def\mathhexbox@#1#2#3{\relax
\ifmmode\mathpalette{}{\m@th\mathchar"#1#2#3}%
\else\leavevmode\hbox{$\m@th\mathchar"#1#2#3$}\fi}
\fi

\newdimen\@mydim \newdimen\@shiftdim
\newbox\@myboxa \newbox\@myboxb

\def\figcaption{\refstepcounter\@captype
  \@dblarg{\@figcaption\@captype}}
\long\def\@figcaption#1[#2]#3{\par\addcontentsline{\csname
  ext@#1\endcsname}{#1}{\protect\numberline{\csname
  the#1\endcsname}{\ignorespaces #2}}\begingroup
    \@parboxrestore
    \normalsize
  \@makefigcaption{\csname fnum@#1\endcsname}{\ignorespaces #3}
  \par\endgroup}

\long\def\@makefigcaption#1#2{
 \vskip 10pt
 \setbox\@myboxa\hbox{FIG. \thefigure.\kern 0.65em}
   \@mydim=\wd\@myboxa
 \setbox\@myboxb\hbox{\small #2} \advance\@mydim by \wd\@myboxb
\ifdim \@mydim > \hsize \advance\@mydim by -\wd\@myboxb
  \renewcommand\baselinestretch{1}\small\normalsize
{\parskip=0pt\noindent\unhbox\@myboxa\par}
\vskip-8.5pt
\noindent\@shiftdim=\hsize \advance\@shiftdim by -\@mydim
\hskip\@mydim\vbox{
\hsize=\@shiftdim{\noindent\unhbox\@myboxb}}
\else \hbox to \hsize{\hfil\box\@myboxa\box\@myboxb\hfil}
\fi\vskip-6pt}

\def\tblcaption{\refstepcounter\@captype
  \@dblarg{\@tblcaption\@captype}}
\long\def\@tblcaption#1[#2]#3{\par\addcontentsline{\csname
  ext@#1\endcsname}{#1}{\protect\numberline{\csname
  the#1\endcsname}{\ignorespaces #2}}\begingroup
    \@parboxrestore
    \normalsize
  \@maketblcaption{\csname fnum@#1\endcsname}{\ignorespaces #3}
  \par\endgroup}

\long\def\@maketblcaption#1#2{
 \vskip 10pt
 \setbox\@myboxa\hbox{TABLE \@Roman\c@table.\kern 0.65em}
\@mydim=\wd\@myboxa
 \setbox\@myboxb\hbox{#2} \advance\@mydim by \wd\@myboxb
\ifdim \@mydim > \hsize \advance\@mydim by -\wd\@myboxb
  \renewcommand\baselinestretch{1}\small\normalsize
\noindent\hangindent\@mydim\hangafter=1
\unhbox\@myboxa\unhbox\@myboxb\par
\else \hbox to \hsize{\hfil\box\@myboxa\box\@myboxb\hfil}
\fi}

\def\ps@headings{\let\@\mkboth\markboth
\def\@oddfoot{\rm 
\hfil \rm 
}}
\def\@evenfoot{{\rm 
\hfil \rm 
}}

\def\@evenhead{{\makebox[\textwidth]
{{{\small\bf \thepage}\hfill\footnotesize\rm 
LIGHT-FRONT DYNAMICS AND LIGHT-FRONT QCD
\hfill VOL. 32}}}}

\def\@oddhead{{\makebox[\textwidth]
{{\footnotesize\rm VOL. 32 \hfill {\footnotesize\rm 
WEI-MIN ZHANG
}\hfill {\small\bf \thepage}}}}}
\makeatother

\long\def\@makefntext#1{\parindent 0em
      \noindent \hbox to 0.3em{\hss$^{\@thefnmark}$}#1}

\begin{document}
\baselineskip = 15.5pt
\parskip0\baselineskip
\newskip\eqnskip
\eqnskip 0.5\baselineskip
\def\nB{{\bf B}\hspace{-0.24cm} /}
\def\idd{D\hspace{-0.23cm} /}
\def\ivv{v\hspace{-0.188cm} /}
\def\ipp{p\hspace{-0.188cm} /}
\def\iqq{q\hspace{-0.188cm} /}
\def\iep{\epsilon \hspace{-0.17cm} /}
\def\iva{\varepsilon \hspace{-0.175cm} /}
\def\dv{dv\hspace{-0.28cm}{\raisebox{1.5ex}{$\sim$}}}
\def\uA{A\hspace{-0.225cm}{\raisebox{-0.85ex}{\scriptsize $\sim$}}}
\def\uB{B\hspace{-0.225cm}{\raisebox{-0.85ex}{\scriptsize $\sim$}}}
\def\iz{Z\kern-0.45em Z}
\def\inn{I\kern-0.35em N}
\def\ip{I\kern-0.3em P}
\def\iq{I\kern-0.5em Q}
\def\ir{I\kern-0.35em R}
\def\xx{{\em x\kern-0.35em x}}
\def\ch{\raisebox{.4ex}{$\chi$}}
\def\ds{\displaystyle}
\def\skipback{\vskip -\parskip}

%

\baselineskip=15.5pt
\setcounter{page}{717}
\thispagestyle{empty}

\begin{center}
{\large {\bf
Light-Front Dynamics and Light-Front QCD
}}{\footnote{Based on the lectures presented in ``the Second Workshop
on Particle Physics Phenomenology'', May 19-21, 1994, Kenting, Taiwan.}}
\end{center}
\vspace{0.3pt}

\begin{center}
Wei-Min Zhang \\
\vspace{2pt}
{\small\em
Institute of Physics, Academia Sinica, \\
\vspace{-4pt}
Taipei, Taiwan 115, R.O.C.
} \\
{\footnotesize (Received July 12, 1994) \\
\vspace{-4pt}
(Published in Chinese J. Phys. {\bf 32}, 717 (1994))}
\end{center}
\vspace{10pt}

\baselineskip=15.5pt

\begin{quotation}
\small
In this article we review the basic formulation of light-front field
theory and light-front phenomena in
strong interaction. We also explore various approaches to the
understanding of these phenomena and the associated problems
of hadronic bound states based on QCD (quantum chromodynamics) on the
light-front.
\end{quotation}

\begin{quotation}
\small
\noindent PACS. 11.10.Ef$\;\;\!$ -- Lagrangian and Hamiltonian approach. \\
\noindent PACS. 11.10.Gh -- Renormalization. \\
\noindent PACS. 11.10.$\,$St$\;\;\!\!$ -- Bound state problems.
\end{quotation}
\baselineskip=15.5pt

\footnotetext%
{\footnotesize
\vspace{.05in}
\begin{flushright}
\hspace{5.3cm} \hfill {\small \bf 717} \hfill $\copyright$ 1994 THE PHYSICAL
SOCIETY \\
\hfill OF THE REPUBLIC OF CHINA
\end{flushright}
}

\vspace{0.3cm}

\begin{center}
{\normalsize \bf I. INTRODUCTION:
A HISTORICAL DEVELOPMENT OF LIGHT-FRONT DYNAMICS
AND SOME COMMENTS}
\vspace{3pt}
\end{center}

\hspace{0.3cm} In this article, I will try to address some aspects of
nonperturbative QCD.  The theory of nonperturbative QCD
(at the hadronic scale) has indeed not been well defined at the present
time. Firstly, how much we can understand about the real theory of the strong
interaction from the canonical QCD structure is totally unknown.
Secondly, there are in the literature many models or effective theories
that are inspired by QCD, and yet they cannot be derived directly from
QCD, although they are more or less successful in describing hadronic
phenomenology. Undoubtedly, to discuss the problems of nonperturbative
QCD is not a simple task. Yet the fact that little progress has been
made for the past twenty years on the study of nonperturbative
QCD (i.e., the explicit solution of the dynamics of quarks and gluons
in the strong coupling regions) may force us to think about whether one is able
to extend the traditional approach of the perturbative field theory
to the description of a strong coupling theory.
The covariant perturbative framework lacks the simple Schr\"{o}dinger
picture of quantum mechanics, which may be the obstacle to further
development of a nonperturbative covariant framework. In this
article I am going to introduce a ``new'' dynamical framework,
namely {\em light-front dynamics}, which can overcome the above
obstacle and has many advantages for exploring nonperturbative
field theory{\footnote{One may note that this framework
also has many disadvantages compared to the traditional covariant
framework if we apply it to perturbation theory, as we will
see later.}} that may point to a new direction in
the investigation of the relativistic dynamics of the
strong interaction and nonperturbative QCD in the future.

\hspace{0.3cm} In field theory textbooks {\em light-front dynamics}
may not be a familiar term for most of us. However, this topic is
actually not new at all.  In every step of the
development of strong interaction theory light-front
dynamics has indeed played a crucial, if not the essential,
role for our understanding of hadronic physics. Therefore, before
I list the main problems that I will discuss in this
paper, I would like to
give a brief review of the historical development of
light-front dynamics and the role it played in the study of the
strong interaction and QCD over the past forty-five years
since its invention.  However, I must emphasize that all
I discuss in this article is only a part of the problems
in the strong interaction and in light-front dynamics that
currently interests me.  A complete description of the
theory of the strong interaction and the
light-front field theory is certainly beyond
my knowledge. If the reader is interested in this topic, he or she
may find a relatively complete list of references in [1].

\hspace{0.3cm} Light-front dynamics, simply speaking, is a description of the
evolution of a relativistic system along a light-front direction.

\hspace{0.3cm} The light-front time-space variables, defined as
$x^{\pm}=t\pm x$
where $x$ is one of the three components of the space variables, were
indeed invented a long time ago, even before the birth of modern
physics. They have been widely
used in solving various problems in mathematical physics for the last
hundred years. A typical example that one might remember from an
undergraduate course is how to solve the wave equation (for simplicity,
we only consider one-time and one-space variable for illustration
here):
\def\theequation{1.1}
\begin{equation}
  \biggl ( \ds\frac{\partial^2}{\,\partial t^2\,}-v^2
  \ds\frac{\partial^2}{\,\partial x^2\,}\biggr ) \Psi (t,x) = 0 \, ,
\end{equation}
where $v$ is the velocity which is a constant. It is
well-known that, to solve the above equation, it is most
convenient to introduce the new variables: $\xi =t+x/v$
and $\eta =t-x/v$. With these variables, the wave equation is
reduced to $(\frac{\partial^2}{\,\partial \xi \partial \eta\,})\Psi =0$ and its
solution must have the form $\Psi (t,x)=f(t+x/v)+g(t-x/v)$. Physically, this
solution indicates that the system moves
along the line $t=\pm x/v$ with the velocity $v$ in the $t-x$ space.
The light-front variables $x^\pm$ correspond to the case where the
velocity $v=c$ (the speed of light) with the unis $c=1$.  In such a
case, we say that the system moves on the null-plane.

\hspace{0.3cm} The physical implication of the use of the light-front
coordinates can indeed be manifested from the solution of the above wave
equation.  First, we consider a small $v$ (compared to the speed of light)
in Eq. (1.1) which, physically, corresponds to nonrelativistic
dynamics.  In this situation, the system moves along the line $x=\mp vt$ which
does not deviate from the line $x=0$ very much (see Fig. 1a).
As a result, it may be convenient to describe the evolution of the
system in terms of the usual time variable $t$ as the time direction.
Such a choice of the time direction is the one people are most familiar
with because this is the world we live in now.   However, if
$v\sim c$ (corresponding to relativistic dynamics), the system
indeed moves near the line $x^+ = t + x = 0$ or $ x^- = t - x =0$
in the $t-x$ plane (see Fig. 1b).
In this case, if we still use the variable $t$ as the time direction
along which the system evolves, the phenomena
we measure in laboratories must behave very differently from the
picture of nonrelativistic dynamics that we intuitively think
about. Yet, if we
choose a new ``time'' direction, namely in terms of the light-front
variable $x^+$ (or $x^-$), structurally the time evolution of a
relativistic system may look very similar to a nonrelativistic system
in the ordinary time-space coordinate system.  Consequently, with the
light-front variables, relativistic dynamics may become simple
and transparent.

\hspace{0.3cm} Forty-five years ago
Dirac first introduced the concept of the light-front form for
relativistic dynamics, which he called the front form. He defined it
as follows: the three-dimensional surface in space-time formed by a
plane wave front advancing with the velocity of light, for
example $x^+=t+x^3=0$, is called a {\em front}, where $x^3$ is the
third component in the three-dimensional space.  For a dynamical
theory, a form of dynamics
which is associated with the sub-group of the inhomogeneous Lorentz group that
leaves the front invariant may appropriately be called the {\em front
form} [2]. The front form is now commonly called the
light-front framework or simply the light-front. In the literature,
it is sometimes also called the null-plane.

\hspace{0.3cm} Dirac's attempt to introduce the light-front form was to look
for
the most convenient Hamiltonian formulation for a relativistic system.
A relativistic system is required to$\,$ satisfy

\begin{figure}[hptb] 
\vspace{6cm}
\figcaption{
The illustration of light-front dynamics.
}
\end{figure}

\pagebreak

\noindent
two principles:
it is invariant under infinitesimal inhomogeneous Lorentz
transformations and it is a Hamiltonian system.
The first requirement arises from the relativistic principle
and the second is given by quantum mechanics.
Then a dynamical system is characterized by the ten generators, $P^{\mu},
M^{\mu \nu}$, of the Poincare group, which are determined by the
Hamilton equations (Heisenberg equations in quantum mechanics).
Based on this philosophy,
Dirac found that there are three possible forms for relativistic dynamics:
i). The instant form of dynamics: the dynamical variables referring to
physical conditions at some instant of time.{\footnote{An instant is a flat
three-dimensional surface containing only directions which lie outside
the light-front for example, $t=0$.}} In this case the
subgroup that leaves an instant invariant are the translations and
rotations while the dynamics is determined by $P^0$ and the boost
operators $M^{0i}$. According to Dirac, the instant form has the
advantage of being the one people are most familiar with, and, that is
all. ii). The point form of dynamics: a form of dynamics associated with
the subgroup of the inhomogeneous Lorentz
transformations that leaves a point, say the origin $x_{\mu}=0$, invariant.
In this case, the homogeneous Lorentz transformation is simple (kinematic)
but all the four-vector momenta $P^{\mu}$ are dynamically dependent. The
advantage of the point form is perhaps the dynamical separation between
$P^{\mu}$ and $M^{\mu \nu}$.
iii). The front form of dynamics that has been defined in the previous
paragraph. The front form has the advantage that of the ten generators of
the Lorentz group, only three (instead of four in the other forms)
are dynamically dependent (as we will discuss later in detail). Furthermore,
there is no square root for the
Hamiltonian, which might simplify the dynamical structure since a square
root relation between the energy and the momentum cannot provide
a simple picture of the bound state Schr\"{o}dinger equation.

\hspace{0.3cm} At the time Dirac proposed the light-front Hamiltonian
dynamics modern quantum field theory underwent a revolutionary change,
namely a covariant perturbative field theory was established
[3-5]. This formally beautiful
and practically useful theory has been used so successfully in the
description of all the known physical phenomena
in quantum electrodynamics (QED).
At that time, Dirac's ideas about the front
form of relativistic dynamics were completely ignored. This should not
surprise us because, if
we can formulate and compute
everything in a covariant form, we never need to  dynamically solve
the Poincare algebra.  This was perhaps the main reason why the above mentioned
ideas of Dirac have not recieved enough attention
since their invention.  Feynman's covariant formulation has dominated
the investigation of field theory for the past forty-five years.
Undoubtedly it has so many advantages for our understanding of the
microscopic processes with such a novel picture and provides a tremendous
simplification of the practical computations.
However, one might recall, that, since
it was invented covariant field theory has achieved little progress
for nonperturbative theory, specially for the relativistic bound
state problem.  More precisely speaking, the quantum field theory
we use today may be better to be explicitly called the perturbative
field theory. Nonperturbative theory requires a complete
description of the whole physical state space.  Yet, in the
covariant form, in order to keep the covariance, one has to
work in a state space that is much bigger than the original physical
state space. This makes the nonperturbative picture
physically unclear and computationally much more complicated
within the framework of the covariant form.

\hspace{0.3cm} Note that, even for perturbation theory, the covariant formalism
involves too many unphysical processes in
practical computations in order to keep its covariant form.
A typical example from the early days of field theory is
the Z-graph (see Fig. 2) which occurs in the amplitude of various fermion
scattering processes. This diagram contains a
state that violates the Pauli principle that cannot be explicitly excluded in
a covariant calculation. In principle, the contribution of
this unphysical state is cancelled by the contribution of the
same unphysical state involved in the vacuum diagram which is, however,
factorized out in the calculation.  In order to avoid such subtle
problems, Weinberg developed a new formulation for perturbative field
theory in the infinite momentum frame in 1966 [7],
which is equivalent to the field theory quantized
on the light-front.{\footnote{It must be pointed out that the infinite
momentum frame itself is not the same as the light-front. The former
is a special Lorentz frame while the later is a coordinate system
of space-time in which all Lorentz frames, from the rest
frame to the infinite momentum frame, can be described.}}
 In the infinite momentum (as well as in the light-front)
perturbation theory, all diagrams that involve particle production from
the vacuum must vanish, thereby the Z-graphs are automatically
excluded.{\footnote{Strictly speaking, this statement is only
conditionally true which we will discuss later.}}

\hspace{0.3cm} In fact, the development of perturbative field theory in the
infinite momentum frame given by Weinberg was motivated very much
by the works of Fubini and Furlan [8] and by Dashen
and Gell-Mann [9] on the so-called fixed mass sum rules
derived from current algebras.
{}From the current algebras of the vector and axial vector currents,
one can easily derive many sum rules for various current matrix
elements that can be measured in experiments.  However, in an arbitrary
frame, these sum rules may mix the space-like and time-like processes
and depend on the intermediate state energies so
that the momentum transfer in these matrix elements in the sum rules
varies term by term.  As a result, these sum rules are difficult to
be used to test theory with experiments.  Fubini and Furlan found that
if one boosts the target into the infinite momentum frame, the above
difficulty is solved and the sum rules are reduced to those with a fixed
momentum transfer ($q^2$), namely the fixed-mass sum rules.

\begin{figure}[hptb] 
\vspace{3.2cm}
\figcaption{
The Z-diagram.
}
\vspace{0.6cm}
\end{figure}

\hspace{0.3cm} Meanwhile, the sum rules become very useful if they are
approximately satisfied by some set of single particle and resonant
intermediate states that provide us with a large number of algebraic
relations among coupling constants and form factors.
In the infinite momentum frame the nonzero matrix elements of currents
should span an $SU(6)$ representation, as was noticed by Dashen and
Gell-Mann [9]. Thus, the main contribution to the sum
rules comes mostly from the single particles and resonance hadronic
states that are classified by the $SU(6)$ spectrum generating
algebra (the current quark model).

\hspace{0.3cm} After the above early application to the study of the strong
interaction,
the most exciting progress made for light-front dynamics was
perhaps the parton phenomena observed in deep inelastic scattering
in the very late 60's. The parton picture, as everyone knows, was
introduced by Feynman [10,11] for interpreting the
scaling behavior discovered by Bjorken [12] from the deep
inelastic scattering experiments.  Although most of us will immediately
think of the partonic picture in the infinite momentum frame, when we talk
about parton phenomena, it is in fact much easier to see its theoretical
underpinning on the light-front.  The cross
section for deep inelastic scattering is proportional to the so-called
hadronic tensor which is a Fourier transformation of the hadronic
matrix element of the commutator of two currents:
\def\theequation{1.2}
\begin{equation}
  W^{\mu \nu}(P,q)= \ds\frac{1}{\,4\pi\,}\ds\int d^4xe^{-iqx}\bigl \langle
  PS|[j^{\mu}(x),j^{\nu}(0)]|PS \bigr \rangle \, .
\end{equation}
In the rest (Lab) frame with high energy and high momentum
transfer where the transfer momentum is assumed to be
along the $z$-direction, it is easy to check that
the dominant contribution in the above integral comes from the
region $x^+=t+z\sim 0$. In other words, in high momentum
processes, the main contribution to the hadronic tensor comes
from the light-front commutator of two currents. On the other hand,
causality, $x^2=x^+x^--x^2_{\bot}\geq 0$, shows that the
light-front commutator is not vanishing only if $x_{\bot}=0$.
Obviously the commutator of the two currents on the light-front
can only depend on $x^-$, the light-front longitudinal
variable, from which one can easily derive the Bjorken scaling
that the structure functions extracted from the hadronic tensor
are independent of the hadronic mass scale.  Gross and Treiman
further proved that the dominant contribution to the hadronic tensor
is from the most singular part of the light-front commutators which
is mainly given by the free theory [13]. This provides a theoretical
foundation for Feynman's parton model.  Therefore one usually
refers to the parton phenomena the light-cone dynamic of the strong
interaction.

\hspace{0.3cm} The manifestation of light-front dynamics from the parton
picture in deep inelastic scattering led to many attempts to reformulate
field theory on the light-front in the very early 70's. Typical works
are by Drell, Levy and Yan [14-16]
and by Bjorken, Kogut and Soper [17], who attempt to
derive the parton structure directly from field theory [18].
Meanwhile, Chang and Ma began to study the canonical structure of field
theory on the light-front [19]; Leutwyler {\em et al}. attempted
to develop a light-front quantization approach to field theory [20];
Kogut and Soper reformulated QED
on the light-front [21]; Rohrlich {\em et al}. explored the relation
of the initial value problem with boundary conditions on the
light-front for QED [22,23], and
Chang and Yan extensively discussed various field theories
quantized on the light-front using Schwinger's action principle
and compared the light-front noncovariant perturbative structure
to the traditional covariant formulation [24-27].
Also there were many
other concerns in the light-front field theory at that
time (for examples, see [28,29]).
These investigations provide a basis for conventional canonical
light-front field theory.  However, the nonperturbative aspects
of field theory on the light-front had not been addressed
at that time.

\hspace{0.3cm} On the other hand, the parton phenomena also led to further
extensive exploration of current algebras on the light-front
and the associated sum rules.  For example, the fixed-mass
sum rules in the infinite momentum frame
can be easily and naturally obtained from the light-front
current algebras, which were provided by Jackiw and his
collaborators [30,31].  It was also found that
with the light-front current algebras some anomaly structures,
such as Schwinger terms, may be easy to address [32,27].
The most important development for strong interaction along
this line is
the development of a fundamental theory for the strongly
interacting basic constituents (quarks and gluons), i.e., QCD.
Now everybody believes that QCD is the only acceptable
fundamental theory for the strong interaction, but
maybe not everyone knows that QCD was initially formulated
by Gell-Mann and Fretsch [35] {\em on the light-front}.
At the time many particle theorists focused on the field theory
interpretation of the parton phenomena, Gell-Mann and many others
still concentrated on looking for a possible fundamental theory for
the quark model which is very successful in the description of the
classification of hadronic states. In the current quark model,
an essential
equation to determine the dynamics of the model is the
quark equation of motion on the light-front.
Gell-Mann and Fretszh wrote down the quark equation
of motion exactly, based on $SU(3)$ gauge theory on the
light-front with the light-front gauge, in their paper of 1972.

\hspace{0.3cm} Of course, QCD is accepted
as the fundamental theory
of strong interactions due mainly to the discovery of
its asymptotic freedom behavior by Gross and Wilczek
and by Politzer [36,37],
which provides the theoretical basis for the parton
phenomena in high energy physics.  Asymptotic freedom
allows us to use perturbation theory to do a QCD
calculation for physical processes near the light cone
$x^2\sim 0$ (the common
statement that perturbative QCD is for short distance
dynamics is indeed a misleading comment).  Obviously,
the natural framework for exploring the parton phenomena
with QCD is the formulation of QCD on the light-front
with the light-front gauge. The remarkable properties
of the light-front formulation are that
the conventional vector and axial vector
currents of quarks remain gauge invariant in QCD, and the concept
of parton distribution functions naturally occurs [38].

\hspace{0.3cm} The development of a light-front formulation for QCD is as
follows. Tomboulis first quantized Yang-Mills theory on the light-front
in 1973 [39].  Cornwell and Crewther discussed possible
singularites arising from the choice of light-front gauge [40,41].
The first QCD calculation with the light-front gauge was given
by Gross and Wilczek in 1974 [42], who used it to
check the gauge dependence of the one-loop anomalous dimension
and the $\beta$ function.  The canonical quantization of QCD
on the light-front was provided by Casher [43]
and by Bardeen {\em et al}. [44,45] in 1976.  Thorn addressed
the problems of asymptotic freedom$\,$ and$\,$ the$\,$ confinement$\,$
mechanism
$\,$ within$\,$ light-front$\,$ QCD$\,$ in$\,$ the$\,$ late$\,$ 70's$\,$ [46].
Lepage and Brodsky developed the Hamiltonian
formulation of light-front QCD
and explored the asymptotic behavior of hadrons and the perturbation
expansion of various exclusive processes
[47-49]. Franke {\em et al}. first addressed
the problem of light-front gauge fixing and the associated nontrivial
QCD structures [50]. G{\l}azek
discussed the quark condensate in light-front QCD [51].
Meanwhile, due to the scaling evolution of the hadronic
structure functions suggested by Altarelli and Parisi
[52], a systematic development of
the factorization theorem for various inclusive processes
and a practical calculation scheme of the various anomalous
dimensions within light-front QCD were also provided in the early
80's [53-55]. Note that the factorization
theorem for deep inelastic scattering was initially proven by
the use of operator product expansions which are only valid
near the light-cone [56]. In the mid 80's, there
were also a lot of discussions about the trouble caused by
the light-front gauge singularity and its possible solutions in perturbation
theory [57-60].
In conclusion, due to the light-cone behavior of
high energy dynamics, perturbative light-front QCD,
as the foundation of the parton picture, has been widely used
in various investigations of high energy processes.

\hspace{0.3cm} Here I shall make some comments.
Because of the two conventional requirements of field
theory, namely, covariance and unitarity, the development of
non-abelian gauge theory with gauge fixing  is not at all simple.
In the usual covariant path integral quantization,
the unphysical gauge components violate unitarity and
therefore one has to introduce ghost fields to cancel the
unphysical states in the theory [61].  A covariant
formalism has several advantages for many aspects, yet the physical
picture becomes obscure once unphysical particles are introduced.
On the other hand, in canonical quantization one has to choose
a physical gauge, such as the axial gauge or light-front gauge, where
unitarity is automatically satisfied but covariance is
no longer manifest.  Although we know that the natural picture
of QCD for parton phenomena is based on the light-front gauge,
many still prefer
to use the covariant formulation. The main complaint against
the light-front gauge is that the usual prescription (principal
value prescription) for
the gauge singularity prohibits any continuation
to Euclidean space (Wick rotation) and hence power counting
for Feynman loop integrals. As a result, non-local counterterms
have to be introduced, which break the multiplicative
renormalizability [41,59].

\hspace{0.3cm} However, even in covariant perturbation theory, formal
multiplicative renormalizability may not
be significantly useful in the calculation of hard scattering
coefficients.  This is because gauge invariant composite operators
in the operator product expansion are not multiplicatively
renormalizable [62], and the gauge
variant operators will induce gauge invariant counterterms that do
contribute to gauge invariant matrix elements.  Although it was
believed that the mixing problem had been solved [63],
the long-standing discrepancy between the covariant and
light-front gauge calculations for the second-order anomalous
dimension of the gluon composite operators has recently renewed
attention to the renormalizability of QCD
[64,65]. In fact, the mixing of gauge
invariant operators is an intrinsic property that exists in any
covariant formulation of QCD.  The above mixing problem of gauge
invariant operators apparently disappears in the physical gauges and
perturbative calculations become straightforward.  However, a new
spurious mixing of ultraviolet and infrared divergences associated
with the gauge singularity does occur in the light-front gauge using the
principal value prescription.
One has to be careful to see whether or not the spurious mixing
affects physical quantities. In the light-front gauge,
it has been shown that the most severe
infrared divergences caused by the $A_a^+=0$ gauge choice with
the principal value prescription are
indeed cancelled in the gauge invariant sectors [53].
Thus, in the perturbation region, the light-front gauge singularity in QCD may
not be a severe problem.

\hspace{0.3cm} Although QCD manifests asymptotic freedom, the main feature
of QCD as the fundamental theory of the strong interaction
should lie in its nonperturbative behavior.  In the past twenty
years, there has been tremendous work on nonperturbative
QCD, and some nontrivial structures, such as the instanton
[66], were revealed.
Unfortunately, the real picture of nonperturbative QCD
that leads to quark confinement and chiral symmetry breaking,
which are believed to be the two essential mechanisms for form hadron
formation from QCD, is still lacking.
Roughly speaking, all methods that we use today for exploring
nonperturbative QCD were mainly developed by Wilson in the late 60's
and the early 70's.  These are the lattice gauge theory [67],
operator product expansions [56] and the renormalization group
approach [68,69].  The lattice
gauge theory is the only computable approach in the market for
the nonperturbative study of QCD.  In the past one and a half decades,
extensive work on lattice gauge theory has led to significant
progress in many research avenues; however, accurate information
on the bound state of light quarks is still not available [70].
On the other hand, a phenomenological approach of nonperturbative
QCD, namely QCD sum rule [71], has become a reliable approach for
various hadronic structures.  The theoretical basis of QCD
sum rule is the operator product expansion which
separates the perturbative and nonperturbative contributions of
QCD in hadrons near the light-cone.  In fact, what one has
learned from QCD sum rules about QCD is nothing more than the
perturbative part of QCD.  The nonperturbative aspect in
QCD sum rules is still purely phenomenological, including
the parameterization of the quark and gluon condensate (which
must be directly calculated from QCD for a full understanding
of nonperturbative QCD but so far no one knows how to compute them).

\hspace{0.3cm} Succinctly speaking, the aim of nonperturbative QCD is to find
the hadronic bound states directly from QCD.  However,
the only truly successful approach to bound states in field theory has
been Quantum Electrodynamics (QED), with its combination of
nonrelativistic quantum mechanics to handle bound states and
perturbation theory to handle relativistic effects
[72-75].  Lattice Gauge
Theory is maturing but has yet to rival QED's comprehensive success.
There are three barriers which prohibit an approach to Quantum
Chromodynamics (QCD) that is analogous to QED.  They are: (1)
the unlimited growth of the running coupling constant $g$ in the
infrared region, which invalidates perturbation theory; (2)
confinement, which requires potentials that diverge at long distances
as opposed to the Coulombic potentials of perturbation theory; (3)
spontaneous chiral symmetry breaking, which does not occur in
perturbation theory.  The last two barriers are essentially
related to the nonperturbative structure of the QCD vacuum.

\hspace{0.3cm} In contrast to the gloomy picture of the strong interaction in
QCD, however, is that of the Constituent Quark Model (CQM),
the most successful model in the description of hadronic
dynamics.  In CQM, only a minimum number of constituents
required by the symmetries is used to build each hadron
[76].  Moreover, Zweig's rule leaves little role for the
production of extra constituents [77].  Instead,
the rearrangement of pre-existing constituents dominates the physics [78-80].
Yet the CQM has never been reconciled with QCD --- not even
qualitatively.  Note though that QCD was initially proposed on the
light-front in order to develop a fundamental theory for the quark
model [35]. In 1989, Wilson suggested that QCD on the
light-front may provide a natural framework to solve QCD
nonperturbatively for hadrons because the use of the light-front
coordinates may allow us to develop the same techniques that solve
QED for solving the physics of CQM [81]. To do so,
he proposed a new idea
for constructing the low-energy QCD theory for hadrons based on
light-front power counting which differs from what we are familiar
with in the equal-time framework.  Combined with the
two-component formulation of the
light-front QCD [82] and a new scheme for Hamiltonian
renormalization [83,84], it seems to provide a
newer avenue for exploring nonperturbative QCD dynamics and hadrons
than that used for solving QED.  This has been presented
in a recent paper by Wilson, Walhout, Harindranath, Zhang, Perry,
and Glazek [85].

\hspace{0.3cm} At this moment, I cannot predict whether this new approach will
give the final solution for QCD or whether this approach will
be successful for a strong coupling theory.  Light-front dynamics
may shorten the distance from our hopes for QCD to a real
picture of the hadronic world.  What I can comment on is that in
the past twenty years various phenomenological understandings
of nonperturbative QCD, such as the constituent quark
model for the heavy quark system, the heavy quark symmetry
for the heavy-light quark system and the effective chiral
theory for the light quark system, could guide us in the right
direction.  Optimistically, we should believe that any new
attempt inspired by a successful understanding of phenomenology
always contains a possibility of providing a real answer to
the unsolved problems.

\hspace{0.3cm} After this long historical overview of the development of
light-front dynamics and light-front QCD in terms of my personal
point of view, I now give an outline of this review.
In the next section, I will present a brief description of the
basic structure of the light-front form and the free field
theory on the light-front.  Then in Sec. III, I will discuss
the light-front phenomena from which I hope  a picture is provided
of what we really need to understand from QCD on the light-front
for strong interaction phenomena, i.e.
the light-front bound states for hadrons.  In Sec. IV, I
will discuss the general structure of light-front bound states
and their advantages in describing hadronic physics,
and some phenomenological light-front wave functions
for hadrons that have been used for successfully describing
many hadronic properties.  In Sec. V, as a
basis for exploring the light-front dynamics of QCD,
canonical light-front QCD and its various properties and
problems are discussed.  In Secs. VI and VII, I will introduce
new ideas that were originally proposed by Wilson and his
collaborators for finding light-front QCD bound states
for the hadrons, and I will also discuss possible practical schemes
to realize these ideas and thereby to provide a possible
real theory of QCD for hadrons.

\vspace{6pt}
\begin{center}
{\normalsize \bf II. LIGHT-FRONT FORM}
\end{center}
\vspace{3pt}

\vspace{6pt}
\noindent
{\bf II-1. Lorentz Transformations on the Light-Front}

\hspace{0.3cm} One of the useful properties in light-front formulation of
relativistic dynamics is its convenient
Lorentz transformation structure [2].  In this section,
we shall introduce some basic properties of the Lorentz transformation on
the light-front.

\hspace{0.3cm} The light-front coordinates of space-time $(+,-,-,-)$ are
defined as
\def\theequation{2.1}
\begin{equation}
  x^{\pm} = x^0\pm x^3\, , \quad \quad x_{\bot}^i = x^i\, , \quad \quad (i=1,2)
\, ,
\end{equation}
where we choose $x^+$ as the ``time'' direction along which the system
evolves.  Thus $x^-$ and $x_{\bot}$ naturally become the longitudinal
and transverse coordinates.  All other 4-vectors in space-time
are defined in the same form as Eq. (2.1).  The inner product of any
two four-vectors is given by
$a_{\mu}b^{\mu}=\frac{1}{\,2\,}(a^+b^-+a^-b^+)-a_{\bot}\cdot b_{\bot}$, where
the metric tensor for Eq. (2.1)
is $g^{+-}=g^{-+}=1/2$, $g^{ij}=-\delta^{ij}$ and the other components
are zero.  The light-front time and space
derivatives are $\partial^-=2\frac{\partial}{\,\partial x^+\,}$,
$\partial^+=2\frac{\partial}{\,\partial x^-\,}$,
$\partial^i=\frac{\partial}{\,\partial x_i\,}=-\frac{\partial}{\,\partial
x^i\,}$
(see the definition
of the metric tensor), and the four-dimensional volume element
$d^4x=\frac{1}{\,2\,} dx^+dx^-d^2 x_{\bot}$.

\hspace{0.3cm} {\em 1. Poincar\'{e} algebra}.
A relativistic dynamical system must be Lorentz invariant. The
inhomogeneous Lorentz transformation is generated by the Poincar\'{e}
algebra:
\def\theequation{2.2}
\begin{equation}
\begin{array}{l}
  [P^{\mu},P^{\nu}] = 0, \quad [M^{\mu\nu},P^{\rho}] = i (
  -g^{\mu \rho} P^{\nu}+g^{\nu \rho}P^{\mu})\, , \\[\eqnskip]
  [M^{\mu \nu},M^{\rho \sigma}]=i(-g^{\mu \rho}M^{\nu \sigma}
  + g^{\nu \rho} M^{\mu \sigma}-g^{\mu \sigma}M^{\rho \nu}
  + g^{\nu \sigma} M^{\rho \mu}) \, ,
\end{array}
\end{equation}
where $P^{\mu}$ is the four-vector momentum, and $M^{\mu \nu}$ describes
the rotational and boost transformations. In the instant form, the
rotational and boost operators are given as $M^{ij}=\epsilon_{ijk}J^k$
and $M^{0i}=K^i$, respectively, with
\def\theequation{2.3}
\begin{equation}
  [J^i,J^j] = i\epsilon_{ijk} J^k\, , \quad [J^i,K^j] = i\epsilon_{ijk}
  K^k\, , \quad [K^i,K^j]=-i\epsilon_{ijk}K^k \, .
\end{equation}

\hspace{0.3cm} In field theory these generators are constructed from the
energy momentum density $\Theta^{\mu \nu}$.  For a given Lagrangian ${\cal L}
(\phi_i)$, the energy-momentum tensor in the instant form is
defined by{\footnote{If ${\cal L}$
contains gauge fields, we must add to $\Theta^{\mu \nu}$ an additional
term $\partial_{\rho}(F^{\mu \rho} A^{\nu})$ to ensure that it is gauge
invariant, symmetric and traceless.}}

\def\theequation{2.4}
\begin{equation}
  \Theta^{\mu\nu} ={ \partial {\cal L} \over \partial(\partial_{\mu}
  \phi_i)} \partial^{\nu} \phi_i - g^{\mu \nu} {\cal L} \, .
\end{equation}
Then,
\def\theequation{2.5}
\begin{equation}
  P^{\mu} = \ds\int d^3 x \Theta^{0\mu}\, ,\quad M^{\mu \nu} = \ds\int d^3 x
  [x^{\mu} \Theta^{0\nu} - x^{\nu} \Theta^{0\mu}] \, .
\end{equation}
In light-front coordinates, the four-vector momentum operators
and the boost and rotational operators are defined by:
\def\theequation{2.6}
\begin{equation}
\begin{array}{lll}
  P^{\pm} & = & P^0 \pm P^3 \, , \quad P_{\bot}^i = P^i \quad (i=1,2) \, ,
\\[\eqnskip]
  E^1 & = & K^1 + J^2\, ,\quad E^2 = K^2 - J^1\, ,\quad K^3 \, , \\[\eqnskip]
  F^1 & = & K^1 -J^2\, , \quad F^2 = K^2 + J^1\, , \quad J^3 \, .
\end{array}
\end{equation}
In terms of the energy-momentum density,
$K^3=-\frac{1}{\,2\,}M^{+-}$, $E^i=M^{+i}$ and $J^3=M^{12}$, $F^i=M^{-i}$,
where
\def\theequation{2.7}
\begin{equation}
  P^{\mu} = \ds\frac{1}{\,2\,}\ds\int dx^-d^2x_{\bot} \Theta^{+\mu}\, , \quad
  M^{\mu\nu} = \ds\frac{1}{\,2\,} \ds\int dx^-d^2x_{\bot} [x^{\mu}
  \Theta^{+\nu} - x^{\nu} \Theta^{+\mu}]  \, .
\end{equation}
Physically, the light-front transverse boost ($E^i$) is a combination
of a pure Lorentz boost and a usual rotation, and so is the
light-front transverse rotation ($F^i$).

\hspace{0.3cm} {\em 2. Boost transformation}.
The most useful property of the Lorentz transformation on the
light-front is the kinematics of its boost transformation.  The
subgroup that leaves the light-front $x^+ = 0$ invariant is the
translations $(P^+,P^1,P^2)$ and the Lorentz transformations
generated by $J^3$, $K^3$, $E^i$. A boost transformation,
\def\theequation{2.8}
\begin{equation}
  L(\beta ) = \exp\{-i\beta_i E^i\} \exp\{-i\beta_3 K^3\} \, ,
\end{equation}
transforms the light-front momenta operator as follows:
\def\theequation{2.9}
\begin{equation}
  L(\beta ) P^+ L^{-1}(\beta ) = P^+ e^{\beta_3}\, , \quad \quad
  L(\beta ) P_{\bot} L^{-1}(\beta ) = P_{\bot} + P^+ \beta_{\bot}
  e^{\beta_3}\, .
\end{equation}
Consider a state in the rest frame: $|P\rangle$, where $P=(M,0,0,0)$
in the usual instant coordinates. The light-front momentum
$P^+=M$, $P_{\bot}=0$. Let the transformation parameters be
$\beta_{\bot}=p_{\bot}/p^+$, $\beta_3=\ln (p^+/M)$, then we have
\def\theequation{2.10}
\begin{equation}
  L(p)|P\rangle =|P'\rangle \, , \quad \quad P'^+ = p^+, ~P'_{\bot}
  = p_{\bot} \, ,
\end{equation}
which boosts the state in the rest frame to the frame with momentum $P'=p$.
The boost for a state with an arbitrary momentum $A$
is given by
\def\theequation{2.11}
\begin{equation}
  L(P)|A\rangle =|A'\rangle \, , \quad \quad A'^+ = (P^+/M) A^+ \, ; \quad
\quad
  A'_{\bot}=A_{\bot} + (A^+/M)P_{\bot} \, .
\end{equation}

\hspace{0.3cm} The remaining three generators of the Poincare algebra do not
keep the light-front plane invariant.
The operator $P^-$ transforms the plane $x^+=0$ to the plane $x^+=\tau$
while $F^i$ rotates the plane $x^+=0$ around the surface of the light-cone
$x^2=0$. In other words, once we specify the initial data on the
light-front $x^+=0$, the three operators, $P^-$ and $F^i$, determine the
evolution
of the system away from $x^+=0$.  Obviously, $x^+=\tau$ is the
evolution parameter, namely the light-front time variable.

\vspace{15.5pt}
\noindent
{\bf II-2. Basic Formulation of Light-Front Field Theory}

\hspace{0.3cm} In this section we discuss some basic light-front structures in
the
field theory by looking at the canonical light-front quantization for
a free scalar, a free fermion and a free gauge theory.  The motivation
is to show how light-front field theory differs from instant
field theory.

\hspace{0.3cm} {\em 1. Scalar field}.
Consider a free scalar field theory,
\def\theequation{2.12}
\begin{equation}
  {\cal L} = \ds\frac{1}{\,2\,} (\partial_{\mu} \phi \partial^{\mu} \phi
  - m^2 \phi^2) \, ,
\end{equation}
the equation of motion on the light-front is given by
\def\theequation{2.13}
\begin{equation}
  (\partial^- \partial^+ - \partial^2_{\bot} + m^2 ) \phi = 0 \, .
\end{equation}
The light-front canonical theory can be obtained by defining the conjugate
momentum of $\phi$ referred to the light-front time $x^+$:
\def\theequation{2.14}
\begin{equation}
  \pi_{\phi}(x) = \ds\frac{\partial{\cal L}}{\,\partial (\partial^- \phi )\,}
  = \ds\frac{1}{\,2\,}\partial^+ \phi (x)\, .
\end{equation}
It is worth noting that $\pi_{\phi}$ is not a light-front time
derivative of $\phi$, which implies that the light-front conjugate
momentum is not an independent dynamical degree of freedom.
The equal light-front
time canonical commutation relation is then given by{\footnote{Since the
conjugate momentum is not an independent dynamical degree of freedom,
the light-front dy-namical system is always a constrained Hamiltonian
system for which the naive canonical quantization is generally not valid,
and one should use the Dirac quantization procedure [86]. However,
here the con-straints can simply be solved so that the naive canonical
approach, or more generally the phase space quantization approach
is good enough to provide the basic commutation relations without any
loss of the generality [87,88].}}

\def\theequation{2.15}
\begin{equation}
  \bigl [ \phi (x),\pi_{\phi} (y)\bigr ]_{x^+=y^+} = i \ds\frac{1}{\,2\,}
  \delta (x^- - y^-) \delta^2(x_{\bot} - y_{\bot})\, .
\end{equation}
If we define
\def\theequation{2.16}
\begin{equation}
  \biggl ( \ds\frac{1}{\,\partial^+\,} \biggr ) f(x^-)
  = \ds\frac{1}{\,4\,}\ds\int_{-\infty}^{\infty}dx^{'-}\epsilon (x^- - x^{'-})
  f(x^{'-})\, ,
\end{equation}
where $\epsilon (x)=1,0,-1$ for $x>0, =0, <0$, then
\def\theequation{2.17}
\begin{equation}
  \bigl [ \phi (x),\phi (y)\bigr ]_{x^+=y^+} = - \ds\frac{i}{\,4\,}
  \epsilon (x^--y^-)\delta^2(x_{\bot} - y_{\bot}) \, .
\end{equation}

\hspace{0.3cm} The light-front Hamiltonian of the free scalar field is given by
\def\theequation{2.18}
\begin{equation}
\begin{array}{lll}
  H_{LF}=P^- & = & \ds\int dx^- d^2 x_{\bot} (\pi_{\phi} \partial^- \phi -
  {\cal L}) \\[1.5\eqnskip]
  & = & \ds\int dx^- d^2 x_{\bot}\ds\frac{1}{\,2\,}
  \bigl ( (\partial_{\bot} \phi)^2 + m^2 \phi^2\bigr ) \, .
\end{array}
\end{equation}
A solution of the free scalar field on the light-front in terms of
momentum space is
\def\theequation{2.19}
\begin{equation}
  \phi (x) = \ds\int \ds\frac{dk^+ d^2 k_{\bot}}{\,2(2\pi)^3 [k^+]\,}
  \bigl [ a(k)e^{-ikx} + a^{\dagger} (k) e^{ikx}\bigr ] \, , \quad \quad
  k^+ \geq 0 \, ,
\end{equation}
with the commutation relation:
\def\theequation{2.20}
\begin{equation}
  \bigl [ a(k),a^{\dagger}(k')\bigr ] = 2 (2\pi)^3 k^+ \delta (k^+
  -k^{'+})\delta^2(k_{\bot} - k'_{\bot}) \, .
\end{equation}
In momentum space the light-front Hamiltonian becomes
\def\theequation{2.21}
\begin{equation}
  H_{LF} = \ds\int \ds\frac{dk^+ d^2 k_{\bot}}{\,2(2\pi)^3[k^+]\,}
  \ds\frac{\,k^2_{\bot}+m^2\,}{[k^+]} a^{\dagger}(k) a(k)\, ,\quad \quad
  k^+ \geq 0 \, ,
\end{equation}
where an infinite zero-point energy has been ignored. Thus, the light-front
energy $k^-$ for the scalar physical state $|k\rangle =a^{\dagger}(k)|0\rangle$
is
\def\theequation{2.22}
\begin{equation}
  H_{LF}|k\rangle = k^-|k\rangle \longrightarrow
  k^- = \ds\frac{\,k^2_{\bot}+ m^2\,}{[k^+]} \, .
\end{equation}
This is the light-front energy dispersion relation, a result of
$k^2=k^+k^- - k_{\bot}^2=m^2$.

\hspace{0.3cm} The motivation for presenting such a detailed discussion of the
free scalar theory here is to show the difference between
the light-front field theory and the instant
field theory. There are three major differences we can see from
the above discussion.  i) On the light-front, the canonical
conjugate momentum is not a dynamical
degree of freedom.  In other words, the number of the dynamical degrees
of freedom is reduced by half compared to the
instant formulation.  As a result, the scalar field variables themselves
do not commute with each other on the equal light-front time plane (see
Eq. (2.17)).
ii) The energy dispersion relation does not involve a square root
structure which may help us to formulate the relativistic bound states
in terms of a simple nonrelativistic-like Schr\"{o}dinger picture.
iii) There is a new singularity in the light-front formulation which exists
even in the free theory and which is hidden in the operator
$\frac{1}{\,\partial^+\,}$.
In momentum space it corresponds to the singularity at
$k^+=0$ in the form $\frac{1}{\,k^+\,}$.  As we have mentioned
in the introduction and as we will discuss later in detail the
$k^+=0$ singularity is the most difficult problem to handle in
the light-front field theories but it also plays an essential
role in determining the nontrivial structure of the theory on
the light-front.  The definition of Eq. (2.16) is to
remove this singularity.  The notation $\frac{1}{\,[k^+]\,}$ denotes the
Fourier transformation of Eq. (2.16):
\def\theequation{2.23}
\begin{equation}
  \ds\frac{1}{\,[k^+]\,} = \ds\frac{1}{\,2\,}\biggl [
  \ds\frac{1}{\,k^++i\epsilon\,} + \ds\frac{1}{\,k^+-i\epsilon\,}\biggr ] \, ,
\end{equation}
which corresponds to a principle value prescription. Since
Eq. (2.16) is an integral it allows an arbitrary constant
in defining the operator ${1/\partial^+}$. Different definitions
of ${1/\partial^+}$ can lead to different commutation relations
of Eq. (2.17) and different prescriptions for $\frac{1}{\,k^+\,}$. However,
Eq. (2.16) has two advantages: it uniquely
determines the initial and boundary conditions
for the light-front field formulation [22], and it
explicitly eliminates the $k^+=0$ modes from the theory [88].
The latter property plays an important role in simplifying the
vacuum structure for light-front bound states.  We shall come back to
address these properties in Sec. V where we shall discuss
them in nonabelian gauge theory.  Keeping these
differences in mind, we now discuss the fermion field
formulation on the light-front.

\hspace{0.3cm} {\em 2. Fermion field}.
For a free massive fermionic field with spin $\frac{1}{\,2\,}$ the symmetric
Lagrangian is
\def\theequation{2.24}
\begin{equation}
  {\cal L} = \overline{\psi} (i \gamma \cdot
  \stackrel{\leftrightarrow}{\partial} - m) \psi \, ,
\end{equation}
which leads to the Dirac equation
\def\theequation{2.25}
\begin{equation}
  (i\gamma \cdot \partial - m) \psi = 0\, , \quad \quad \overline{\psi}
  (i\gamma \cdot \partial + m) = 0 \, .
\end{equation}
In the canonical procedure the light-front conjugate momentum
of $\psi$ and $\psi^{\dagger}$ are defined by
\def\theequation{2.26}
\begin{equation}
  \pi_{\psi} = \ds\frac{\partial {\cal L}}{\,\partial (\partial^-\psi )\,}
  =i\ds\frac{1}{\,4\,}\overline{\psi}\gamma^+\, ,\quad \quad
  \pi_{\psi^{\dagger}} = \ds\frac{\partial {\cal L}}{\,\partial (\partial^-
\psi^{\dagger})\,}
  =-i\ds\frac{1}{\,4\,}\gamma^0\gamma^+ \psi \, ,
\end{equation}
where $\gamma^{\pm}=\gamma^0\pm \gamma^3$. From the above expression,
we see that if we define
\def\theequation{2.27}
\begin{equation}
  \psi_{\pm}=\Lambda_{\pm} \psi \, , \quad \quad \Lambda_{\pm} =
\ds\frac{1}{\,2\,}
  \gamma^0\gamma^{\pm}\, ,
\end{equation}
where $\Lambda_{\pm}$ are the light-front fermionic projectors:
$\Lambda_+^2 = \Lambda_+$, $\Lambda_-^2=\Lambda_-$, $\Lambda_+\Lambda_-=0$,
then we have $\psi =\psi_++\psi_-$ and
$\pi_{\psi}=\frac{i}{\,2\,}\psi^{\dagger}_+$,
$\pi_{\psi^{\dagger}}=-\frac{i}{\,2\,}\psi_+$. In other
words, the fermion field on the light-front can be divided into the
up $(+)$ and down $(-)$ components, while the canonical conjugate momentum
of $\psi$ only relates to the up component.

\hspace{0.3cm} The most interesting property for the light-front fermion field
is that the Dirac equation becomes two coupled equations:
\def\theequation{2.28}
\begin{equation}
  i\partial^- \psi_+ = (i \alpha_{\bot} \cdot \partial_{\bot}
  + \beta m) \psi_-\,  ,
\end{equation}
\def\theequation{2.29}
\begin{equation}
  i\partial^+ \psi_- = (i \alpha_{\bot} \cdot \partial_{\bot}
  + \beta m) \psi_+ \, .
\end{equation}
The equation of motion for $\psi_-$ turns out to be a
constraint equation since it does not contain any light-front time
derivative. Using the definition of Eq. (2.16), we can express
$\psi_-$ in terms of $\psi_+$ and then substitute it into the
equation of motion for $\psi_+$, obtaining
\def\theequation{2.30}
\begin{equation}
  i\partial^- \psi_+(x) = \ds\frac{1}{\,i4\,} \ds\int_{-\infty}^{\infty}
  dx^{'-}\epsilon(x^- - x^{'-})(-\partial^2_{\bot}+m^2)
  \psi_+(x^+,x^{'-},x_{\bot}) \, .
\end{equation}
or
\def\theequation{2.31}
\begin{equation}
  (\partial^+\partial^- - \partial_{\bot}^2 +m^2) \psi_+ = 0 \, .
\end{equation}
If we introduce the following $\gamma$ matrix representation:
\def\theequation{2.32}
\begin{equation}
\begin{array}{l}
  \gamma^0 = \left [
\begin{array}{cc}
  0 & -i \\
  i & 0
\end{array}
  \right ] \, , ~~\gamma^3 = \left [
\begin{array}{cc}
  0  & i \\
  i & 0
\end{array}
  \right ] \, , ~~\gamma^i = \left [
\begin{array}{cc}
  -i\sigma^i & 0 \\
  0 & i\sigma^i
\end{array}
  \right ] \, , \\[2.5\eqnskip]
  \gamma^5 = \left [
\begin{array}{cc}
  \sigma^3 & 0 \\
  0 & -\sigma^3
\end{array}
  \right ] \, ,
\end{array}
\end{equation}
then
\def\theequation{2.33}
\begin{equation}
   \Lambda_+ = \left [
\begin{array}{cc}
  1 & 0 \\
  0 & 0
\end{array}
  \right ] \, , ~~\Lambda_- = \left [
\begin{array}{cc}
  0 & 0 \\
  0 & 1
\end{array}
  \right ] \, ,
\end{equation}
the Dirac spinor is reduced to
\def\theequation{2.34}
\begin{equation}
  \psi = \left [
\begin{array}{c}
  \varphi \\
  \upsilon
\end{array}
  \right ] \, , ~~\psi_+ = \left [
\begin{array}{c}
  \varphi \\
  0
\end{array}
  \right ] \, ,~~\psi_- = \left [
\begin{array}{c}
  0 \\
  \upsilon
\end{array}
  \right ] = \left [
\begin{array}{c}
  0 \\
  \frac{1}{\,\partial^+\,}(\sigma_{\bot}\partial_{\bot}+m)\varphi
\end{array}
  \right ] \, ,
\end{equation}
In other words, the fermion field on the light-front can be expressed
by a two-component spinor $\varphi$ which satisfies the Klein-Gordon
equation, and all $\gamma$ matrices can be
reduced to the Pauli matrix.  The relativistic fermion particle can
then be described in terms of a nonrelativistic spin $\frac{1}{\,2\,}$
particle on the light-front.  This is a very useful property of the
fermion field on the light-front. The canonical commutation relation
can be simply obtained,
\def\theequation{2.35}
\begin{equation}
  \bigl [ \varphi(x),\varphi^{\dagger}(y')\bigr ]_{x^+=y^+}
  = \delta(x^--y^-)\delta^2(x_{\bot} - y_{\bot}) \, .
\end{equation}

\hspace{0.3cm} It must be emphasized that the above two-component light-front
field still describes relativistic spin-1/2 particles, which are
intrinsically different from spin-1/2 non-relativistic particles.
In other words, it contains quarks and antiquarks. To see
how antiquarks can be described in a two-component representation, we need
to construct the charge conjugate for $\varphi$. By using the condition
$C\gamma^{\mu T}C^{-1}=-\gamma^{\mu}$, it is easy to find that the charge
conjugation operator in the $\gamma$-representation of Eq. (2.32) is
\def\theequation{2.36}
\begin{equation}
  C=i \gamma^3 \gamma^1 = \left [
\begin{array}{cc}
  0 & -i \sigma^1 \\
  i\sigma^1 & 0
\end{array}
  \right ] \, .
\end{equation}
Hence,
\def\theequation{2.37}
\begin{equation}
  \psi^c = \eta C \gamma^0 \psi^* \longrightarrow
  \varphi^c = \eta \sigma^1 \varphi^* \, ,
\end{equation}
where $\eta$ is a phase factor which will be set to unity in this
paper. Eq. (2.37) determines the definition of fermion and
antifermion in a two-component representation on the light-front.

\hspace{0.3cm} The light-front Hamiltonian of the free fermion field is given
by
\def\theequation{2.38}
\begin{equation}
  H_{LF} = \ds\int dx^-d^2x_{\bot}(\pi_{\psi}\partial^-\psi-{\cal L})
  = \ds\int dx^-d^2x_{\bot}
  \varphi^{\dagger} \ds\frac{\,(-\partial_{\bot}^2+m^2)\,}{i\partial^+}
  \varphi \, .
\end{equation}
The solution of the free fermion equation of motion is
\def\theequation{2.39}
\begin{equation}
  \varphi (x) = \ds\sum_{\lambda}\ch_{\lambda}\ds\int
\ds\frac{\,dp^+d^2p_{\bot}\,}{2(2\pi)^3}
  \bigl [ b(p,\lambda)e^{-ipx}+d^{\dagger}(p,-\lambda)e^{ipx}\bigr ] \, , ~~
  p^+ \geq 0 \, ,
\end{equation}
where $\ch_{1/2} =\bigl ( {1\atop 0}\bigr )$,
$\ch_{-1/2}=\bigl ( {0\atop 1}\bigr )$
are the spin eigenstates, and the fermionic and antifermionic annihilation
and creation operators satisfy the following commutation relations:
\def\theequation{2.40}
\begin{equation}
\begin{array}{lll}
  \bigl [ b(p,\lambda),b^{\dagger}(p',\lambda')\bigr ] & = &
  \bigl [ d(p,\lambda),d^{\dagger}(p',\lambda')\bigr ] \\[\eqnskip]
  & = & 2(2\pi)^3\delta_{\lambda,\lambda'}
  \delta(p^+-p^{'+})\delta^2(p_{\bot}-p'_{\bot}) \, .
\end{array}
\end{equation}
In momentum space, the light-front Hamiltonian is
\def\theequation{2.41}
\begin{equation}
\begin{array}{l}
  H_{LF} = \ds\sum_{\lambda} \ds\int \ds\frac{\,dp^+d^2p_{\bot}\,}{2(2\pi)^3}
  \ds\frac{\,p^2_{\bot}+m^2\,}{[p^+]}\bigl [ b^{\dagger}(p,\lambda)
  b(p,\lambda )+d^{\dagger}(p,\lambda )d(p,\lambda )\bigr ] \, , \\[2\eqnskip]
  \quad \quad \quad p^+\geq 0 \, ,
\end{array}
\end{equation}
where again an infinite zero-point energy has been ignored. Thus, the
light-front energy $p^-$ for a fermion physical state
$|p,\lambda \rangle =b^{\dagger}(p,\lambda )|0\rangle$ is
\def\theequation{2.42}
\begin{equation}
  H_{LF}|p,\lambda \rangle = p^-|p,\lambda \rangle
  \longrightarrow p^- = \ds\frac{\,p^2_{\bot}+m^2\,}{p^+} \, .
\end{equation}
The same result can be obtained for the antifermions. This is the light-front
energy dispersion relation for the fermion field which is the same as that
for the scalar field.

\hspace{0.3cm} {\em 3. Gauge field}.
In the last part of this section, we consider a free gauge field:
\def\theequation{2.43}
\begin{equation}
  {\cal L} = - \ds\frac{1}{\,4\,}F_{\mu \nu}F^{\mu \nu}\, , \quad \quad
  F_{\mu \nu} =\partial_{\mu}A_{\nu}-\partial_{\nu}A_{\mu} \, .
\end{equation}
The light-front canonical momentum of $A^{\mu}$ is defined by
\def\theequation{2.44}
\begin{equation}
  E^{\mu} = \ds\frac{\partial {\cal L}}{\,\partial (\partial^-A_{\mu})\,}
  = - \ds\frac{1}{\,2\,}F^{+\mu} \, .
\end{equation}
It immediately shows that $E^+=0$, which implies that $A^-$ is not a
physically independent field variable. With these canonical
conjugate momenta, the equations of motion, $\partial_{\mu}F^{\mu \nu}=0$, are
reduced to the light-front Maxwell's equations:
\def\theequation{2.45}
\begin{equation}
\begin{array}{l}
  \ds\frac{1}{\,2\,}\partial^+E^--\partial_{\bot}\cdot E_{\bot}=0 \, , \quad
  \partial^- E^- = \partial^1 B^2-\partial^2 B^1 \, , \\[1.5\eqnskip]
  \partial^- E^1 = \ds\frac{1}{\,2\,}\partial^+B^2+\partial^2 B^- \, , \quad
  \partial^- E^2 = -\ds\frac{1}{\,2\,}\partial^+B^1-\partial^1 B^- \, ,
\end{array}
\end{equation}
where $\vec{E}=(E^-,E^1,E^2)$, and
$\vec{B}=(B^-=F^{12},B^1=F^{2-},B^2=F^{-1})$.

\hspace{0.3cm} Gauge invariance allows us to choose a special gauge. The most
convenient gauge choice in light-front gauge theory is the light-front
gauge
\def\theequation{2.46}
\begin{equation}
  A^+ = A^0 + A^3 =0 \, .
\end{equation}
With this gauge, the canonical momentum is reduced to
$E^-=-\frac{1}{\,2\,}\partial^+ A^-$, $E^i=-\frac{1}{\,2\,}\partial^+A^i$.
Thus the first light-front Maxwell equation (Gauss Law) turns out to be a
constraint equation,
\def\theequation{2.47}
\begin{equation}
  \ds\frac{1}{\,2\,}\partial^+ A^- = \partial^i A^i \, ,
\end{equation}
from which the $A^-$ component can be explicitly determined in terms of the
physical (transverse) gauge field.  Therefore,
a gauge theory on the light-front can
be expressed explicitly with the physical gauge field $A^i$:
\def\theequation{2.48}
\begin{equation}
  H_{LF} = \ds\int dx^-d^2x_{\bot}\, \ds\frac{1}{\,2\,}(\partial^i
  A_{\bot}^j)^2 \, .
\end{equation}

\hspace{0.3cm} The canonical commutation relation for the physical gauge field
is the same as that of the scalar field:
\def\theequation{2.49}
\begin{equation}
  \bigl [ A^i(x),A^j(y')\bigr ]_{x^+=y^+} = - \ds\frac{i}{\,4\,}\delta_{ij}
  \epsilon (x^- - y^-)\delta^2(x_{\bot}-y_{\bot}) \, .
\end{equation}
The solution of the free light-front gauge field in momentum space is
\def\theequation{2.50}
\begin{equation}
  A^i(x) = \ds\sum_{\lambda} \ds\int
\ds\frac{dq^+d^2q_{\bot}}{\,2(2\pi)^3[q^+]\,}
  \bigl [ \epsilon^i_{\lambda}a(q,\lambda)e^{-iqx}+h.c\bigr ] \, , \quad
  q^+ \geq 0 \, ,
\end{equation}
where $\epsilon^i_{\lambda}$ is the gauge field polarization vector:
$\epsilon^i_1 =\frac{1}{\,\sqrt{2\,}\,}(1,i)$,
$\epsilon^i_{-1}=\frac{1}{\,\sqrt{2\,}\,}(1,-i)$, and the canonical commutation
relation for vector boson annihilation and creation operators is:
\def\theequation{2.51}
\begin{equation}
  \bigl [ a(q,\lambda ),a^{\dagger}(q',\lambda')\bigr ] =2(2\pi)^3
  q^+\delta_{\lambda ,\lambda'}\delta(q^+-q^{'+})
  \delta^2(q_{\bot}-q'_{\bot}) \, .
\end{equation}
In momentum space the light-front Hamiltonian has the same
structure as the other fields,
\def\theequation{2.52}
\begin{equation}
  H_{LF} = \ds\sum_{\lambda} \ds\int \ds\frac{\,dq^+d^2q_{\bot}\,}{2(2\pi)^3}
  \ds\frac{q^2_{\bot}}{\,[q^+]\,}a^{\dagger}(q,\lambda )
  a(q,\lambda ) \, , \quad q^+ \geq 0 \, .
\end{equation}
The light-front energy $q^-$ for the physical gauge state
$|q,\lambda \rangle =b^{\dagger}(q,\lambda )|0\rangle$ is then
\def\theequation{2.53}
\begin{equation}
  H_{LF}|q,\lambda \rangle =q^-|q,\lambda \rangle
  \longrightarrow q^- = \ds\frac{\,q^2_{\bot}\,}{q^+} \, .
\end{equation}
Namely, $q^2=0$.

\hspace{0.3cm} It may be worth noting that there is a remarkable property of
the
light-front free field theory discussed in this section: both fermion
and gauge fields are reduced to a two-component representation.  Later
we will see that this is still true even for QCD.  In QCD,
since the gauge fixing $A_a^+=0$ removes the unphysical degrees of
freedom, unitarity is automatically satisfied and no ghost field
is needed, the price to pay being the loss of manifest covariance.
The choice of the $A_a^+=0$ gauge promises that QCD involves
only two-component gauge fields and two-component quark
fields in light-front coordinates.  In early stages of the
development of light-front QED, Bjorken, Kogut and Soper used
the two-component formulation to discuss various physical QED
processes [17].  In the middle-80's a
two-component Feynman perturbation theory with the Mandelstam-Leibbrandt
prescription [57,58] was developed by
Capper {\em et al}. [59]
and Lee {\em et al}. [60] for pure Yang-Mills theory. It has been
shown that there are many advantages in such a two-component light-front
field theory.  The extension of the two-component theory
to the study of light-front canonical quantization (i.e. $x^+$-ordered
theory) of QCD has been explored just recently [82].
Practically, one will see that the two-component
formulation indeed provides a transparent physical picture. In order
to understand clearly the origin of light-front singularities in
various physical processes and to simplify practical calculations,
all discussions in this article are based on the two-component
formulation.

\hspace{0.3cm} With the above basic formulation of the light-front field
theory,
we now discuss physical phenomena of the strong interaction
on the light-front.

\vspace{6pt}
\begin{center}
{\normalsize \bf III. LIGHT-FRONT PHENOMENA}
\end{center}
\vspace{3pt}

\hspace{0.3cm} It has been well-known as early as the 60's that the partially
conserved axial current (PCAC) provides the dominatant theoretical underpinning
for the strong interaction.  Although QCD has now been commonly accepted
as the fundamental strong interaction theory,
we are still unable to solve hadronic structures from QCD in the low-energy
scale due to the lack of a practically computable nonperturbative approach
for field theory.  In the past three decades, most hadronic structures
at low-energy can only be extracted from various phenomenological matrix
elements of the currents in the hadronic bound states and the sum rules
that these matrix elements satisfy by the use of current algebras.
In other words, after about thirty years of investigation of the strong
interaction for quarks, it may still
be true that of all the theoretical methods which have been invented to
describe hadrons, only the algebra of conserved and partially conserved
quark currents has led to {\em precise connections} between the underlying
fundamental degrees of freedom and the observed hadron properties.
Typical examples are: i), the low-energy theorem for hadrons developed
from current algebras in the middle of the 60's which underlies almost all
of the theoretical understanding
of the low-energy strong interaction; ii), the parton dynamics
for the deep inelastic scattering invented in the late 60's which
has dominated the study of high-energy hadronic physics in the last two and
a half decades; and iii), the QCD sum rules of the quark current correlations
for meson and baryon structures investigated in the 80's, which are currently
a main approach of the phenomenological nonperturbative QCD descriptions
for hadrons. In fact, it is particularly of interest to see that the
connections between the fundamental degrees of freedom and hadronic
observables via currents are naturally manifested when we use light-front
coordinates. Furthermore,
the most important feature of the above connection in the light-front
coordinates is that it directly addresses the fundamental interactions
between the underlying degrees of freedom (quarks and gluons) in terms
of physical observables that
may further provide an alternative but more transparent approach to
the study of nonperturbative QCD dynamics for hadrons.  We will discuss
this feature later. In this section I will review some phenomenological
descriptions of strong interaction on the light-front.

\vspace{15.5pt}
\noindent
{\bf III-1. Fixed-mass Sum Rules and Light-Front Current Algebra}

\hspace{0.3cm} The earliest application of light-front dynamics to strong
interaction
phenomena is the derivation of fixed-mass sum rules from current
algebra in the infinite momentum frame.  An introductory discussion
can be found in the famous book by Adler and Dashen [89].
Without any loss of generality, let us consider the matrix element
of the equal-time commutator of two vector currents sandwiched between the
spin-averaged hadronic states:
\def\theequation{3.1}
\begin{equation}
  \langle \beta({\bf p}) | [J_+({\bf q}), J_-(-{\bf q}) ]| \beta
  ({\bf p})\rangle = 2 \langle \beta({\bf p}) | J_3
  | \beta ({\bf p}) \rangle \, ,
\end{equation}
where ($+,-,3$) denote the isospin algebra indices,
$J_a(q)=\int d^3xe^{-i{\bf q\cdot x}}J^0_a(x)$ and
\def\theequation{3.2}
\begin{equation}
  \bigl [ J_+^0(x),J_+^0(y)\bigr ]_{x^0=y^0} = 2 J_3^0(x)
  \delta^3 ({\bf x-y}) \, .
\end{equation}
The sum rule can be obtained by inserting a complete set of intermediate
states $\{ |n\rangle\}$ into the left-hand side of Eq. (3.1):
\def\theequation{3.3}
\begin{equation}
  (2\pi )^3\ds\sum_n\bigl ( |\langle \beta ({\bf p})|J_+^0(0)|
  n\rangle |^2-|\langle \beta({\bf p})|J_-^0(0)|
  n\rangle |^2\bigr ) \delta^3 ({\bf p + q - p_n}) =
  4p^0I_3(\beta ) \, ,
\end{equation}
where the momentum transfer $q^2 = (p-p_n)^2$.  The difficulty with the
above sum rules is that $q^2$ depends on the energies of the intermediate
states. These matrix elements can be measured or estimated from
PCAC for certain values of $q^2$, while in the sum over states $q^2$
increases without limit as the mass of $|n\rangle$ increases.
There is no obvious condition for the convergence of
this summation. Consequently, its applicability is not clear
since, in practice, one has to truncate the sum over states.

\hspace{0.3cm} The difficulty can actually be removed if we boost the hadronic
states to the infinite momentum frame, as was first discovered by
Fubini and Furlan [8]. In the infinite momentum
frame, $|{\bf p}|\rightarrow \infty$. Also, for simplification,
it is convenient to impose the condition ${\bf p\cdot q}=0$.
Then
\def\theequation{3.4}
\begin{equation}
  q^0 = (\sqrt{({\bf p+q})^2+m_n^2\,} \, - \sqrt{{\bf p}^2+m_{\beta}^2\,}\,)^2
  \stackrel{|{\bf p}| \rightarrow \infty}{\longrightarrow} 0 \, ,
\end{equation}
so that $q^2$ becomes independent of the mass of the intermediate state
and is fixed: $q^2=-{\bf q}^2$. All matrix elements in the sum
rule have then a fixed momentum transfer. This leads to the so-called
fixed-mass sum rule,
\def\theequation{3.5}
\begin{equation}
  \ds\int_0^{\infty}d\nu W_2(\nu, q^2= -{\bf q}^2) = 4I_3(\beta ) \, ,
\end{equation}
where $\nu =p\cdot q$, and $W_2$ is defined by
\def\theequation{3.6}
\begin{equation}
\begin{array}{lll}
  C^{\mu \nu}(p,q) & = & (2\pi )^3\ds\sum_n \langle \beta ({\bf p})|
  J_+^{\mu}(0)|n\rangle \langle n|J_-^{\nu}(0)|
  \beta ({\bf p})\rangle \delta^3(p+q-p_n) \\[1.5\eqnskip]
  & = & - g^{\mu \nu}W_1+p^{\mu}p^{\nu}W_2+q^{\mu}q^{\nu}W_3
  + \ds\frac{1}{\,2\,}(p^{\mu}q^{\nu}+p^{\nu}q^{\mu})W_4
\end{array}
\end{equation}
in which $W_i$ are the structure functions measured in experiments.
This is perhaps the first application of the infinite momentum frame
to the strong interaction phenomena.{\footnote{The above fixed-mass
sum rules can also be obtained in the framework of a dispersive
formalism with an unsubtracted assumption [89].}}

\hspace{0.3cm} However, one may note that in deriving the fixed-mass sum rule,
one has in fact used the assumption that it is legal to take the
limit inside the integral.  The validity of the
above assumption is questionable due to the possible illegitimate
interchange of limit and integral.  It was shown that the
fixed mass sum rules are equivalent to the appropriate light-front
commutators and to the $|{\bf p}| \rightarrow \infty$
technique [90,91].
One can derive the various fixed mass sum rules from the
light-front commutators without using the unreliable
$|{\bf p}|\rightarrow \infty$ [31].

\hspace{0.3cm} The light-front current algebra was given by Cornwell and Jackiw
[30] based on the basic commutation relation
for the light-front fermionic field:
$\{\psi_+(x),\psi_+^{\dagger}(y)\} =\delta (x^--y^-)\delta^2
(x_{\bot}-y_{\bot})$.
For the $SU(3)$ currents, we obtain
\def\theequation{3.7}
\begin{equation}
\begin{array}{rl}
  \bigl [ V_a^+(x),V_b^+(y)\bigr ]_{x^+=y^+} = \!\! & if_{abc} V_c^+(x)
  \delta(x^- - y^-) \delta^2 (x_{\bot} - y_{\bot}) \\[\eqnskip]
  & -\ds\frac{i}{\,4\,}\delta_{ab} \partial^+_x
  \partial^+_y \bigl [ \varepsilon (x^- - y^-) \delta^2 (x_{\bot}
  - y_{\bot})S\bigr ] \, ,
\end{array}
\end{equation}
\def\theequation{3.8}
\begin{equation}
\begin{array}{l}
  \bigl [ V_a^+(x),V_b^-(y)\bigr ]_{x^+=y^+} \\[\eqnskip]
  \hspace{0.8cm} = if_{abc} V_c^-(x)
  \delta(x^- - y^-) \delta^2 (x_{\bot} - y_{\bot}) \\[\eqnskip]
  \hspace{1.2cm} -\ds\frac{1}{\,4\,}(if_{abc} + d_{abc})\bigl \{ \partial^+_x
  [\varepsilon(x^- - y^-) \delta^2 (x_{\bot}-y_{\bot})V_c^-(x|y)]
\\[1.5\eqnskip]
  \hspace{1.2cm} -\ds\frac{1}{\,2\,}\partial^i_x \bigl [ \varepsilon
  (x^- - y^-)\delta^2 (x_{\bot}-y_{\bot})(V_c^i(x|y)
  + i\varepsilon^{ij} A_{jc}(x|y)] \bigr \} \\[1.5\eqnskip]
  \hspace{1.2cm} +\biggl \{ \ds\frac{i}{\,16\,}\bigl [ \varepsilon(x^- - y^-)
\delta^2
  (x_{\bot} - y_{\bot})\overline{\psi}(x)\gamma^+\gamma^-\Lambda_{ab}\psi (y)
  - h.c.\biggr \} \\[1.5\eqnskip]
  \hspace{1.2cm} +\ds\frac{i}{\,8\,}\delta_{ab}\partial^i_x \partial^i_x
  [\varepsilon(x^- - y^-)\delta^2 (x_{\bot} - y_{\bot})S] \, ,
\end{array}
\end{equation}
\def\theequation{3.9}
\begin{equation}
\begin{array}{l}
  \!\!\!\! \bigl [ V_a^+(x),V_b^i(y)\bigr ]_{x^+=y^+} \\[\eqnskip]
  \hspace{0.5cm} = if_{abc} V_c^i(x)
  \delta (x^- - y^-)\delta^2(x_{\bot}-y_{\bot}) \\[\eqnskip]
  \hspace{0.9cm} -\ds\frac{1}{\,8\,}(if_{abc}+d_{abc})
  \bigl \{ \partial^+_x[\varepsilon (x^- - y^-)\delta^2 (x_{\bot}
  -y_{\bot}) (V_c^i(x|y) - i \varepsilon^{ij} A_{jc} (x|y))] \\[1.5\eqnskip]
  \hspace{0.9cm} +\partial_{xj} [\varepsilon
  (x^- - y^-)\delta^2 (x_{\bot} - y_{\bot})(-\delta^{ij}
  V_c^i(x|y) + i\varepsilon^{ij} A_{c}^+(x|y))] \bigr \} \\[\eqnskip]
  \hspace{0.9cm} +\biggl \{ \ds\frac{i}{\,16\,}[\varepsilon (x^- - y^-)
  \delta^2 (x_{\bot}-y_{\bot})\overline{\psi}(x)\gamma^+\gamma^i\Lambda_{ab}
  \psi (y)-h.c. \biggr \} \\[1.5\eqnskip]
  \hspace{0.9cm} + \ds\frac{i}{\,4\,}\delta_{ab}\partial^+_x
  \partial^i_x \partial^i_y[\varepsilon (x^- - y^-)
  \delta^2 (x_{\bot}-y_{\bot})S] \, ,
\end{array}
\end{equation}
where $V_a^{\mu}$ is the usual $SU(3)$ vector current:
\def\theequation{3.10}
\begin{equation}
  V_a^{\mu}(x) = \overline{\psi}(x)\gamma^{\mu}\ds\frac{1}{\,2\,}\lambda_a
  \psi (x)\, ,
\end{equation}
and $V_a^{\mu}(x|y)$ and $A_a^{\mu}(x|y)$ are the bilocal generalization of the
vector and axial vector currents:
\def\theequation{3.11}
\begin{equation}
  V_a^{\mu}(x|y) = \overline{\psi}(x)\gamma^{\mu}\ds\frac{1}{\,2\,}
  \lambda_a \psi (y) \, ,
\end{equation}
\def\theequation{3.12}
\begin{equation}
  A_a^{\mu}(x|y) = \overline{\psi}(x)\gamma^{\mu} \gamma^5
  \ds\frac{1}{\,2\,}\lambda_a \psi (y)\, ,
\end{equation}
with the internal symmetry matrices that satisfy the relation
$\lambda_a \lambda_b=(if_{abc}+d_{abc})\lambda_c$.

\hspace{0.3cm} The fixed mass sum rules based on light-front current
algebra are now derived from the tensor
\def\theequation{3.13}
\begin{equation}
  C_{ab}^{\mu \nu}(p,q)= \ds\int d^4 x e^{ipx} \bigl \langle p s| [
V_a^{\mu}(x),
  V_b^{\nu}(0)]|ps\bigr \rangle \, .
\end{equation}
The time reversal invariance, Lorentz invariance, parity conservation and
current conservation allow us to rewrite
\def\theequation{3.14}
\begin{equation}
\begin{array}{rl}
  C_{ab}^{\mu \nu} (p,q) = & \biggl ( -g^{\mu \nu}
  + \ds\frac{\,q^{\mu}q^{\nu}\,}{q^2}\biggr ) W_L^{ab} \\[1.5\eqnskip]
  & +\biggl ( p^{\mu}p^{\nu}
  -\ds\frac{\nu}{\,q^2\,}(p^{\mu}q^{\nu}+p^{\nu}q^{\mu})
  +g^{\mu \nu}\ds\frac{\,\nu^2\,}{q^2}\biggr ) W_2^{ab} \\[1.5\eqnskip]
  & + i\varepsilon^{\mu \nu \alpha \beta}s_{\alpha}
  q_{\beta}W_3^{ab}+i\varepsilon^{\mu \nu \alpha \beta}
  p_{\alpha} q_{\beta} q\cdot s W_4^{ab}\, .
\end{array}
\end{equation}
By setting $q^+=0$ and integrating over $q^-$ for the $+\nu$ components
of $C_{ab}^{\mu \nu}$ [31], we have
\def\theequation{3.15}
\begin{equation}
\begin{array}{l}
  \ds\frac{1}{\,2\pi \,}\ds\int_{-\infty}^{\infty}d\biggl (
\ds\frac{\nu}{\,p^+\,}
  \biggr ) C_{ab}^{+\nu}(p,q)
  \Bigl |_{^{q^+=0}_{q^-=(\nu + p_{\bot}\cdot q_{\bot})/p^+}} \\[2\eqnskip]
  \hspace{0.8cm} = \ds\int dx^- d^2 x_{\bot} e^{-ip_{\bot} \cdot x_{\bot}}
  \bigl \langle ps|[V_a^{\mu}(x),V_b^{\nu}(0)]_{x^+=0}|ps\bigr \rangle \, ,
\end{array}
\end{equation}
from which one can immediately obtain the following fixed mass sum rules:
\def\theequation{3.16}
\begin{equation}
  \ds\int_0^{\infty}d\nu W_2^{[ab]}(q^2,\nu )=\pi f_{abc} \Gamma_c \, ,
\end{equation}
\def\theequation{3.17}
\begin{equation}
  \ds\int_0^{\infty}d\nu W_3^{[ab]}(q^2,\nu ) =\ds\frac{1}{\,2\,}\pi
  f_{abc}\ds\int_0^{\infty}d\alpha \overline{A}_c^1 (0,\alpha ) \, ,
\end{equation}
\def\theequation{3.18}
\begin{equation}
  \ds\int_0^{\infty}d\nu W_4^{[ab]}(q^2,\nu ) = \ds\frac{1}{\,2\,}\pi
  f_{abc} \ds\int_0^{\infty}d\alpha \overline{A}_c^2(0,\alpha ) \, ,
\end{equation}
\def\theequation{3.19}
\begin{equation}
  \ds\int_0^{\infty}d\nu W_L^{[ab]}(q^2,\nu ) = 0 \, ,
\end{equation}
\def\theequation{3.20}
\begin{equation}
  \ds\int_0^{\infty} d\nu W_4^{(a,b)}(q^2,\nu ) = 0 \, ,
\end{equation}
\def\theequation{3.21}
\begin{equation}
  \ds\int_0^{\infty}d\nu (\nu /-q^2)W_2^{(a,b)}(q^2,\nu )
  =\ds\frac{1}{\,2\,}\pi d_{abc}\ds\int_0^{\infty}d\alpha
  \overline{V}_c^1(0,\alpha) \, , \quad \quad q^2\leq 0 \, ,
\end{equation}
where $W_i^{(ab)}$ and $W_i^{[ab]}$ are the symmetric and antisymmetric
decomposition of $W_i^{ab}$ in $ab$: $W_i^{ab} = W_i^{(ab)} + iW_i^{[ab]}$,
and $\overline{A}^i_c$ and $\overline{V}_c^i$ are defined as follows:
\def\theequation{3.22}
\begin{equation}
\begin{array}{l}
  \bigl \langle p|A_a^{\mu}(x|0)-A_a^{\mu}(0|x)|p\bigr \rangle \\[\eqnskip]
  \hspace{0.8cm} =2i\bigl ( s^{\mu}\overline{A}_a^1(x^2,x\cdot p)+p^{\mu}
  x\cdot s\overline{A}_a^2(x^2, x\cdot p)+x^{\mu}
  x\cdot s A_a^3 (x^2, x \cdot p)\bigr ) \, , \\[\eqnskip]
  \bigl \langle p|V_a^{\mu}(x|0)-V_a^{\mu}(0|x)|p\bigr \rangle \\[\eqnskip]
  \hspace{0.8cm} =2i\bigl ( p^{\mu}\overline{V}_a^1(x^2,x\cdot p)+x^{\mu}
  \overline{V}_a^2(x^2,x\cdot p)\bigr ) \,  .
\end{array}
\end{equation}
Eq. (3.16) is the Dashen-Gell-Mann and Fubini sum rule, and
Eq. (3.17) the so-called Beg sum rule. The sum rules of Eqs. (3.16-21)
provide a large number of relations between the various hadronic
coupling constants and form factors that can be measured experimentally
[31]. There are also many other sum rules among the
matrix elements of various currents that can be derived in the same way.
These are the earliest applications of light-front dynamics to
the strong interaction phenomena.  Theoretically, to justify these sum
rules, one needs to develop the fundamental theory (QCD) on the
light-front and to find the light-front hadronic states, which allow
us to directly compute these matrix elements from first principles.

\vspace{15.5pt}
\noindent
{\bf III-2. Light-Front Axial Charge and Its Dynamics}

\hspace{0.3cm} In this subsection, I will discuss the properties of the axial
current and axial charge matrix elements in hadronic states, which are the most
important parts for the understanding of hadronic structure.

\hspace{0.3cm} Consider the axial current $A_a^{\mu}$. The ordinary axial
charge
$Q_a^5$ and the light-front axial charge $Q_{aL}^5$ are defined
respectively as follows:
\def\theequation{3.23}
\begin{equation}
  Q_a^5(t) = \ds\int d^3 x A_a^0(x) \, ,
\end{equation}
\def\theequation{3.24}
\begin{equation}
  Q_{aL}^5 (x^+) = \ds\frac{1}{\,2\,} \ds\int dx^-d^2 x_{\bot}
  A_a^+ (x) \, .
\end{equation}
Thus, the matrix elements of $Q_a^5$ and $Q_{aL}^5$ in two hadronic
states, $|\alpha \rangle$ and $|\beta \rangle$, are related to the
matrix elements of the axial current as follows:
\def\theequation{3.25}
\begin{equation}
  \bigl \langle \beta |Q_a^5 (t)|\alpha \bigr \rangle = (2\pi)^3 \delta^3
  ({\bf p}_{\beta}-{\bf p}_{\alpha}) \bigl \langle \beta |A_a^0
  (t,0)| \alpha \bigr \rangle \, ,
\end{equation}
\def\theequation{3.26}
\begin{equation}
  \bigl \langle \beta |Q_{al}^5(x^+)|\alpha \bigr \rangle
  = (2\pi)^3\delta (p_{\beta}^+ -p_{\alpha}^+)\delta^2 ({\bf p}_{\beta \bot}
  -{\bf p}_{\alpha \bot})\bigl \langle \beta |A_a^+ (x^+,0)
  |\alpha \bigr \rangle \, .
\end{equation}
The low-energy hadronic phenomena ensure that the axial current is
partially conserved (PCAC):
\def\theequation{3.27}
\begin{equation}
  \partial_{\mu} A_a^{\mu} = f_{\pi} m_{\pi}^2 \pi_a  \, ,
\end{equation}
where $f_{\pi}$ is the pseudoscalar (here I simply call it the pionic)
decay constant, $m_{\pi}$ the
pionic mass and $\pi_a$ the pionic interpolating field with $a=1,\cdots 8$
for $SU(3)$ flavor. PCAC implies that the matrix element of the axial
current is dominated by the pionic mass pole [92,93]:
\def\theequation{3.28}
\begin{equation}
  \bigl \langle \beta |A_a^{\mu}|\alpha \bigr \rangle_{m_{\pi}} =
  -if_{\pi}\bigl \langle \beta \pi_a |\alpha \bigr \rangle
  \ds\frac{q^{\mu}}{\,q^2-m_{\pi}^2\,} + \bigl \langle \beta |A_a^{\mu}|
  \alpha \bigr \rangle_{N} \, ,
\end{equation}
where $\bigl \langle \beta |A_a^{\mu}|\alpha \bigr \rangle_{N}$ denotes the
nonpole matrix element.  Using the PCAC and the identity,
$\bigl \langle \beta |\pi_a|\alpha \bigr \rangle
=-\frac{1}{\,q^2-m_{\pi}^2\,}\langle \beta \pi_a|\alpha \rangle$, we have
\def\theequation{3.29}
\begin{equation}
  \bigl \langle \beta |A_a^{\mu}|\alpha \bigr \rangle = \biggl ( g^{\mu}_{\nu}
  - \ds\frac{q^{\mu}q_{\nu}}{\,q^2-m_{\pi}^2\,} \biggr ) \bigl \langle
  \beta |A_a^{\nu}(t,0)|\alpha \bigr \rangle_N\, , \quad \quad
  q=p_{\beta}-p_{\alpha} \, ,
\end{equation}
which leads to
\def\theequation{3.30}
\begin{equation}
  \bigl \langle \beta |Q_a^5(0)|\alpha \bigr \rangle = (2\pi)^3 \delta^3
  ({\bf p}_{\beta}-{\bf p}_{\alpha})
  \ds\frac{-m_{\pi}^2}{\,q^2-m_{\pi}^2\,} \bigl \langle \beta |A_a^0(0)|\alpha
  \bigr \rangle_N \, ,
\end{equation}
\def\theequation{3.31}
\begin{equation}
  \bigl \langle \beta |Q_{al}^5(0)|\alpha \bigr \rangle = (2\pi)^3 \delta
  (p_{\beta}^+ - p_{\alpha}^+)\delta^2 ({\bf p}_{\beta \bot}
  -{\bf p}_{\alpha \bot}) \bigl \langle \beta |A^+_a (0) | \alpha
  \bigr \rangle_N \, .
\end{equation}

\hspace{0.3cm} Obviously, the matrix element of $Q_a^5$ in two hadronic states
describes the pionic transition processes. But in the instant form the matrix
element is singular and is zero in the chiral limit $m_{\pi} \rightarrow 0$;
therefore it does not provide us with much information about
dynamical chiral symmetry breaking while on the light-front the
matrix element is well defined and is independent of the chiral limit.
Furthermore, the equation of motion for the axial charge is
\def\theequation{3.32}
\begin{equation}
  \ds\frac{\,\partial Q_{aL}^5\,}{dx^+} = \ds\frac{1}{\,i\,}[Q_{aL}^5,
  H_{LF}] = D \, ,
\end{equation}
where $H_{LF}=P^-$ is the QCD light-front Hamiltonian, and $D$ is
determined by the above commutation. Since the light-front momentum
$P^+$, $P_{\bot}$ are conserved quantities and commute with $Q_{aL}^5$,
this leads to
\def\theequation{3.33}
\begin{equation}
  [Q_{aL}^5,P^+P^- -P_{\bot}^2] = [Q,M^2] = iP^+ D \, ,
\end{equation}
where $M^2$ is a mass-square operator. Sandwiching the above equation
between two hadronic states, we have
\def\theequation{3.34}
\begin{equation}
  (m_{\alpha}^2-m_{\beta}^2)\langle \beta |Q_{aL}^5 | \alpha
  \rangle = ip^+ \langle \beta |D| \alpha \rangle \, .
\end{equation}

\hspace{0.3cm} Combining Eqs. (3.31) and (3.34), we have made a
simple but direct connection between the
measurable hadronic phenomena and the fundamental QCD interaction.
Obviously, the basic theoretical aspect that we need to explore
here is $H_{LF}$ and the corresponding bound states.

\vspace{15.5pt}
\noindent
{\bf III-3. Parton Distribution Functions and Fragmentation Functions}

\hspace{0.3cm} Probing the hadronic structure functions via deep inelastic
scattering experiments has dominated the study of the strong interaction
and hadronic physics over the last twenty-five years since
 the discovery of the scaling rule of the structure functions
by Bjorken [12] and the interpretation of the
scaling behavior via the parton picture introduced by
Feynman [10] in 1969.  The parton picture emerges
most naturally on the light-front.  In the last subsection,
we have discussed the sum rules for the hadronic structure functions
which one extracts from cross sections measured in deep
inelastic scattering.  Theoretically, using the operator product
expansion, the hadronic structure functions can be separated into
hard and soft parts near the light-cone. The hard part is now known as the
so-called hard scattering coefficient in QCD, which
is the short distance contribution of quarks and gluons
to the structure functions.  The soft part measures
the low-energy (nonperturbative) properties of quarks and
gluons in the parent hadron. When QCD is quantized on the
light-front with the light-front gauge, the soft contribution
can be identified with the parton distribution functions, i.e., the
number density of partons as a function of the fraction of the
light-front longitudinal momentum of the parent hadron.
In the last two decades, QCD has been used
extensively to compute the hard scattering coefficients that are
relevant for the scale evolution of hadronic structure functions.
However, the QCD based exploration of the nonperturbative part,
the parton distribution functions themselves, is still in a very
preliminary stage.  The current high-energy experiments, such as the
$e^-p$ HARE experiment and $p\bar{p}$ collisions at FNL and
CERN, will provide more accurate information about the parton
distribution functions over a broader range of the momentum
transfer in these processes. Currently, the most interesting
problem in the study of parton distribution functions is to look for the
origin of proton spin and the small-$x$ physics, as well as various
inclusive heavy meson decay properties.  Meanwhile, the study of the hadron
productions in high energy processes, namely the fragmentation
functions which measure the probability of finding a hadron in
a high energy parton, is also one of the currently most interesting
subjects. In this subsection, I will discuss the formulation of
these problems on the light-front.

\hspace{0.3cm} {\em 1. Distribution functions}.
In deep inelastic scattering, the cross section is proportional to
the hadronic tensor which is defined by
\def\theequation{3.35}
\begin{equation}
  W^{\mu \nu} (p,q) = \ds\frac{1}{\,4\pi\,} \ds\int d^4 x e^{ipx} \bigl \langle
  ps|[J^{\mu}(x),J^{\nu}(0)]|ps\bigr \rangle \, .
\end{equation}
We can rewrite $W^{\mu \nu}$ in terms of $p$, $q$, $s$ and the invariant
tensors $g^{\mu\nu}$ and $\varepsilon^{\mu \nu \alpha \beta}$ under the
constraints of
parity, time-reversal invariance and current conservation:
\def\theequation{3.36}
\begin{equation}
\begin{array}{rl}
  W^{\mu \nu} (p,q) = & W_1 \biggl ( -g^{\mu \nu} +
\ds\frac{\,q^{\mu}q^{\nu}\,}{q^2}
  \biggr ) + \ds\frac{\,W_2\,}{\,M^2\,}\biggl (
  p^{\mu}-\ds\frac{\nu}{\,q^2\,}q^{\mu}\biggr ) \biggl ( p^{\nu}
  - \ds\frac{\nu}{\,q^2\,}q^{\nu}\biggr ) \\[1.5\eqnskip]
  & + iW_3 \varepsilon^{\mu \nu \alpha \beta} q_{\alpha} s_{\beta}
  + iW_4 \varepsilon^{\mu \nu \alpha \beta} q_{\alpha}
  (q\cdot ps_{\beta} -s\cdot qp_{\beta}) \, ,
\end{array}
\end{equation}
where the $W_i$'s are Lorentz invariant structure functions which determine
the underlying hadrons.
Based on the factorization theorem, the hadronic tensor can be
separated into hard and soft parts:
\def\theequation{3.37}
\begin{equation}
  W^{\mu \nu} (p,q) = \ds\sum_{\tau =2} \biggl ( \ds\frac{\,\Lambda\,}{Q}\biggr
)^{\tau-2}
  \ds\sum_i \ds\int d\{x_i\} C_{i \tau}^{\mu \nu}\bigl ( x_B,\alpha (Q^2),
  \{x_i\}\bigr ) f_{i\tau}\biggl ( \{x_i\},
\ds\frac{\,Q^2\,}{\,\Lambda^2\,}\biggr ) \, ,
\end{equation}
where $C_{i\tau}^{\mu \nu}$ is the hard parton scattering coefficient
which can be calculated from perturbative QCD, while $f_{i\tau}$ is
defined as the parton correlation function in a hadron, which is associated
with the nonperturbative aspect of QCD.  The notation $\tau$ represents
the twist index,
$x_i$ denotes the momentum fraction carried by the $i$-th parton, $x_B$ is
the Bjorken scale, $\alpha(Q^2)$ the QCD running coupling constant,
$Q^2$ the momentum transfer by the photon, and $\Lambda$ the scale
of the strong interaction. On the light-front, $f_{i\tau}$
is reduced to the {\em parton distribution function}:
\def\theequation{3.38}
\begin{equation}
  f_{i \tau} \biggl ( \{x_i\},\ds\frac{\,Q^2\,}{\,\Lambda^2\,}\biggr ) \propto
  \ds\int \ds\frac{\,d\lambda\,}{\,2\pi \,}e^{i\lambda x} \bigl \langle ps|
  \psi_{\pm}^{\dagger}(0)\Gamma^i\psi_{\pm}(n\lambda )|ps\bigr \rangle \, ,
\end{equation}
where $\Gamma^i=1$, $\gamma^5$, $\gamma_{\bot}\gamma^5$, $n\lambda =x^-$
which means that the two field operators are separated in the longitudinal
direction.
The study of parton dynamics, which has dominated the investigation
of the strong interaction in both experiments and QCD in the past
twenty years, is indeed based on the above picture of
factorization and parton distribution functions.  There are a huge
number of papers in the literature devoted to this aspect.  A nice
discussion and derivation may be found in Ellis, Furmanski and
Petronzio's work [54], and a review paper given by Collins, Soper
and Sterman could also be a good reference on this topic [55].
Jaffe

\pagebreak

\begin{table}[hptb] 
\tblcaption{
$\mbox{ }$
}

\vspace{0.5cm}
\begin{tabular}{lcccl}\hline\hline
\vspace{-8pt}
\\
                          & $\!\!$Twist-2: $O(1)$ & $\!\!$Twist-3: $O(1/Q)$ &
$\!\!$Twist-4: $O(1/Q^2)$ & \\[\eqnskip]\hline
\vspace{-8pt}
\\
  $\Gamma^i$              & $\!\!$$+ +$           & $\!\!$$+ -$             &
$\!\!$$- -$               & $\!\!$Light-front projection \\[\eqnskip]
  1                       & $\!\!$$f_1(x)$        & $\!\!$$e(x)$            &
$\!\!$$f_4(x)$            & $\!\!$Target spin average \\[\eqnskip]
  $\gamma^5$              & $\!\!$$g_1(x)$        & $\!\!$$h_2(x)$          &
$\!\!$$g_3(x)$            & $\!\!$Target helicity asymmetry  \\[\eqnskip]
  $\gamma_{\bot}\gamma^5$ & $\!\!$$h_1(x)$        & $\!\!$$g_T(x)$          &
$\!\!$$h_3(x)$            & $\!\!$Target helicity flip \\
\vspace{-8pt}
\\ \hline \hline
\end{tabular}
\vspace{1cm}
\end{table}

\noindent
and Ji have recently classified various parton distribution functions
based on the fermionic light-front up and down components and helicity
on the light-front.  These parton distribution functions can be
naturally classified in terms of their helicity (spin) dependence
and twists on the light-front, as we can see from Table I.
For the detailed definition see Ref. [94].

\hspace{0.3cm} {\em 2. Fragmentation functions}.
Fragmentation functions were introduced by Feynmen to describe
hadron production from the underlying hard parton processes [11].
Collins and Soper showed later [95] that, in
QCD, parton fragmentation functions can be defined as matrix
elements of quark and gluon field operators at light-front
separations.  Physically, the quark and gluon fragmentation
functions are the probability to find a hadron in a hard parton
in hard processes, such as the lepton-hadron and hadron-hadron colliders.
Explicitly, the quark fragmentation is defined
by
\def\theequation{3.39}
\begin{equation}
  f_{A/i}(z) =\ds\frac{z}{\,36\pi\,}\ds\int dx^- e^{-i p^+ x^-/z}
  \mbox{Tr}\bigl \langle 0 | \phi_+ (0) | H(p) s \bigr \rangle \bigl \langle
  H(p) s | \phi^{\dagger}_+ (x^-) | 0 \bigr \rangle \, ,
\end{equation}
while the gluon fragmentation is
\def\theequation{3.40}
\begin{equation}
  f_{A/g}(z) = \ds\frac{-z}{\,32\pi k^+}\ds\int dx^- e^{-i k^+ x^-}
  \bigl \langle 0 | F^{+\mu}(0)|H(p)s \bigr \rangle \bigl \langle
  H(p)s|F^+_{~\mu}(x^-)|0 \bigr \rangle \, ,
\end{equation}
where Tr traces the color and Dirac components of quarks, $|H(p)s\rangle$
is the hadronic states with the momentum $p$ and the helicity $s$ on the
light-front.
Here we have also taken the light-front gauge $A^+=0$, $F^{\mu \nu}$ is the
usual gauge field strength tensor.

\hspace{0.3cm} {\em 3. Inclusive decay of heavy-mesons}.
Recently it was found that, using the heavy quark symmetry and operator
product expansion, heavy meson decay spectra can also be reformulated
in terms of parton distribution functions [97].
Typical examples are the radiative and semileptonic $B$-meson decay
spectra which are given by
\def\theequation{3.41}
\begin{equation}
\begin{array}{l}
  \ds\frac{\,d\Gamma (B\rightarrow X_s \gamma)\,}{dE_{\gamma}} =
  \ds\int dk^+ \ds\frac{\,d\Gamma_{parton}\,}{dE_{\gamma}}
  (E_{\gamma},m_b^*)\bigl ( f(k^+)+O(\Lambda/m_b)\bigr ) \, , \\[1.5\eqnskip]
  \ds\frac{\,d\Gamma (B\rightarrow X_ul\overline{\nu})\,}{dE_l} =
  \ds\int dk^+ \ds\frac{\,d\Gamma_{parton}\,}{dE_l}(E_l,m_b^*)
  \bigl ( f(k^+)+O(\Lambda/m_b)\bigr ) \, ,
\end{array}
\end{equation}
where $\frac{\,d\Gamma_{parton}\,}{dE}$ is the hard $b$-parton decay spectra.
The universal parton distribution function inside the $B$ meson, $f(k^+)$,
is defined on the light-front:
\def\theequation{3.42}
\begin{equation}
  f(k^+) = \ds\frac{1}{\,2\pi\,}\ds\int dx^- e^{i k^+ x^-}
  \bigl \langle B(v)|h_+^{\dagger}(0)h_+(x^-)|B(v) \bigr \rangle \, ,
\end{equation}
with $h_+$ is the light-front up component of the heavy quark field and
$|B(v)\rangle$ is the light-front $B$-meson states [96].

\hspace{0.3cm} In conclusion, we can see that light-front dynamics
has played a very important role in the study of the strong interaction
phenomena in the past thirty years, especially for parton physics.
Based on the kinematics, symmetry and unitarity of physical
strong interaction processes, most of the measurable hadronic structural
properties have been reduced to matrix elements of light-front field
operators in the light-front hadronic bound states plus perturbative
QCD corrections.  The latter are calculatable from the fundamental theory.
While these nonperturbative hadronic matrix elements (the form factors,
the coupling constants, the parton distribution functions and
parton fragmentation functions) have been understood quite well
from the experimental data and some phenomenological model studies,
unfortunately, we still lack a reliable approach to calculate these
physical quantities from QCD. Obviously, in order to completely understand the
strong interaction phenomena from the fundamental theory, it is natural
to require the knowledge of the light-front QCD bound states for
hadrons.  In the next section,
I will first discuss the structure of light-front wave functions for
hadronic bound states.  Then, in the subsequent sections, we shall look for
a possible approach to determine these hadronic bound states from QCD
on the light-front.

\vspace{6pt}
\begin{center}
{\normalsize \bf IV. STRUCTURE OF LIGHT-FRONT BOUND STATES}
\end{center}
\vspace{3pt}

\hspace{0.3cm} In terms of the standard language of field theory, relativistic
bound states and resonances are identified by the occurrence of poles in Green
functions. Although the information extracted from this approach
provides a definition of physical particles, it does not
provide their detailed properties, namely their wave functions
which determine all the structures of the particles. In order to
understand the dynamics of hadrons, we need to find the explicit form of
the hadronic wave functions.

\hspace{0.3cm} However, the wave functions of bound states in field theory
have not been well established.  One may define the
wave functions as the eigenstates of $P^0$ and may determine them
by solving the eigenequation of $P^0$.  But $P^0$ is a square root
function of the momentum and mass which does not give us a clear
picture of the Schr\"{o}dinger eigenstate equation in quantum mechanics.
The well-defined framework for finding the relativistic bound
states is to solve the Bethe-Salpeter equation.  However, the
Bethe-Salpeter equation only provides the amplitude of a Fock
sector in the bound states so that it cannot be normalized.
In other words, the Bethe-Salpeter amplitudes
do not have the precise meaning of wave functions for particles.

\hspace{0.3cm} In principle, the wave function of a relativistic bound state
can always be written
as an operator function of the particle creation operators acting
on the vacuum of the theory.  However, for many theories that
we are interested in, especially for QCD, the vacuum is
very complicated.  Thus, it is even more difficult to formally
write down a relativistic bound state wave function.

\hspace{0.3cm} In this section, I will discuss how these difficulties may be
solved when we look at the problem on the light-front.

\vspace{15.5pt}
\noindent
{\bf IV-1. Light-Front Vacuum}

\hspace{0.3cm} In the equal-time framework, the vacuum of QCD is crucial for
a realization of chiral symmetry breaking.  From the vacuum the axial charges
$Q_5^a$ $(a=1,\cdots 8)$ can create pseudoscalar particles
which are the lowest bound states in the strong interaction regime that
have been used as the building blocks in the development of the
effective chiral field theory for low-energy hadron dynamics.
Undoubtedly, the complicated vacuum is the most important object to
be understood for the nonperturbative aspect of QCD.  As we have already
argued, it is also a starting point for the construction of hadronic wave
functions.  However, the understanding of the true QCD vacuum
is still very limited, although a lot of informative work has been
carried out in the past two decades based on the instanton phenomena
[66] and the QCD sum rule [71].

\hspace{0.3cm} In the light-front coordinates, a particle's momentum is
divided into the longitudinal component (along the light-front
time direction $x^+=t+x^3$) and the transverse components
(perpendicular to the light-front time direction). The longitudinal
momentum of each particle, $p^+=p^0+p^3$, cannot be negative
since the energy of a physical state always dominates its momentum.
As a result, the light-front vacuum for any interacting field
theory can only be occupied by the particles with zero-longitudinal
momentum, namely
\def\theequation{4.1}
\begin{equation}
  |vac \rangle_{LF} = f(a^{\dagger}_{k^+=0}) | 0 \rangle \, ,
\end{equation}
so that $P^+|vac\rangle_{LF}=0$, where $P^+=\sum_ik_i^+$.
At this point, the light-front vacuum is still nontrivial. In the
past several years, many tried to solve the so-called zero-mode
(the particles with $k^+=0$) problem to obtain a nontrivial
light-front vacuum but not much useful progress has been made since then
[98-107].

\hspace{0.3cm} To construct hadronic bound states in the Fock space of quarks
and gluons, it is natural to ask whether one can simplify the vacuum
so that we can easily construct the hadronic states by the use of
the Fock space expansion with a trivial vacuum.  It is obvious that if
we can remove the basic constituents with zero longitudinal light-front
momentum, the vacuum of the full interacting theory is the same as
the free field theory, namely
\def\theequation{4.2}
\begin{equation}
  |vac\rangle_{LF} = |0\rangle \, .
\end{equation}

\hspace{0.3cm} To remove these particles with zero longitudinal momentum,
we can either use a prescription that requires the field variables
to satisfy the antisymmetric boundary condition in the light-front
longitudinal direction [88] or deal with a cutoff
theory that imposes $p^+>\epsilon$ on the momentum expansion of
each field variable, where $\epsilon$ is a small number [85].
Thus, the positivity of longitudinal momentum with such a prescription
or an explicit cutoff ensures that the light-front vacuum must
be trivial, as has been shown above.  Now the wave function of a
relativistic bound state can be expressed as an ordinary Fock state
expansion:
\def\theequation{4.3}
\begin{equation}
  |\Psi \rangle = f(a^{\dagger}) | 0 \rangle \, .
\end{equation}
For QCD, $a^{\dagger}$ should be the quark, antiquark and gluon creation
operators with nonzero longitudinal momentum, and $f(a^{\dagger})$ must be a
color singlet operator.

\hspace{0.3cm} Of course, in this case, the light-front axial charge, denoted
by $Q_{5L}^a$, vanishes the vacuum state. One might ask that,
once the vacuum is simple, what happens to the nontrivial
effect of the theory associated with the nontrivial vacuum.
At least we believe that confinement and chiral
symmetry breaking in QCD must be a result of the nontrivial vacuum.
In the next subsections, we will take two simple examples, the sigma
model and 1+1 QCD, to show how the zero-mode particles are removed
but the nontrivial vacuum effects and confinement mechanism can be
recovered or manifested.  But before we explore the nontrivial
vacuum effects with the trivial vacuum structure, I would like to
discuss first the light-front bound state equation.

\vspace{15.5pt}
\noindent
{\bf IV-2. Light-Front Bound State Equation}

\hspace{0.3cm} Once the light-front vacuum becomes trivial, the light-front
bound states
for various hadrons can be expanded in terms of the Fock space.
Explicitly, a hadronic bound state labeled by $\alpha$ with the total
longitudinal and transverse momenta $P^+$ and $P_{\bot}$, and the helicity
(the total spin along the longitudinal direction) $\lambda$ can be
expressed as follows:
\def\theequation{4.4}
\begin{equation}
  |\alpha ,P^+,P_{\bot},\lambda \rangle = \ds\sum_{n,\lambda_i}
  \ds\int' \ds\frac{\,dx_id^2k_{\bot i}\,}{2(2\pi)^3}
  |n,x_iP^+,x_iP_{\bot}+k_{\bot i},\lambda_i \rangle
  \Phi_{n/\alpha}(x_i,k_{\bot i},\lambda_i)\, ,
\end{equation}
where $n$ represents $n$ constituents contained in the state
$|n,x_iP^+,x_iP_{\bot}+k_{\bot i},\lambda_i\rangle$, $\lambda_i$ is the
helicity of the $i$-th constituent, and $\int'$ denotes the integral
over the space:
\def\theequation{4.5}
\begin{equation}
  \ds\sum_i x_i =1\, , \quad \quad \mbox{and} \quad \quad
  \ds\sum_i k_{\bot i} = 0 \, ,
\end{equation}
here $x_i$ is the fraction of the total longitudinal momentum that the
$i$-th constituent carries, and $k_{\bot i}$ is its relative transverse
momentum with respect to the center of mass frame:
\def\theequation{4.6}
\begin{equation}
  x_i = \ds\frac{\,p_i^+\,}{\,P^+\,} \, , \quad \quad
  k_{i\bot}=p_{i\bot}-x_i P_{\bot} \, ,
\end{equation}
with $p_i^+$, $p_{i\bot}$ being the transverse and longitudinal momentum of
the $i$-th
constituent, and $\Phi_{n/\alpha}(x_i,k_{\bot i},\lambda_i)$ the amplitude
of the Fock state $|n,x_iP^+,x_iP_{\bot}+k_{\bot i},\lambda_i \rangle$ with the
normalization condition:
\def\theequation{4.7}
\begin{equation}
  \ds\sum_{n,\lambda_i} \ds\int' \ds\frac{\,dx_id^2k_{\bot i}\,}{2(2\pi)^3}
  |\Phi_{n/\alpha}(x_i,k_{\bot i},\lambda_i)|^2 = 1 \, .
\end{equation}

\hspace{0.3cm} The eigenstate equation that the wave functions obey on the
light-front is obtained from the operator Einstein equation
$P^2=P^+P^--P_{\bot}^2=M^2$:
\def\theequation{4.8}
\begin{equation}
  H_{LF}|\alpha ,P^+,P_{\bot},\lambda \rangle =
  \ds\frac{\,P_{\bot}^2+M_{\alpha}^2\,}{P^+}|\alpha ,P^+,P_{\bot},\lambda
  \rangle \, ,
\end{equation}
where $H_{LF}=P^-$ is the light-front Hamiltonian. Futhermore, since
the boost on the light-front only depends on kinematics, we can consider
the bound state in the rest frame $(P_{\bot}=0$, $P^+=M_{\alpha})$.
Thus, the eigenstate equation simply becomes:
\def\theequation{4.9}
\begin{equation}
  H_{LF}|\alpha ,P^+,P_{\bot}=0,\lambda \rangle = M_{\alpha}
  |\alpha ,P^+,P_{\bot}=0,\lambda \rangle \, ,
\end{equation}
which is the familiar Schr\"{o}dinger equation in ordinary quantum mechanics.
On the light-front, boosting a bound state in the rest frame to any other
frame is quite simple, and is dynamically independent, as we have shown
in Eq. (2.10).  Thus, once we find the bound state in the rest frame,
we can completely understand the particle structure in any frame. Yet
this is not true in the instant form. In the instant form, although the
bound state equation in the rest frame has the same form,  the
solutions in the rest frame are not easily boosted to other Lorentz
frames due to the dynamical dependence of the boost transformation.
Therefore, in each different Lorentz frame, one needs to solve the
bound state equation of $P^0$ to obtain the corresponding wave
functions.  This is perhaps the reason why we have not established
a reliable approach to construct relativistic wave functions in the
instant field theory in terms of the Schr\"{o}dinger picture. This
obstacle is obviously removed on the light-front.

\hspace{0.3cm} To see the explicit form of the light-front bound state
equation,
let us consider a meson wave function (for instance, a pion, [47]).
The light-front bound state equation can be expressed as:
\def\theequation{4.10}
\begin{equation}
\begin{array}{l}
  \biggl ( m_{\pi}^2-\ds\sum_i \ds\frac{\,k_{i\bot}^2+m_i^2\,}{x_i}\biggr )
  \left (
\begin{array}{c}
  \Psi_{q\bar{q}} \\
  \Psi_{q\bar{q}g} \\
  \vdots
\end{array}
  \right ) \\[2.5\eqnskip]
  \hspace{0.8cm} = \left (
\begin{array}{ccc}
  \langle q\bar{q}|H_{int}|q\bar{q} \rangle & \langle
q\bar{q}|H_{int}|q\bar{q}g\rangle & \cdots \\
  \langle q\bar{q}g|H_{int}|q\bar{q} \rangle & \cdots & \\
  \vdots & &
\end{array}
  \right ) \left (
\begin{array}{c}
  \Psi_{q\bar{q}} \\
  \Psi_{q\bar{q}g} \\
  \vdots
\end{array}
  \right ) \, .
\end{array}
\end{equation}
Of course, to exactly solve the above equation for the whole Fock space is
still impossible. Currently, two approaches have been developed.  One
is given by Brodsky and Pauli [108-111],
the so-called discretize light-front approach, the other by Perry,
Harindranath and Wilson [112-115], based
on the old idea of the Tamm-Dancoff approach [116,117] that
truncates the Fock space to only include these Fock states
with a small number of particles.  Furthermore, if one can eliminate
all the high order Fock space sectors (approximately) by an effective
two-body interaction kernel,
the light-front bound state equation is reduced to the light-front
Bethe-Salpeter equation:
\def\theequation{4.11}
\begin{equation}
  \biggl ( m_{\pi}^2-\ds\frac{\,k_{\bot}^2+m^2\,}{\,x(1-x)\,}\biggr )
  \Psi_{q\bar{q}}(x,k_{\bot})= \ds\int \ds\frac{\,dyd^2k'_{\bot}\,}{2(2\pi)^3}
  V_{eff}(x,k_{\bot},y,k'_{\bot})\Psi_{q\bar{q}}(y,k'_{\bot}) \, .
\end{equation}
The advantage for Eq. (4.10) with the Tamm-Dancoff approximation
is that it provides a reliable way to study the contribution of Fock
states which contain more particles step by step by increasing the size
of truncated Fock space, while the Bethe-Salpeter equation lacks such an
ability.

\vspace{15.5pt}
\noindent
{\bf IV-3. Spontaneous Symmetry Breaking on the Light-Front}

\hspace{0.3cm} Now, we turn back to the question that if the vacuum on the
light-front becomes trivial, how are we to realize the physics of the
spontaneous
chiral symmetry breaking and the associated Goldstone bosons which are all the
theoretical underpinning for our understanding of the low-energy
theorem. We argue that
eliminating the longitudinal zero modes leads to a trivial light-front
vacuum, as a result, the theory must become explicitly symmetry
breaking by a proper modification of the Hamiltonian.  To be
explicit, let us consider the sigma model as an example.
The sigma model is one of the few models in field theory that can
clearly demonstrate the mechanism of spontaneously symmetry breaking.

\hspace{0.3cm} The Lagrangian density of the sigma model is given by
\def\theequation{4.12}
\begin{equation}
  {\cal L} = \ds\frac{1}{\,2\,}\bigl ( (\partial^{\mu}\sigma)^2
  +(\partial^{\mu}\pi)^2\bigr ) -V(\sigma^2+\pi^2) \, ,
\end{equation}
where
\def\theequation{4.13}
\begin{equation}
  V(\sigma^2+\pi^2)=-\ds\frac{1}{\,2\,}\mu^2(\sigma^2+\pi^2)
  +\ds\frac{\,\lambda\,}{4}(\sigma^2+\pi^2)^2 \, .
\end{equation}
It is well-known that this Lagrangian has a continuous rotational symmetry
in the $\sigma-\pi$ plane.  This symmetry is spontaneously
broken since the vacuum expectation value of the field variables, for
example the $\sigma$ field, does not vanish (the vacuum is nontrivial).
As a result the $\pi$ is a Goldstone boson.

\hspace{0.3cm} Now we want to define the theory on the light-front such
that the vacuum is trivial but the massless $\pi$ can be determined in an
alternative way (i.e., it is not realized from the mechanism of
spontaneous symmetry breaking).

\hspace{0.3cm} Since the original theory contains a zero mode in the $\sigma$
field, we can remove the zero modes by shifting the $\sigma$ field to
$\sigma'$ such that $\sigma'$ does not contain zero modes:
$\sigma'=\sigma -\sigma_v$.  The resulting potential in terms of
$(\sigma',\pi$) is
\def\theequation{4.14}
\begin{equation}
\begin{array}{rl}
  \!\!\!\!\! V(\sigma',\pi )= \!\!\! & \lambda \biggl ( \ds\frac{1}{\,4\,}
  (\sigma^{'2}+\pi^2)+\sigma_v\sigma'(\sigma^{'2}+\pi^2)+\ds\frac{3}{\,2\,}
  \sigma_v^2\sigma^{'2}+\ds\frac{1}{\,2\,}\sigma^2_v\pi^2+\sigma_v^3\sigma'
  +\sigma_v^4\biggr ) \\[1.5\eqnskip]
  & -\mu^2\bigl ( \sigma^{'2}+\pi^2+2\sigma_v\sigma'+\sigma_v^2 \bigr ) \, .
\end{array}
\end{equation}
With this potential, the vacuum is trivial.  Now the
question is, what is the value of $\sigma_v$ which cannot
be determined in the theory with the zero modes removed.
However, current conservation can be used to fix
$\sigma_v$. The continuous transformation of the original potential
leads to a conserved current:
$j^{\mu}=\sigma \partial^{\mu}\pi -\pi \partial^{\mu}\sigma$. After the shift,
it becomes:
\def\theequation{4.15}
\begin{equation}
  j^{\mu} = \sigma' \partial^{\mu} \pi - \pi \partial^{\mu} \sigma'
  + \sigma_v \partial^{\mu} \pi \, .
\end{equation}
The dropping of the zero longitudinal momentum modes however does not
affect local current conservation, $\partial_{\mu}j^{\mu}=0$.
It is easy to check that current conservation results in the value
of $\sigma_v=\sqrt{\frac{\,\mu^2\,}{\,2\lambda\,}\,}$.  The potential then
is reduced to
\def\theequation{4.16}
\begin{equation}
  V(\sigma',\pi ) = \ds\frac{1}{\,2\,}\mu^2 \sigma^{'2}
  +\ds\frac{\,\lambda\,}{4}(\sigma^{'2}+\pi^2)^2
  +\sqrt{\ds\frac{\,\lambda \mu^2\,}{2}\,}\,(\sigma^{'2}+\pi^2)\sigma' \, ,
\end{equation}
which shows that the pion is massless.
One can now find that the charge $Q=\frac{1}{\,2\,}\int dx^-d^2x_{\bot}j^+$ no
longer commutes with the Hamiltonian.  Thus, in the
theory with zero modes dropped, the Hamiltonian explicitly breaks
the symmetry and the vacuum is trivial but the pion is still massless.

\hspace{0.3cm} It may be noted that the above analysis is also true for the
instant formulation but the conclusion is different.  In the instant form,
we can follow the same route (by a shift of the $\sigma$ field) to determine
the symmetry breaking and the massless $\pi$ but it has nothing to
do with the simplification of the vacuum.  The shift in the instant
form does remove the zero modes from the theory but the instant
vacuum is not only occupied by the zero modes.

\hspace{0.3cm} Of course, the above demonstration of symmetry breaking, with
a trivial vacuum on the light-front, cannot be directly applied to QCD.
The situation with QCD on the light-front is much worse than the
sigma model.  In QCD, the operator that is expected to have a non-zero vacuum
expectation value due to spontaneous symmetry breaking is
the composite operator $\overline{\psi}\psi$ where $\psi$ is
the quark field.  There is no obvious mechanism for this expectation
value to show up in the QCD Hamiltonian since the QCD Hamiltonian
does not apparently contain squares or higher power terms of
$\overline{\psi}\psi$. This is perhaps the reason that all
these beautiful but naive pictures of spontaneously symmetry
breaking are not manifest in QCD.  However, the trivial
vacuum with an explicitly symmetry breaking Hamiltonian on the
light-front may provide a new avenue to explore the physics
associated with Goldstone particles. Therefore, for QCD,
the light-front zero
modes are not the essential problem we need to solve. The
problem is how we can obtain a correct effective Hamiltonian
after the elimination of zero modes.

\vspace{15.5pt}
\noindent
{\bf IV-4. Bound States in 1+1 QCD and Confinement}

\hspace{0.3cm} Perhaps a clear understanding of quark confinement is best seen
in the quark
dynamics of 1+1 QCD on the light-front [118].  Here we will
demonstrate the confinement mechanism  on the light-front in this
theory with a trivial vacuum.  The Lagrangian density for
1+1 QCD is given by
\def\theequation{4.17}
\begin{equation}
  {\cal L}_{1+1} = -\ds\frac{1}{\,4\,}F_a^{\mu \nu}F_{a \mu \nu}
  +\overline{\psi}(i\gamma \cdot D-m) \psi \, ,
\end{equation}
where $F_a^{\mu
\nu}=\partial^{\mu}A_a^{\nu}-\partial^{\nu}A_a^{\mu}-gf^{abc}A_b^{\mu}A_c^{\nu}$
and $D^{\mu}=\partial^{\mu}+igT^aA_a^{\mu}$, with $\mu ,\nu =+$ or $-$. If we
take the light-front gauge, $A_a^+=0$, the Lagrangian can be rewritten as
\def\theequation{4.18}
\begin{equation}
  {\cal L}_{1+1} = i\psi_+^{\dagger} \partial^- \psi
  +i\psi_-^{\dagger} \partial^+ \psi_- + g j^+_a A_a^-
  - \ds\frac{1}{\,4\,}(\partial^+ A^-)^2 \, ,
\end{equation}
where $\psi_-$ and $A_a^-$ are the light-front constraint variables
that satisfy the following constraint equations:
\def\theequation{4.19}
\begin{equation}
  \psi_- = -im \beta \ds\frac{1}{\,\partial^+\,}\psi_+ \, ,
\end{equation}
\def\theequation{4.20}
\begin{equation}
  \partial^+ A^-_a = 2g \ds\frac{1}{\,\partial^+\,}j^+_a \, , \quad \quad
  j^+_a=2\psi_+^{\dagger} T^a \psi_+ \, .
\end{equation}
Then the light-front 1+1 QCD Hamiltonian is simply given by
\def\theequation{4.21}
\begin{equation}
  H_{1+1} = \ds\int dx^- \biggl \{ m^2 \psi^{\dagger}_+
  \ds\frac{1}{\,i\partial^+\,}\psi_+ -2g^2j_a^+\ds\frac{1}{\,\partial^{+2}\,}
  j_a^+ \biggr \} \, .
\end{equation}
The confinement can be seen explicitly if we use the definition of
Eq. (2.16) for the operator $\frac{1}{\,\partial^+\,}$. In such a
definition, we have removed the zero longitudinal modes.
The Hamiltonian becomes
\def\theequation{4.22}
\begin{equation}
\begin{array}{rl}
  H_{1+1}= \!\!\! & \ds\frac{1}{\,4\,}\ds\int dx^+d{x'}^-\biggl \{ -im^2
  \psi^{\dagger}_+ (x^-) \varepsilon (x^- - {x'}^-)\psi_+({x'}^-)
\\[1.5\eqnskip]
  & -g^2j_a^+(x^-)|x^--{x'}^-|j_a^+ ({x'}^-)\biggr \}
  +\biggl ( \ds\lim_{\lambda \rightarrow \infty} \lambda
  \ds\frac{\,g^2\,}{4}\biggl ( \ds\int dx^-j^+_a(x^-)\biggr ) \biggr )^2 \, .
\end{array}
\end{equation}
In this Hamiltonian the first term is the free quark mass term, the second
term is a linear confinement potential between quarks, and the last
term is a singular (divergent) term due to the elimination of
the zero modes. It is this term that forbids the observation
of a single quark.  To see how the confinement mechanism is manifested
in this theory, let us consider the color non-singlet single quark state
and the color-singlet two quark bound state.

\hspace{0.3cm} Let us first consider a single quark state:
$|P^+,c\rangle =b_c^{\dagger}(P^+)|0\rangle$, where $b_c^{\dagger}(p^+)$ is the
quark creation operator with momentum $p^+$ and color index $c$;
\def\theequation{4.23}
\begin{equation}
  \psi_{c+}(x^-)= \ds\int \ds\frac{\,dp^+\,}{4\pi}\bigl [ b_c(p^+)e^{ip^+x^-/2}
  +d^{\dagger}_c(p^+)e^{-ip^+x^-/2}\bigr ] \, ,
\end{equation}
satisfies the light-front fermion commutation relations.
It is easy to check that the solution of the single quark
bound state equation,
\def\theequation{4.24}
\begin{equation}
  P^-|P^+,c\rangle = H_{1+1}|P^+,c\rangle = \ds\frac{\,M^2\,}{\,P^+\,}|P^+,c
  \rangle \, ,
\end{equation}
is
\def\theequation{4.25}
\begin{equation}
  M^2 = m^2 - \ds\frac{\,g^2\,}{\,4\pi\,}C_f + \ds\frac{\,g^2\,}{8}C_f
  \ds\frac{\,P^+\,}{\epsilon} \, ,
\end{equation}
with $C_f=T^aT^a=\frac{3}{\,4\,}$,
$\epsilon =\frac{\,2\pi\,}{\lambda}\rightarrow 0 (\lambda \rightarrow \infty)$.
Obviously $M^2\rightarrow \infty$ $(\epsilon \rightarrow 0)$
which implies that no single quark can be measured. In other words,
the quark is confined.  This confinement mechanism is obviously an
effect of the original nontrivial vacuum since the term
$\frac{1}{\,\epsilon\,}$ originates from the elimination of the $k^+=0$
singularity.

\hspace{0.3cm} Now we consider the quark-antiquark color singlet state:
\def\theequation{4.26}
\begin{equation}
  |\Psi (P)\rangle = \ds\int \ds\frac{\,dp^+\,}{\,4\pi\,} \Phi
  (p^+)b_c^{\dagger}(p^+)d_c^{\dagger}(P^+-p^+)|0 \rangle \, .
\end{equation}
The eigenstate equation of the above bound state is
\def\theequation{4.27}
\begin{equation}
  \biggl ( M^2-\ds\frac{m^2}{\,x(1-x)\,}\biggr ) \Phi(x)
  = \ds\frac{\,g^2\,}{\pi}C_f \ds\int \ds\frac{\,dy\,}{\,4\pi\,}
  \ds\frac{\,\Phi(y)-\Phi(x)\,}{(x-y)^2\,} \, .
\end{equation}
Here the confinement of two quarks is given by a linear potential
(the right hand side of the above equation).
The singular term in Eq. (4.22) does not contribute
to the above bound state equation since it is proportional
to the color charge operator, and $M^2$ should be finite.

\hspace{0.3cm} The above equation does not have an analytic
solution.  The discussion about the general form of the wave function
can be found in Ref. [118]. The approach can be applied
to 1+1 QED where an additional property, namely the gauge anomaly,
can be addressed canonically [119]. However, I must
emphasize that all the
conclusions about quark confinement obtained here disappear when we
consider the real theory of the strong interaction, namely 3+1 QCD.
As we will see later, in 3+1 QCD there are physical gauge degrees
of freedom (the transverse gauge field) which
cancel the effect of confinement from the linear potential due to the
elimination of the unphysical gauge degrees of freedom.  Nevertheless,
here it shows that eliminating the zero longitudinal momentum modes
leads to a trivial vacuum but the confinement mechanism can be manifested
in the Hamiltonian.  Undoubtedly, the manifestation of confinement
in 3+1 QCD with a trivial vacuum must be much more complicated
and we will address it later.

\hspace{0.3cm} It may be worth mentioning that
in the last few years, many studies on  nonperturbative features of
light-front dynamics were focused on the 1+1 field theory.  Typical
examples are: the discretized light-front quantization approach for the
bound states in the 1+1 field theory developed by Pauli and Brodsky [108,120],
spontaneous symmetry breaking and vacuum
structure in the 1+1 Yukawa model first considered by Harindranath
and Vary [121], the light-front Tamm-Dancoff approach
for bound state Fock space truncation discussed
by Perry {\em et al}. [112,113], and other extensive discussions
of various 1+1 field theory models of these aspects [122-128].

\vspace{15.5pt}
\noindent
{\bf IV-5. Phenomenological Hadronic Bound States on the Light-Front}

\hspace{0.3cm} At the present time, how to solve for the bound states,
discussed
above in this section, from 3+1 QCD is still unknown.  Hence,
it may be useful to have some insights
into the light-front behavior of the meson and baryon wave functions
which have been constructed phenomenologically in describing hadrons.
In fact, the phenomenological light-front meson and baryon bound
states have been studied extensively in the last few years,
this is called the relativistic quark model which was initially proposed
by Terent'ev in 1976 [129,130].
The motivation is to provide a simple relativistic
constituent quark model for mesons and baryons that
can yield a consistent description of the
hadronic properties for both low and high $Q^2$.

\hspace{0.3cm} The general construction of the phenomenological wave functions
is based on the non-relativistic constituent quark model.  The
constituent quark model has been very successful in the description
of hadronic structure based on a very simple structure, namely
that all mesons consist of a quark and antiquark pair and the baryons
are made of three constituent quarks, their wave functions satisfy
the $SU(6)$ classification and Zweig's rule which suppresses
particle production in favor of rearrangement of constituents
for hadrons [76-78,157,80]. However,
such a simple picture is very difficult to understand from
QCD, due to its nonrelativistic assumption and due to our belief
that the QCD vacuum must be very complicated so that hadrons must contain
an infinite number of quark-antiquark pairs and gluons.

\hspace{0.3cm} The light-front wave functions describe the relativistic
hadronic structure with a nonrelativistic form.  Furthermore, the simple
vacuum on the light-front
ensures the validity of the Fock state expansion of hadronic
wave functions.  With the assumption of existence of constituent
quarks (with masses of hundreds of MeVs), the
leading approximation to hadronic wave functions that consist of a
quark-antiquark pair for mesons and a three-quark cluster for baryons should
be a reasonable starting point.

\hspace{0.3cm} However, it is not so easy to identify the light-front hadronic
wave functions with hadronic states. The difficulty is their spin structure.
On the light-front, we are unable to kinematically
construct the hadronic wave functions with fixed spin.
The light-front wave functions discussed in the last section are
labeled by helicity not spin. In these calculations of the
parton distribution and fragmentation functions, the hadronic bound
states are defined or classified in terms of the helicity
(=chirality). However, when we use the light-front
wave functions to compute the hadronic structural quantities, such as
hadronic decay form factors and coupling constants,
we must have states with a definite spin. A general solution
to the spin problem on the light-front has not been found. However,
phenomenologically, the helicity part of the wave functions on the
light-front can be transformed to a light-front spin wave function part
via the so-called Melosh transformation (which is exact only
for free quark theory) such that the hadronic states
can be projected (approximately) from the set of light-front wave
functions with fixed helicity [131,132]. Here,
I list the detailed meson and baryon light-front wave functions that have been
used to calculate various hadronic quantities in the past few years.

\hspace{0.3cm} The general form of the phenomenological light-front hadronic
bound states has a similar structure to the constituent quark
model states: for meson states (Eq. (4.4) with only the $q\bar{q}$
Fock space sector),
\def\theequation{4.28}
\begin{equation}
\begin{array}{l}
  |P^+,P_{\bot},SS_3\rangle \\[\eqnskip]
  \hspace{0.8cm} = \ds\int \ds\frac{\,dxd^2k_{\bot}\,}{16\pi^3}
  \ds\sum_{\lambda_1\lambda_2}\Psi_m^{SS_3}(x,k_{\bot},\lambda_1,
  \lambda_2)|x,k_{\bot},\lambda_1;1-x,-k_{\bot},\lambda_2 \rangle \, ,
\end{array}
\end{equation}
and for baryon states (Eq. (4.4) with the three quark Fock space
sector),
\def\theequation{4.29}
\begin{equation}
\begin{array}{rl}
  |P^+,P_{\bot},SS_3\rangle = \!\!\! & \ds\sum_{\lambda_i} \ds\int \prod_{i}^2
  \ds\frac{\,dx_id^2k_{1\bot}\,}{16\pi^3}\Psi_b^{SS_3}(x_i,k_{i\bot},
  \lambda_i) \\[1.5\eqnskip]
  & \times |x_1,k_{1\bot},\lambda_1;x_2,k_{2\bot},\lambda_2;
  1-x_1-x_2,-(k_{1\bot}+k_{2\bot}),\lambda_3 \rangle \, ,
\end{array}
\end{equation}
where $\Psi^{SS_3}$ is the amplitude of the corresponding $q\bar{q}$ or
three quark sector (the wave function of the quark model):
\def\theequation{4.30}
\begin{equation}
  \Psi^{SS_3}={\cal F}\, \Xi^{SS_3}(k_{i\bot},\lambda_i)
  \Phi (x_i,k_{i\bot}) \, ,
\end{equation}
with ${\cal F}$ the flavor part of the wave function which is
the same as in the constituent quark model, and $\Xi$ and $\Phi$
are the spin and space parts that depend on the dynamics.
By using the Melosh transformation,
\def\theequation{4.31}
\begin{equation}
  R_M(k_{i\bot},m_i) = \ds\frac{\,m_i+x_iM_0-i\vec{\sigma}\cdot (\vec{n}\times
\vec{k}_{i\bot})\,}
  {\sqrt{(m_i+x_iM_0)^2+k_{i\bot}^2\,}} \, ,
\end{equation}
where $\vec{n}=(0,0,1)$, $\vec{\sigma}$ is the Pauli spin matrix,
$m_i$ the $i$-th constituent quark mass, and $M_0$ satisfies
\def\theequation{4.32}
\begin{equation}
  M_0^2 = \ds\sum_i \ds\frac{\,k_{i\bot}^2+m_i^2\,}{x_i} \, ,
\end{equation}
the light-front spin wave function is then given by
\def\theequation{4.33}
\begin{equation}
\begin{array}{l}
  \Xi_m^{SS_3}(k_{\bot},\lambda_1, \lambda_2) \\[\eqnskip]
  \hspace{0.5cm} = \ds\sum_{s_1,s_2}\langle \lambda_1|R_M^{\dagger}
  (k_{\bot},m_1)|s_1\rangle \langle \lambda_2|R_M^{\dagger}(-k_{\bot},m_2)
  |s_2 \rangle \biggl \langle \ds\frac{1}{\,2\,}s_1,\ds\frac{1}{\,2\,}
  s_2\biggl | SS_3\biggr \rangle \, ,
\end{array}
\end{equation}
for mesons; for baryons the spin part is rather complicated for a
detailed construction, see Ref. [133]. The momentum
part of the wave function is
\def\theequation{4.34}
\begin{equation}
  \Phi_m(x_i,k_{i\bot}) = {\cal N}_m\exp (-M_0^2/2\beta^2_m) \, ,
\end{equation}
for mesons and for baryons
\def\theequation{4.35}
\begin{equation}
  \Phi_b(x_i, k_{i\bot}) = {\cal N}_b
\ds\frac{1}{\,(1+M_0^2/\beta_b^2)^{3.5}\,} \, ,
\end{equation}
where ${\cal N}$ is the normalization constant and $\beta$ is a parameter
fixed by the data.  The exponential function for the momentum
wave function of baryons has also been used but Schlumpf found that the
pole-type wave
function is much better fit to the data with high momentum transfer
[133].

\hspace{0.3cm} These phenomenological light-front wave functions have been
widely used to calculate hadronic form factors and coupling constants;
the results are in pretty good agreement with experiments
for a very broad range of momentum transfer; (see Fig. 3 and 4),
and are much better than the nonrelativistic constituent
quark model and other phenomenological descriptions [134-137,133].

\hspace{0.3cm} There is also another phenomenological light-front wave function
derived by Chernyak from QCD sum rules [138]. Such a wave
function is valid for relatively high momentum transfer, and recently
it has also been applied to describe various hadronic processes
[97,139,140].

\hspace{0.3cm} However, all these are just a phenomenological understanding of
light-front hadronic wave functions.  If these wave functions
are very close to the real wave functions of hadrons, they
must be solved dynamically from QCD.  In the remaining sections of
this article, I will discuss QCD on the light-front.

\vspace{20pt}

\pagebreak

\begin{figure}[hptb] 
\vspace{5.7cm}
\figcaption{
Proton form factor calculated by the use of the phenomenological
light-front proton wave function [133]. The solid line is given by the
pole-type
wave function while the dashed line is based on the Gaussian-type wave
function.
}
\end{figure}

\begin{figure}[hptb] 
\vspace{6.5cm}
\figcaption{
Pion form factor calculated by the use of the phenomenological
light-front pion wave function [134]. The different lines correspond to
different
Parameters used in the phenomenological wave function.
}
\vspace{0.6cm}
\end{figure}

\vspace{6pt}
\begin{center}
{\normalsize \bf V. LIGHT-FRONT QCD}
\end{center}
\vspace{3pt}

\hspace{0.3cm} As we have mentioned several times, an intuitive physical
picture of high-energy processes is provided by the partonic
interpretation [11].
We have already seen that such a picture emerges most naturally
in the light-front canonical quantization with the light-front gauge
$A^+_a=A^0_a+A^3_a=0$. In the last two decades,
QCD with either the covariant gauge or the light-front gauge has been used
extensively to compute the hard scattering coefficients that are
relevant for the scale evolution of hadronic structure functions.
However, the QCD based exploration of the nonperturbative part, such
as the parton distribution functions, is still in a very preliminary
stage. One may also note that many practical calculations for short
distance QCD have been performed in the standard Feynman perturbation
theory with the light-front gauge [42,141,142,54,143].
In this scheme, Feynman rules
are derived by the use of the path integral approach with a
(light-front) gauge-fixing term [40] and various
hard scattering coefficients are calculated via either the operator
product expansion [42,141] or a Feynman diagrammatic approach [142,54,143].
Despite the successful perturbative calculations, it is still
not clear whether one can extend the Feynman approach to
nonperturbative studies, such as the QCD calculation of parton
distribution functions,
which requires the full information of the hadronic bound states.

\hspace{0.3cm} Current attempts to explore nonperturbative QCD on the
light-front,
as we have emphasized throughout this paper,
are based on the Hamiltonian theory, which is defined by
quantizing the theory on a light-front plane {\em via}
equal-$x^+$ (the light-front time) commutation relations,
as was first developed by Kogut and Soper for QED [21].
In the light-front QCD Hamiltonian theory,
the hadronic bound states may be obtained by
diagonalizing a light-front QCD Hamiltonian in a truncated
quark-gluon Fock space based on the old ideas of Tamm and
Dancoff [116,117]. In this formalism the low-energy
hadronic structures, namely the parton distribution functions,
parton fragmentation functions, various hadronic form factors
and strong interaction coupling constants, can be addressed
directly from QCD. Meanwhile, the short distance behavior can
also be studied in the same framework, i.e., in the old-fashioned
perturbation theory or explicitly in the $x^+$-ordered perturbative
QCD Hamiltonian theory based on equal-$x^+$ commutation relations
[47]. Indeed, the interpretation of high-energy processes
via the parton picture had led to extensive investigations
of perturbative field theory in the infinite momentum frame in
the early 70's, as we have mentioned in the introduction. At that
time Drell, Levy
and Yan [15] and Bjorken, Kogut and Soper [17]
had already pointed out that, in the old-fashioned
theory, the physical picture for various real physical processes
becomes much clearer.  The parton picture is just a typical example.

\hspace{0.3cm} Yet, beyond applications to the exclusive processes
given by Lepage and Brodsky in the early 80's [47],
the $x^+$-ordered Hamiltonian QCD has not been explored and
extensively utilized in the last decade. Only very recently, loop
calculations in the $x^+$-ordered perturbative QED and QCD
theory have been performed [144-147,82,148].
It is seen that there are severe divergences in light-front QCD which
are associated with light-front gauge singularities. However,
a systematic computational method, which involves various
regularization and renormalization schemes for light-front
singularities and ultraviolet transverse divergences, has not been
well established.  In fact, although the light-front gauge is known to
be a convenient gauge in practical QCD calculations for short distance
behavior, there are persistent concerns about its utility because of
its ``singular'' nature.  The study of nonperturbative QCD on the
light-front for hadronic bound states requires one to gain
{\em a priori} systematic control of such gauge singularities.
Clearly, it will be very helpful to first understand the canonical
structure of light-front QCD and the origin of various singular
problems before we go on to explore the nonperturbative properties
light-front QCD.

\hspace{0.3cm} In this section I will discuss the canonical form of light-front
QCD, the origin of the light-front gauge singularity,
the light-front two-component formulation of QCD, the
old-fashioned perturbation theory, various severe infra-red (IR)
divergences occurring in the old-fashioned light-front Hamiltonian
calculations for QCD and the associated nontrivial QCD structures.
I will also discuss some of the difficulties caused by the light-front
gauge singularity in applications to both the old-fashioned perturbative
calculations for short distance physics and the upcoming nonperturbative
investigations for hadronic bound states.

\vspace{15.5pt}
\noindent
{\bf V-1. Canonical Light-Front QCD}

\hspace{0.3cm} The QCD Lagrangian is
\def\theequation{5.1}
\begin{equation}
  {\cal L} = - \ds\frac{1}{\,2\,}{\rm Tr}(F^{\mu \nu}F_{\mu \nu})
  + \overline{\psi} (i\gamma_{\mu}D^{\mu}-m)\psi \, ,
\end{equation}
where $F^{\mu
\nu}=\partial^{\mu}A^{\nu}-\partial^{\nu}A^{\mu}-ig[A^{\mu},A^{\nu}]$,
$A^{\mu}=\sum_a A_a^{\mu}T^a$ is a
$3\times 3$ gluon field color matrix and the $T^a$ are the generators
of the $SU(3)$ color group: $[T^a,T^b]=if^{abc}T^c$ and ${\rm Tr}
(T^aT^b)=\frac{1}{2}\delta_{ab}$. The field variable $\psi$
describes quarks with three colors and $N_f$ flavors,
$D^{\mu}=\partial^{\mu}-igA^{\mu}$ is the
symmetric covariant derivative, and $m$ is an $N_f\times N_f$
diagonal quark mass matrix.  The Lagrange equations of motion are
well-known:
\def\theequation{5.2}
\begin{equation}
  \partial_{\mu}F_a^{\mu \nu} +gf^{abc}A_{b\mu}F_c^{\mu \nu}
  + g\overline{\psi}\gamma^{\nu}T^a \psi =0 \, ,
\end{equation}
\def\theequation{5.3}
\begin{equation}
  (i\gamma_{\mu}\partial^{\mu}-m+g\gamma_{\mu}A^{\mu})
  \psi = 0 \, .
\end{equation}

\hspace{0.3cm} {\em 1. Light-front QCD Hamiltonian}.
The canonical theory of QCD on the light-front
is constructed as follows.  The conjugate momenta of the field
variables $\{ A_a^{\mu}(x),\psi(x)\}$ are defined by
\def\theequation{5.4}
\begin{equation}
  E_a^{\mu}(x) = \ds\frac{\partial{\cal L}}{\,\partial (\partial^-A_{a \mu})\,}
  = - \ds\frac{1}{\,2\,}F_{a}^{+\mu}(x) \, ,
\end{equation}
\def\theequation{5.5}
\begin{equation}
  \pi_{\psi} (x) =  \ds\frac{\partial{\cal L}}{\,\partial (\partial^- \psi )\,}
  = \ds\frac{i}{\,2\,}\overline{\psi}\gamma^+ = i\psi_+^{\dagger}(x) \, .
\end{equation}
In terms of the field variables and their conjugates, the Lagrangian
density can be rewritten as follows:
\def\theequation{5.6}
\begin{equation}
  {\cal L} = \biggl \{ \ds\frac{1}{\,2\,}F_a^{+i}(\partial^-A_a^i)
  + i\psi_+^{\dagger}(\partial^- \psi_+)\biggr \}
  - {\cal H} - \bigl \{ A_a^- {\cal C}_a + \psi_-^{\dagger}{\cal C}\bigr \} \,
,
\end{equation}
where
\def\theequation{5.7}
\begin{equation}
\begin{array}{rl}
  {\cal H} = & \ds\frac{1}{\,2\,}(E_a^{-2}+B_a^{-2})
  +\psi_+^{\dagger}\{ \alpha_{\bot}\cdot (i\partial_{\bot}
  +gA_{\bot}) + \beta m\} \psi_-  \\[1.5\eqnskip]
  & + \biggl ( \ds\frac{1}{\,2\,}\partial^+(E_a^- A_a^-)
  - \partial^i (E_a^i A_a^-) \biggr ) \, ,
\end{array}
\end{equation}
and
\def\theequation{5.8}
\begin{equation}
  {\cal C}_{a} = \ds\frac{1}{\,2\,}(\partial^+ E_a^- + g f^{abc}
  A_b^+ E_c^-) - (\partial^i E_a^i + g f^{abc} A_b^i E_c^i)
  + g\psi_+^{\dagger} T^a \psi_+  \, ,
\end{equation}
\def\theequation{5.9}
\begin{equation}
  {\cal C} = (i\partial^+ + g A^+) \psi_- - (i \alpha_{\bot}
  \cdot \partial_{\bot} + g \alpha_{\bot} \cdot A_{\bot}
  + \beta m) \psi_+ \, .
\end{equation}
In Eq. (5.7), we have defined $B_a^-=F_a^{12}$ as the
longitudinal component of the light-front color magnetic field.

\hspace{0.3cm} The reason for writing the Lagrangian in the above form is to
make the Hamiltonian density and also the dynamical variables and
constraints manifest.  Here the time derivatives have explicitly
been separated from the others, from which it immediately follows that only
the transverse gauge fields $A_a^i$ and the up-component
quark fields $\psi_+$ and $\psi_+^{\dagger}$ are dynamical
variables.  The Hamiltonian density ${\cal H}$ contains three parts; the first
part involves
the light-front color electric and magnetic fields; the
second, the usual quark Hamiltonian with coupling to the gauge
field and the last a surface term.  The last terms in this new
form of the QCD Lagrangian indicate
that the longitudinal gauge field $A_a^-$ and the down-component
quark fields $\psi_-$ are only
Lagrange multipliers for the constraints ${\cal C}_{a},{\cal C}=0$.
This is true even for the free field theory, as we have
already seen in Sec. II.
With the use of the equations of motion, one can verify that
the gauge field constraint, ${\cal C}_a = 0$, is in fact
the light-front Gauss law which is an intrinsic property of
gauge theory.  The fermion constraint, ${\cal C} = 0$, is
purely a consequence of using the light-front coordinates.
The existence of constraint terms simply implies that QCD in the
light-front coordinates is a generalized Hamiltonian
system.  These constraints are all secondary,
first-class constraints in the Dirac procedure of quantization
[86].  To obtain a canonical formulation of
light-front QCD for non-perturbative calculations we need to
explicitly solve the constraints, namely to determine the Lagrange
multipliers, to all orders of the coupling constant.  Generally, it
is very difficult to analytically determine the Lagrange
multipliers from the constraints ${\cal C}_a, {\cal C}=0$
since they are coupled by $A_a^+$.  But if we choose the
light-front gauge, $ A_a^+ (x) =0$, the constraints are
reduced to solvable one-dimensional differential equations:
\def\theequation{5.10}
\begin{equation}
  \ds\frac{1}{\,2\,}\partial^+E_a^- = \partial^i E_a^i + g (f^{abc}
  A_b^iE_c^i - \psi_+^{\dagger} T^a \psi_+) \, ,
\end{equation}
\def\theequation{5.11}
\begin{equation}
  i\partial^+ \psi_- = ( i \alpha_{\bot} \cdot \partial_{\bot}
  + g \alpha_{\bot} \cdot A_{\bot} + \beta m) \psi_+ \, .
\end{equation}

\hspace{0.3cm} Now the light-front QCD Hamiltonian can be expressed in terms of
the physical degrees of freedom:
\def\theequation{5.12}
\begin{equation}
\begin{array}{rl}
  H = \!\!\! & \ds\int dx^- d^2 x_{\bot}\biggl \{ \ds\frac{1}{\,2\,}(\partial^i
  A_a^j)^2+gf^{abc}A_a^iA_b^j\partial^iA_c^j
  + \ds\frac{\,g^2\,}{4}f^{abc}f^{ade}A_b^iA_c^jA_d^iA_e^j \\[1.5\eqnskip]
  & + \biggl [ \psi_+^{\dagger}
  \bigl \{ \sigma_{\bot}\cdot (i\partial_{\bot}+gA_{\bot})-im\bigr \}
  \biggl ( \ds\frac{1}{\,i\partial^+\,}\biggr ) \bigl \{ \sigma_{\bot}
  \cdot (i\partial_{\bot}+g A_{\bot})+im\bigr \}
  \psi_+ \biggr ] \\[1.5\eqnskip]
  & + g\partial^iA_a^i \biggl ( \ds\frac{1}{\,\partial^+\,}
  \biggr ) (f^{abc}A_b^j \partial^+ A_c^j + 2 \psi_+^{\dagger}
  T^a \psi_+) \\[1.5\eqnskip]
  & + \ds\frac{\,g^2\,}{2}\biggl ( \ds\frac{1}{\,\partial^+\,}
  \biggr ) (f^{abc} A_b^i \partial^+ A_c^i + 2 \psi_+^{\dagger}
  T^a \psi_+ )\biggl ( \ds\frac{1}{\,\partial^+\,}\biggr ) \\[1.5\eqnskip]
  & \cdot (f^{ade}A_d^j\partial^+ A_e^j + 2 \psi_+^{\dagger}
  T^a \psi_+ )\biggr \} \\[1.5\eqnskip]
  & + {\rm surface terms},
\end{array}
\end{equation}
where the operator $\frac{1}{\partial^+}$ is defined by Eq. (2.16),
and the surface terms are given by the last term in Eq. (5.7).

\hspace{0.3cm} {\em 2. A phase space approach to light-front quantization}.
A self-consistent canonical formulation requires that the resulting
Hamiltonian must generate the correct equations of motion for the
physical degrees of freedom $(A_a^i, \psi_+, \psi^{\dagger}_+)$.
To reproduce the Lagrangian equations of motion, we need to find
consistent commutators for physical field variables. In the
light-front gauge, the (symmetricalized) Lagrangian of Eq. (5.6)
can be rewritten as
\def\theequation{5.13}
\begin{equation}
  {\cal L} = \ds\frac{1}{\,2\,}(\partial^+ A_a^i \partial^- A_a^i
  + i \psi_+^{\dagger} \partial^- \psi_+ - i(\partial^-
  \psi_+^{\dagger}) \psi_+) - {\cal H} \, .
\end{equation}
Thus the light-front QCD phase space is spanned by the field variables,
$A_a^i, \psi_+, \psi_+^{\dagger} $ and their canonical momenta,
${\cal E}_a^i=\frac{1}{2}\partial^+ A_a^i$,
$\pi_{\psi_+} =\frac{i}{2}\psi_+^{\dagger}$,
$\pi_{\psi^{\dagger}_+}=-\frac{i}{2}\psi_+$. The phase space structure
which determines the Poisson brackets of its variables can be found by the
Lagrangian one-form ${\cal L}dx^+$ (apart from a total light-front
time derivative),
\def\theequation{5.14}
\begin{equation}
\begin{array}{rl}
  {\cal L}dx^+ = & \ds\frac{1}{\,2\,}2({\cal E}_a^i dA_a^i + \pi_{\psi_+}
  d\psi_+ + d\psi_+^{\dagger} \pi_{\psi_+^{\dagger}} - A_a^i
  d{\cal E}_a^i - d\pi_{\psi_+}\psi_+ - \psi_+^{\dagger}
  d\pi_{\psi_+^{\dagger}}) \\[1.5\eqnskip]
  & - {\cal H} dx^+  \\[\eqnskip]
  = & \ds\frac{1}{\,2\,} q^{\alpha} \Gamma_{\alpha \beta}
  dq^{\beta} - {\cal H} dx^+ \, ,
\end{array}
\end{equation}
where the first term on the right-hand side is called the canonical
one-form of the phase space, and quark fields are anticommuting
$c$-numbers (Grassmann variables).  Correspondingly, the symplectic
structure or the Poisson bracket of the phase space is given by
\def\theequation{5.15}
\begin{equation}
  \omega = \ds\frac{1}{\,2\,}\Gamma_{\alpha \beta}dq^{\alpha} dq^{\beta} \quad
\;
  {\rm or}\quad \; [q^{\beta},q^{\alpha}]_p = \Gamma_{\alpha \beta}^{-1} \, .
\end{equation}
Canonical quantization is then obtained by replacing the Poisson brackets
by the equal-$x^+$ commutation relations:
\def\theequation{5.16}
\begin{equation}
  [q^{\beta}(x), q^{\alpha}(y)]_{x^+=y^+} = i \Gamma_{\alpha
  \beta}^{-1} \, .
\end{equation}
Explicitly, we have
\def\theequation{5.17}
\begin{equation}
  \bigl [ A_a^i(x),{\cal E}_b^j(y)\bigr ]_{x^+=y^+} = i \ds\frac{1}{\,2\,}
  \delta_{ab}\delta^{ij}\delta^3 (x-y) \, ,
\end{equation}
\def\theequation{5.18}
\begin{equation}
  \bigl \{ \psi_+(x),\pi_{\psi_+}(y)\bigr \}_{x^+=y^+}
  = i \ds\frac{\,\Lambda_+\,}{2}\delta^3 (x-y) \, ,
\end{equation}
\def\theequation{5.19}
\begin{equation}
  \bigl \{ \psi_+^{\dagger}(x),\pi_{\psi_+^{\dagger}}(y)\bigr \}_{x^+=y^+}
  = -i \ds\frac{\,\Lambda_+\,}{2}\delta^3(x-y) \, ,
\end{equation}
or simply
\def\theequation{5.20}
\begin{equation}
   \bigl [ A_a^i(x),A_b^j(y)\bigr ]_{x^+=y^+}
  = -i \ds\frac{\,\delta_{ab}\delta^{ij}\,}{4}\epsilon(x^- - y^-) \delta^2
  (x_{\bot} - y_{\bot}) \, ,
\end{equation}
\def\theequation{5.21}
\begin{equation}
  \bigl \{ \psi_+(x),\psi_+^{\dagger}(y)\bigr \}_{x^+=y^+}
  = i\Lambda_+ \delta^3(x-y) \, .
\end{equation}
{}From these commutation relations it is straightforward to verify that
the Hamiltonian equations of motion are consistent with Eqs. (5.2)
and (5.3):
\def\theequation{5.22}
\begin{equation}
\begin{array}{rl}
  \partial^- \psi_+ = & \ds\frac{1}{\,i\,}[\psi_+,H] \\[1.5\eqnskip]
  = & \biggl \{ igA^- - \bigl \{ \alpha_{\bot} \cdot
  (i\partial_{\bot}+gA_{\bot})+\beta m\bigr \} \\[1.5\eqnskip]
  & \cdot \ds\frac{1}{\,\partial^+\,}
  \bigl \{ \alpha_{\bot}\cdot (i\partial_{\bot}+gA_{\bot})+\beta m\bigr \}
  \biggr \} \psi_+ \, ,
\end{array}
\end{equation}
\def\theequation{5.23}
\begin{equation}
\begin{array}{rl}
  \partial^- A_a^i = & \ds\frac{1}{\,i\,}[A_a^i,H] \\[1.5\eqnskip]
  = & \ds\frac{1}{\,\partial^+\,}[D_{ab}^jF_b^{j i}
  -D_{ab}^iE_b^- - gj_a^i - gf^{abc}A_b^-
  \partial^+ A_c^i] \, ,
\end{array}
\end{equation}
where $D_{ab}^i=\delta_{ab}\partial^i-gf^{abc}A_c^i$.

\vspace{15.5pt}
\noindent
{\bf V-2. Light-Front Gauge Singularity}

\hspace{0.3cm} Note that the above canonical formulation does not completely
define theory for practical computations due to the existence of the gauge
singularity.  The gauge singularity is perhaps the most difficult
problem in non-abelian gauge theory that has not been completely
solved since it was developed. In light-front QCD, it arises from
the elimination of the unphysical gauge degrees of freedom. To eliminate
the unphysical degrees of freedom on the light-front, we need to solve
the constraint equations which depend on the definition of the
operator $1/\partial^+$. In Sec. II, we defined this operator
by Eq. (2.16). In general, we have
\def\theequation{5.24}
\begin{equation}
  \biggl ( \ds\frac{1}{\,\partial^+\,}\biggr ) f(x^-,x^+,x_{\bot})
  = \ds\frac{1}{\,4\,} \ds\int_{-\infty}^{\infty} dx_1^-
  \varepsilon (x^- - x_1^-) f(x_1^-,x^+,x_{\bot})
  + C(x^+,x_{\bot}) \, ,
\end{equation}
where $\varepsilon (x)$ has been given in Eq. (2.16),
and $C(x^+,x_{\bot})$
is a $x^-$ independent constant. The reason for choosing the definition of
Eq. (2.16) is that the light-front initial value problem is associated
with the longitudinal boundary condition. A suitable definition of
$1/\partial^+$ can uniquely
determine the initial value problem at $x^+=0$ for independent field
variables [32].  Rohrlich has shown that such a suitable
definition is to set $C(x^+, x_{\bot}) = 0$ [i.e, Eq. (2.16)]
[22].  This corresponds to choosing an antisymmetric
boundary condition for field variables in the light-front
longitudinal direction.  I must emphasize here that this is not
necessarily the best choice for the operator $\frac{1}{\partial^+}$.
As we will see later this definition still causes many undesirable
problems associated with the gauge singularity, which even exist
in perturbative calculations. It is still an open question whether
one can find a better definition to solve the light-front
gauge ambiguity in Hamiltonian formalism.

\hspace{0.3cm} By using Eq. (2.16) and the following identity [149],
\def\theequation{5.25}
\begin{equation}
  \ds\frac{1}{\,2\,}\ds\int_{-\lambda/2}^{\lambda/2} dx^-
  \varepsilon (x^- - x'^-) \varepsilon (x^- - x''^-)
  = - |x'^- - {x''}^-| + \ds\frac{1}{\,2\,} \lambda \, ,
\end{equation}
where the parameter $\lambda$ denotes the distance between two boundary
points in the longitudinal direction, the constraint gauge and quark
components can be determined:
\def\theequation{5.26}
\begin{equation}
  E_a^- (x) = -\partial^i A^i_a (x) - \ds\frac{\,g\,}{2}
  \ds\int_{-\infty}^{\infty} d{x'}^- \varepsilon (x^- {x'}^-)
  j^+_a (x^+, {x'}^-, x_{\bot}) \, ,
\end{equation}
\def\theequation{5.27}
\begin{equation}
\begin{array}{rl}
  A_a^-(x)= & - \ds\frac{1}{\,2\,}\ds\int_{-\infty}^{\infty}d{x'}^-
  \bigl [ \varepsilon (x^--{x'}^-)\partial^iA_a^i(x^+,{x'}^-,x_{\bot})
\\[1.5\eqnskip]
  & +g|x^--{x'}^-|j^+_a(x^+,{x'}^-,x_{\bot})
  \bigr ]
  + \biggl ( \ds\lim_{\lambda \rightarrow \infty} \lambda \biggr )
  \ds\frac{\,g\,}{4}j^+_a(x^-,x) \, ,
\end{array}
\end{equation}
\def\theequation{5.28}
\begin{equation}
\begin{array}{rl}
  \psi_- (x) = \!\!\! & - \ds\frac{i}{\,4\,} \ds\int_{-\infty}^{\infty}
  d{x'}^-d^2x_{\bot}' \varepsilon (x^- - {x'}^-)
  \delta^2 (x_{\bot} - x'_{\bot}) \\[1.5\eqnskip]
  & \times \bigl [ \alpha_{\bot}\cdot \bigl
(i\partial'_{\bot}+gA_{\bot}(x')\bigr )
  +\beta m\bigr ] \psi_+ (x') \, ,
\end{array}
\end{equation}
where $j^+_a = 2 \psi^{\dagger}_+ T^a \psi_+$.
In this solution, the first surface term at the longitudinal
infinity in Eq. (5.7) vanishes.
Moreover, it is reasonable to assume that the transverse color
electric fields $E_a^i$ as well as $A_a^i$ vanish as $O(r^{-2})$
and $O(r^{-1})$ at $r=|x_{\bot}| \rightarrow \infty$ because
the gauge freedom is totally fixed at transverse infinity.
Thus the second surface term vanishes as well. Now the light-front
QCD Hamiltonian is simply given by
\def\theequation{5.29}
\begin{equation}
\begin{array}{rl}
  H = & \ds\int dx^- d^2 x_{\bot} \biggl \{ \ds\frac{1}{\,2\,} (E_a^{-2}
  +B_a^{-2})+\psi_+^{\dagger}\bigl \{ \alpha^i
  (i\partial^i+gA^i)+\beta m \bigr \} \psi_-
  \biggr \} \\[1.5\eqnskip]
  = & \ds\int dx^- d^2 x_{\bot}\biggl \{ \ds\frac{1}{\,2\,}(\partial^iA_a^j)^2
  +gf^{abc}A_a^iA_b^j\partial^iA_c^j
  +\ds\frac{\,g^2\,}{4}f^{abc}f^{ade}A_a^iA_b^jA_d^i
  A_e^j \\[1.5\eqnskip]
  & + \ds\frac{1}{\,4\,} \ds\int_{-\infty}^{\infty} dx'^-
  \bigl [ 2g\partial^iA_a^i\varepsilon(x^- - x'^-)
  \rho_a(x'^-, x) \\[1.5\eqnskip]
  & -i\psi_+^{\dagger}\bigl \{ \alpha^i
  (i\partial^i+gA^i)+\beta m\bigr \} \varepsilon(x^-
  - x'^-) \bigl \{ \alpha^j(i\partial^j+gA^j)+\beta m \bigr \}
  \psi_+ \bigr ] \\[\eqnskip]
  & - \ds\frac{\,g^2\,}{4} \ds\int_{-\infty}^{\infty} dx'^-
  \rho_a(x^-, x)|x^- - x'^-| \rho_a(x'^-, x)
  \biggr \} \\[1.5\eqnskip]
  & + \biggl ( \ds\lim_{\lambda \rightarrow \infty}\lambda
  \biggr ) \ds\frac{\,g^2\,}{4} \ds\int d^2 x_{\bot} \biggl \{
  \ds\int_{-\infty}^{\infty}dx^- \rho_a(x^-,x) \biggr \}^2 \, .
\end{array}
\end{equation}
{}From Eq. (5.29), we see that, after eliminating the longitudinal gauge
field in the light-front gauge, a color charge instantaneous interaction
emerges in the Hamiltonian, with a linear potential in the longitudinal
direction. In addition, Eq. (5.29) also contains a singular
boundary term
(the last term) associated with the linear instantaneous interaction.
There is no such term in the Coulomb gauge because the Coulomb potential
vanishes at spatial infinity. This term is very similar to that in the $1+1$
QCD case, as a result of eliminating the unphysical gauge degrees of
freedom. However, the role it plays here is totally different from that in
1+1 QCD.
As we will show later, in perturbation theory this term is
regularized by the distribution function of the product of two
principal value prescriptions and leads to the cancellation of
the light-front linear infrared divergences.  For physical states,
the requirement of the finiteness of energy results in asymptotic
equations for the transverse gauge fields, which shows that the
asymptotic transverse gauge fields do not vanish at
longitudinal infinity, as we shall see later.

\hspace{0.3cm} The gauge singularity manifests itself more clearly in momentum
space.
In momentum space, the gauge singularity means that the constraints
Eqs. (5.8) and (5.9) do not uniquely
determine the dependent fields
in terms of physical fields at longitudinal momentum
$k^+=0$.  In other words, simply choosing $A_a^+=0$ with the
definitions of Eq. (2.16) for $\frac{1}{\partial^+}$ does not
completely fix the gauge degrees of freedom  [150].
As we have mentioned many times throughout this paper,
the $k^+=0$ particles are the constituents leading to a complicated
vacuum.  From the momentum representation of Eq. (2.16),
\def\theequation{5.30}
\begin{equation}
\begin{array}{rl}
  \!\!\! \biggl ( \ds\frac{1}{\,\partial^+\,}\biggr )^nf(x^-) = \!\!\!\! &
\biggl (
  \ds\frac{1}{\,4\,}\biggr )^n\ds\int_{-\infty}^{\infty}dx_1^-\cdots dx_n^-
  \epsilon (x^- - x_1^-)\cdots \epsilon (x_{n-1}^-
  -x_n^-)f(x_n^-) \\[1.5\eqnskip]
  & \longrightarrow \biggl [ \ds\frac{1}{\,2\,}\biggl (
  \ds\frac{1}{\,k^++i\epsilon\,}+\ds\frac{1}{\,k^+-i\epsilon\,}\biggr )
  \biggr ]^nf(k^+)=\ds\frac{1}{\,[k^+]^n\,}f(k^+) \, .
\end{array}
\end{equation}
Clearly the $k^+=0$ modes are removed with this
definition, the singularity of $\frac{1}{k^+}$ is regularized,
and the vacuum is guaranteed to be trivial. Of course,
the question now is what happens to the nontrivial QCD structure
within a trivial vacuum?  Unlike the sigma model,
here it is not obvious that we can recover the effect of
zero modes by the use of a shifting technique. However,
although the $k^+=0$ singularity is removed by Eq. (2.16),
the infrared divergences from the small longitudinal momentum,
i.e., surrounding the $k^+=0$ region, are still present in the
above Hamiltonian. These infrared divergences are hidden in
the three point quark-gluon and gluon-gluon vertices, and
are associated with the QCD vacuum properties.
Recently, Wilson pointed out that, due to the different structure
of the light front power counting, we can introduce noncanonical
counterterms to remove the light-front infrared divergences
[152] in light-front QCD, as we will discuss later.
These counterterms may be the sources of the nontrivial
vacuum effect.  In order to understand the behavior of
infrared divergences, we will first discuss how they occur in
perturbation computations.

\vspace{15.5pt}
\noindent
{\bf V-3. Perturbative Light-Front QCD}

\hspace{0.3cm} The light-front gauge singularity discussed above will lead to
severe divergences even in perturbation theory, although
Eq. (2.16) provides a well-defined regulator (a
generalized principal value prescription, Eq. (5.30))
for the small $k^+$ momentum.  In Feynman's
perturbation theory, the use of the principal value
prescription still leads to ``spurious'' poles in
the light-front Feynman integrals, which prohibit any
continuation to Euclidean space (Wick rotation) and hence the
use of standard power counting arguments for Feynman loop
integrals [59].  This causes difficulties in addressing
renormalization of QCD in Feynman perturbation theory with the light-front
gauge. In the last decade there are many investigations attempting
to solve this problem.  One excellent solution is given by Mandelstam
and Leibbrandt, i.e., the Mandelstam-Leibbrandt (ML) prescription
[57,58], which allows continuation to
Euclidean space and hence power counting. It has also been shown
that, with the ML prescription, the {\em multiplicative}
renormalization in the two-component light-front QCD Feynman formulation is
restored [60].

\hspace{0.3cm} Unfortunately, the ML prescription cannot be
applied to equal-$x^+$ quantization because the ML prescription is
defined by a boundary condition which depends on $x^+$ itself
[151] and is not allowed in equal-$x^+$ canonical theory.
Yet, as was pointed out recently by Wilson [152],
light-front power counting differs completely from the power counting
in equal-time quantization in that noncanonical counterterms are
allowed in light-front field theory.  In other words,
multiplicative renormalization is not required in
light-front QCD.  Furthermore, the current attempts to
understand nonperturbative QCD in light-front coordinates is based
on the $x^+$-ordered (old-fashioned) diagrams in which no Feynman
integral is involved. Thus the power counting
criterion for Feynman loop integrals is no longer available in
light-front QCD Hamiltonian calculations. In Hamiltonian perturbation
theory with the principle value prescription, light-front QCD contains
severe linear and logarithmic infrared divergences.
Here I will give some results from the $x^+$-ordered perturbative
loop calculations and renormalization of light-front QCD Hamiltonian
theory up to one-loop [82,148], where the infrared
divergences are systematically analyzed. Since
light-front power counting
allows noncanonical counterterms, a complete understanding of
renormalized light-front QCD may not be worked out within perturbation
theory; new renormalization and regularization approaches
need to be developed, which we will discuss in the next
section.

\hspace{0.3cm} The following calculations are performed in the formulation of
the
light-front Hamiltonian perturbation theory with a two-component
representation [82].  The light-front QCD Hamiltonian
can be rewritten as a free term plus interactions,
\def\theequation{5.31}
\begin{equation}
  H = \ds\int dx^- d^2 x_{\bot} ({\cal H}_0 + {\cal H}_{int} )
  = H_0 + H_I \, .
\end{equation}
It is easy to show that
\def\theequation{5.32}
\begin{equation}
  {\cal H}_0 = \ds\frac{1}{\,2\,}(\partial^iA_a^j)(\partial^i A_a^j)
  +\varphi^{\dagger}\biggl ( \ds\frac{\,-\nabla^2 + m^2\,}{i\partial^+}
  \biggr ) \varphi \, ,
\end{equation}
\def\theequation{5.33}
\begin{equation}
  {\cal H}_{int} = {\cal H}_{qqg} + {\cal H}_{ggg} +
  {\cal H}_{qqgg} + {\cal H}_{qqqq} + {\cal H}_{gggg} \, ,
\end{equation}
and
\def\theequation{5.34}
\begin{equation}
\begin{array}{rl}
  {\cal H}_{qqg} = & g\varphi^{\dagger}\biggl \{ - 2\biggl (
  \ds\frac{1}{\,\partial^+\,}\biggr ) (\partial \cdot
  A_{\bot})+\sigma \cdot A_{\bot}\biggl (
  \ds\frac{1}{\,\partial^+\,}\biggr ) (\sigma \cdot \partial_{\bot}
  +m) \\[1.5\eqnskip]
  & + \biggl ( \ds\frac{1}{\,\partial^+\,}\biggr )
  (\sigma \cdot \partial_{\bot}-m)\sigma \cdot A_{\bot}
  \biggr \} \varphi \, ,
\end{array}
\end{equation}
\def\theequation{5.35}
\begin{equation}
  {\cal H}_{ggg} = g f^{abc} \biggl \{ \partial^i A_a^j A_b^i A_c^j
  + (\partial^i A_a^i) \biggl ( \ds\frac{1}{\,\partial^+\,} \biggr )
  (A_b^j \partial^+ A_c^j) \biggr \} \, ,
\end{equation}
\def\theequation{5.36}
\begin{equation}
\begin{array}{rl}
  {\cal H}_{qqgg} = & g^2 \biggl \{ \varphi^{\dagger} \sigma \cdot
  A_{\bot} \biggl ( \ds\frac{1}{\,i\partial^+\,}\biggr ) \sigma \cdot
  A_{\bot}\varphi \\[1.5\eqnskip]
  & +2\biggl ( \ds\frac{1}{\,\partial^+\,}
  \biggr ) (f^{abc}A_b^i\partial^+A_c^i)\biggl ( \ds\frac{1}{\,\partial^+\,}
  \biggr ) (\varphi^{\dagger}T^a\varphi ) \biggr \} \\[1.5\eqnskip]
  = & {\cal H}_{qqgg1} + {\cal H}_{qqgg2} \, ,
\end{array}
\end{equation}
\def\theequation{5.37}
\begin{equation}
  {\cal H}_{qqqq} = 2g^2 \biggl \{ \biggl ( \ds\frac{1}{\,\partial^+\,}
  \biggr )(\varphi^{\dagger}T^a\varphi ) \biggl ( \ds\frac{1}{\,\partial^+\,}
  \biggr ) (\varphi^{\dagger}T^a\varphi ) \biggr \} \, ,
\end{equation}
\def\theequation{5.38}
\begin{equation}
\begin{array}{rl}
  {\cal H}_{gggg} = \!\!\! & \ds\frac{\,g^2\,}{4}f^{abc}f^{ade}
  \biggl \{ A_b^iA_c^jA_d^iA_e^j+2\biggl ( \ds\frac{1}{\,\partial^+\,}
  \biggr )(A_b^i\partial^+A_c^i)\biggl ( \ds\frac{1}{\,\partial^+\,}
  \biggr ) (A_d^j\partial^+A_e^j)\biggr \} \\[1.5\eqnskip]
  = \!\!\! & {\cal H}_{gggg1} + {\cal H}_{gggg2} \, ,
\end{array}
\end{equation}
where $\varphi$ is the light-front fermion two-component
representation (see Eq. (2.34) discussed in Sec. II).

\hspace{0.3cm} The $x^+$-ordered light-front QCD perturbative theory can be
obtained from the familiar perturbation expansion in quantum mechanics.
The perturbation expansion of a bound state is given by
(in the Rayleigh-Schr\"{o}dinger perturbation theory):
\def\theequation{5.39}
\begin{equation}
  |\Psi \rangle = \ds\sum_{n=0}^{\infty}\biggl ( \ds\frac{Q}{\,E_0-H_0\,}
  (H'_I)\biggr )^n|\Phi \rangle \, ,
\end{equation}
where $|\Phi \rangle$ is a unperturbative state, $Q$ and $H'_I$
are defined by:
\def\theequation{5.40}
\begin{equation}
  Q = |\Phi \rangle \langle \Psi |\, , \quad \quad H'_I=H_I-\Delta E\, , \quad
\quad
  \Delta E=\langle \Phi |H_I|\Psi \rangle \, .
\end{equation}
With this perturbative expansion formula, the mass, wave functions,
and coupling constants renormalizations can be expressed
as follows.  For the convenience of practical calculations,
we consider the expressions in momentum space.

\hspace{0.3cm} i). {\em Wavefunction renormalization}: In momentum space, the
perturbative expansion of a state is given by
\def\theequation{5.41}
\begin{equation}
\begin{array}{rl}
  |\Psi \rangle = & \Biggl \{ |\Phi \rangle
  +{\ds\sum_{n_1}}'\ds\frac{\,|n_1\rangle \langle n_1|H'_I|\Phi \rangle\,}
  {\,p^--p_{n_1}^-+i\epsilon\,} \\[2\eqnskip]
  & + {\ds\sum_{n_1n_2}}' \ds\frac{|n_1\rangle \langle n_1|H'_I|n_2\rangle
\langle n_2|H'_I|\Phi \rangle\,}
  {\,(p^--p_{n_1}^-+i\epsilon)(p^--p_{n_2}^-+i\epsilon )\,}+\cdots \Biggr \} \,
,
\end{array}
\end{equation}
which has not been normalized, where $| n_1 \rangle, | n_2 \rangle,
\cdots $ are properly symmetrized (antisymmetrized) states with
respect to identical bosons (fermions) in the states and $\sum'$ in
Eq. (5.41) sums over all intermediate states except the initial
state $| \Phi \rangle$. The normalized wave function is defined by
$| \Psi' \rangle = \sqrt{Z_{\Phi}} | \Psi \rangle $,
where the factor $Z_{\Phi}$ is the wavefunction renormalization
constant:
\def\theequation{5.42}
\begin{equation}
  Z_{\Phi}^{-1} = \langle \Psi | \Psi \rangle
  = 1 + \ds\sum_{n_1}{'}\ds\frac{|\langle n_1|H'_I|\Phi \rangle|^2\,}
  {\,(p^--p_{n_1}^-+i\epsilon )^2\,}+\cdots \, .
\end{equation}

\hspace{0.3cm} ii). {\em Mass renormalization}.
The mass correction can then be computed from the ``energy-level'' shift,
i.e., the correction to the energy of an on-mass-shell particle.  It
is obvious that the perturbative correction to the light-front energy
($p^-$) is given by
\def\theequation{5.43}
\begin{equation}
\begin{array}{lll}
  \delta p^- & = & \langle \Phi |(H-H_0)|\Psi \rangle
  = \langle \Phi |H_I|\Psi \rangle \\[\eqnskip]
  & = & \langle \Phi |H_I|\Phi \rangle
  +{\ds\sum_{n_1}}'\ds\frac{\,|\langle n_1|H_I|\Phi \rangle
|^2\,}{\,p^--p_{n_1}^-+i\epsilon\,}
  +\cdots \, .
\end{array}
\end{equation}
Using the mass-shell equation $m^2=p^+p^--{\bf p}_{\bot}^2$, and recalling
that $p^+$ and ${\bf p}_{\bot}$ are the conserved light-front kinematical
momenta, we obtain the mass renormalization in the old-fashioned
perturbative light-front field theory:
\def\theequation{5.44}
\begin{equation}
  \delta m^2=p^+ \delta p^- =  p^+ \langle \Phi |H_I
  |\Phi \rangle + p^+ {\ds\sum_{n_1}}'
  \ds\frac{\,|\langle n_1|H_I|\Phi \rangle |^2\,}{\,p^--p_{n_1}^-+i\epsilon\,}
  +\cdots \, .
\end{equation}

\hspace{0.3cm} iii). {\em Coupling constant renormalization}. The coupling
constant renormalization is obtained by the perturbative calculation
of various matrix elements of the vertices in $H_I$.  Consider
a vertex $H_I^i$ that is proportional to the coupling constant
$g$, we have
\def\theequation{5.45}
\begin{equation}
\begin{array}{lll}
  \langle \Psi'_f|H_I^i|\Psi'_i \rangle & \equiv & Z_g
  \sqrt{Z_iZ_f\,}\langle \Psi_f|H_I^i|\Psi_i \rangle  \\[\eqnskip]
  & = & \langle \Phi_f|H_I^i|\Phi_i \rangle
  + {\ds\sum_{n_1}}'\ds\frac{\,\langle \Phi_f|H'_I|n_1\rangle \langle
n_1|H_I^i|\Psi_i\rangle\,}
  {p_f^--p_{n_1}^-+i\epsilon} \\[2\eqnskip]
  & & + {\ds\sum_{n_1}}'\ds\frac{\,\langle \Phi_f|H_I^i|n_1\rangle \langle
n_1|H'_I|\Psi_i\rangle\,}
  {p_i^--p_{n_1}^-+i\epsilon} \\[2\eqnskip]
  & & + {\ds\sum_{n_1,n_2}}'\ds\frac{\,\langle \Phi_f|H'_I|n_1\rangle \langle
n_1|H'_I|n_2\rangle \langle n_2|H_I^i]|\Phi_i \rangle\,}
  {(p_f^--p_{n_1}^-+i\epsilon )(p_f^--p_{n_2}^-+i\epsilon )} \\[2\eqnskip]
  & & + {\ds\sum_{n_1,n_2}}'\ds\frac{\,\langle \Phi_f|H'_I|n_1\rangle \langle
n_1|H_I^i|n_2\rangle \langle n_2|H'_I]|\Phi_i \rangle\,}
  {(p_f^--p_{n_1}^-+i\epsilon )(p_i^--p_{n_2}^-+i\epsilon )} \\[2\eqnskip]
  & & + {\ds\sum_{n_1,n_2}}'\ds\frac{\,\langle \Phi_f|H_I^i|n_1\rangle \langle
n_1|H'_I|n_2\rangle \langle n_2|H'_I]|\Phi_i \rangle\,}
  {(p_i^--p_{n_1}^-+i\epsilon )(p_i^--p_{n_2}^-+i\epsilon )}
  +\cdots \, ,
\end{array}
\end{equation}
where $Z_g$ is the multiplicative coupling constant renormalization,
and $Z_i$ and $Z_f$ are the wavefunction renormalization constants of
the initial and final states.

\hspace{0.3cm} It is also convenient to express the above perturbation
expansion in terms of the diagrammatic approach. All matrix elements of
$H_I^i$ for the light-front QCD vertices
are listed in Table II with the corresponding
diagrams. These are obtained by directly calculating the matrix
elements between free particle states.  The rules for writing the
expression of perturbative expansions from diagrams are as follows:

\begin{enumerate}
\item [{$\bullet$}] Draw all topologically distinct $x^+$-ordered diagrams.

\item [{$\bullet$}] For each internal line, sum over helicity and integrate
using
$\int \frac{dk^+d^2k_{\bot}}{16\pi^3}\theta (k^+)$ for quarks
and $\int \frac{dk^+d^2k_{\bot}}{16\pi^3 k^+}\theta (k^+)$ for
gluons.

\item [{$\bullet$}] For each vertex, include a factor of
$16\pi^3\delta^3(p_f-p_i)$ and a simple matrix element listed in Table I.

\item [{$\bullet$}] Include a factor $(p_i^--\sum_n p_n^-+i\epsilon )^{-1}$
[or $(p_f^--\sum_n p_n^-+i\epsilon )^{-1}]$ for each
intermediate state, where $\sum_n p_n^-$ sum over all on-mass-shell
intermediate particle energies.
\end{enumerate}

\begin{table}[hptb] 
\tblcaption{
The $x^+$-ordered Hamiltonian Diagrammatic Rules
}

\vspace{20cm}
\end{table}

$\mbox{ }$
\vspace{9.5cm}

\begin{enumerate}
\item [{$\bullet$}] Add a symmetry factor $S^{-1}$ for each gluon loop coming
from
the symmetrized boson states.
\end{enumerate}

\hspace{0.3cm} To illustrate the above computation scheme and to see the severe
light-front
infrared divergences, let me list some calculations up to one-loop based
on the $x^+$-ordered diagrammatical approach.

\hspace{0.3cm} i). {\em Quark wavefunction and mass renormalization}. Besides
the
infrared divergence, which is regularized by Eq. (5.30), there
are also ultraviolet divergences for which we use a transverse cut-off
regularization: $|\kappa_{\bot}|\leq \Lambda_{\bot}$. For this
simplest regularization scheme the one-loop light-front quark energy
corrections (for the three diagrams in Fig. 5, respectively) are
\def\theequation{5.46}
\begin{equation}
\begin{array}{rl}
  \delta p_1^- = & - \ds\frac{g^2}{\,8\pi^2\,}C_f\biggl \{
  \ds\frac{\,p^2-m^2\,}{[p^+]}\biggl [ \biggl ( 2\ln
\ds\frac{\,p^+\,}{\epsilon}
  -\ds\frac{\,3\,}{2}\biggr ) \ln \Lambda_{\bot}^2 \\[1.5\eqnskip]
  & - \ds\int_0^1 dx \biggl (
  \ds\frac{2}{\,[x]\,}-2+x\biggr ) \ln f(x)\biggr ] \\[1.5\eqnskip]
  & + \ds\frac{m^2}{\,[p^+]\,}\biggl ( -2 \ln
  \Lambda_{\bot}^2+2\ds\int_0^1 dx\ln f(x)\biggr ) \\[1.5\eqnskip]
  & +\ds\frac{\,\Lambda_{\bot}^2\,}{\,[p^+]\,}\biggl ( \ds\frac{\,\pi
p^+\,}{2\epsilon}
  -1+\ln \ds\frac{\,p^+\,}{\epsilon}\biggr ) \biggr \} \, ,
\end{array}
\end{equation}

\pagebreak

\begin{figure}[hptb] 
\vspace{3.5cm}
\figcaption{
The $x^+$-ordered graphs for the one-loop correction to the
quark mass and wave function renormalization.
}
\vspace{0.6cm}
\end{figure}

\def\theequation{5.47}
\begin{equation}
  \delta p_2^- = \ds\frac{g^2}{\,8\pi^2\,}C_f \ds\frac{\,\Lambda_{\bot}^2\,}
  {\,[p^+]\,}\ln \ds\frac{\,p^+\,}{\epsilon}\, , \quad
  \delta p_3^- = \ds\frac{g^2}{\,8\pi^2\,}C_f \ds\frac{\,\Lambda_{\bot}^2\,}
  {\,[p^+]\,}\biggl ( \ds\frac{\,\pi p^+\,}{2\epsilon}-1 \biggr ) \, ,
\end{equation}
where $f(x)=(xm^2-x(1-x)p^2)$. This shows that, in
the one-loop quark energy correction,
one-gluon exchange gives rise to both linear and logarithmic infrared
divergences.  The instantaneous fermion interaction contribution
(see $\delta p_2^-$ in Fig. 5b) contains only one logarithmic
divergence which cancels the logarithmic divergence in $\delta p_1^-$.
The instantaneous gluon interaction contribution ($\delta p_3^-$
of Fig. 5c) has a linear infrared divergence which precisely
cancels the same divergence in $\delta p_1^-$.  This
cancellation of linear infrared divergences is based on the use
of the regularization for $k^+ \rightarrow 0 $ in Eq. (5.30) [82].

\hspace{0.3cm} The quark mass correction (dropping the finite part) is then
given by
\def\theequation{5.48}
\begin{equation}
  \delta m^2 = p^+ \delta p^- |_{p^2=m^2} = \ds\frac{g^2}{\,4\pi^2\,} C_f
  m^2\ln \ds\frac{\,\Lambda_{\bot}^2\,}{\,m^2\,} \, ,
\end{equation}
which is longitudinally infrared divergence free; and the quark
wavefunction renormalization constant is
\def\theequation{5.49}
\begin{equation}
\begin{array}{rl}
  Z_{2} = \!\! & 1+\ds\frac{\,\partial \delta p^-\,}{\partial p^-}
  \biggr |_{p^2=m^2} \\[1.5\eqnskip]
  = & 1 + \ds\frac{g^2}{\,8\pi^2\,}C_f\biggl \{ \biggl ( \ds\frac{\,3\,}{2}
  -2\ln \ds\frac{\,p^+\,}{\epsilon}\biggr ) \ln \ds\frac{\,\Lambda_{\bot}^2\,}
  {\,m^2\,}+2\ln \ds\frac{\,p^+\,}{\epsilon}
  \biggl ( 1-\ln \ds\frac{\,p^+\,}{\epsilon}\biggr ) \biggr \} \, .
\end{array}
\end{equation}
The wavefunction renormalization contains an additional
type of divergence, the mixing of infrared and ultraviolet
divergences, that does not occur in covariant calculations.
This is the `spurious' mixing associated with the gauge singularity.
It corresponds to the so-called light-front double pole problem in the
Feynman theory with the use of the light-front gauge and
the principal value prescription
that prohibits any continuation to Euclidean space and power counting
in Feynman loop integrals.  In the $x^+$-ordered Hamiltonian
perturbative theory the power counting is different.
The above argument$\,$ of$\,$ power$\,$ counting$\,$ for Feynman loop integrals
may be irrelevant. Furthermore, since the second order
correction

\pagebreak

\begin{figure}[hptb] 
\vspace{7.5cm}
\figcaption{
The $x^+$-ordered graphs for the one-loop correction
to the gluon mass and wave function renormalization.
}
\vspace{0.6cm}
\end{figure}

\noindent
to wavefunctions must
be negative [see Eq. (5.42)],
Eq. (5.49) shows that it is the additional infrared divergence that
gives a consistent answer for wavefunction renormalization.

\hspace{0.3cm} ii). {\em Gluon wave function and mass correction}.  The
one-loop light-front gluon energy corrections are given by  (see Fig. 6)
\def\theequation{5.50}
\begin{equation}
\begin{array}{rl}
  \delta q_1^- = \!\! & -\ds\frac{g^2}{\,8\pi^2\,}C_A\biggl \{
  \ds\frac{q^2}{\,[q^+]\,}\biggl [ \biggl (2\ln \ds\frac{\,q^+\,}{\epsilon}
  -\ds\frac{\,11\,}{6}\biggr ) \ln \Lambda_{\bot}^2 \\[1.5\eqnskip]
  & + \ds\int_0^1dx\biggl (2-x(1-x)-\ds\frac{2}{\,[x]\,}\biggr ) \ln
  (x(1-x)q^2)\biggr ] + \ds\frac{\,\Lambda_{\bot}^2\,}{\,[q^+]\,}
  \ds\frac{\,\pi q^+\,}{2\epsilon} \biggr \} \, ,
\end{array}
\end{equation}
\def\theequation{5.51}
\begin{equation}
\begin{array}{rl}
  \delta q_2^- = & - \ds\frac{g^2}{\,8\pi^2\,}T_fN_f\biggl \{
  \ds\frac{q^2}{\,[q^+]\,}\biggl [ \ds\frac{2}{\,3\,}\ln \Lambda_{\bot}^2
\\[1.5\eqnskip]
  & -\ds\int_0^1dx(2x^2-2x+1)\ln (m^2 -x(1-x)q^2)\biggr ] \\[1.5\eqnskip]
  & + \ds\frac{m^2}{\,[q^+]\,}\biggl [ 2\ln \Lambda_{\bot}^2
  -2\ds\int_0^1dx\ln (m^2-x(1-x)q^2)\biggr ] \\[1.5\eqnskip]
  & +2\ds\frac{\,\Lambda_{\bot}^2\,}{\,[q^+]\,}\biggl (
  \ln \ds\frac{q^+}{\epsilon}-1\biggr ) \biggr \} \, ,
\end{array}
\end{equation}
\def\theequation{5.52}
\begin{equation}
\begin{array}{l}
  \delta q_3^- = - \ds\frac{g^2}{\,4\pi^2\,}
  \ds\frac{\,\Lambda_{\bot}^2\,}{\,[q^+]\,}T_fN_f\ln
\ds\frac{\,k_{\infty}^+\,}{q^+}\, , \\[1.5\eqnskip]
  \delta
q_{4}^-=\ds\frac{g^2}{\,8\pi^2\,}C_A\ds\frac{\,\Lambda_{\bot}^2\,}{\,[q^+]\,}
  \biggl \{ \ds\frac{\,\pi q^+\,}{\,2\epsilon\,}-1+\ln
  \ds\frac{\,k_{\infty}^+\,}{\,\epsilon\,} \biggr \} \, ,
\end{array}
\end{equation}
\def\theequation{5.53}
\begin{equation}
  \delta q_{5}^- = \ds\frac{g^2}{\,16\pi^2\,}C_A
  \ds\frac{\,\Lambda_{\bot}^2\,}{\,[q^+]\,}\ln
  \ds\frac{\,k_{\infty}^+\,}{\epsilon} \, ,
\end{equation}
where $k_{\infty}^+$ is the internal fermion momentum
$k^+\rightarrow \infty$. In Eq. (5.50), there is also  a mass singularity,
which is well known in gauge theory and which is
usually regularized by introducing a small gluon mass ($q^2=\mu_G^2$)
in the energy denominators.  The gluon mass and wavefunction
renormalizations are
\def\theequation{5.54}
\begin{equation}
  \delta \mu_G^2 = - \ds\frac{g^2}{\,4\pi^2\,}
  \biggl \{ T_fN_fm^2\ln \ds\frac{\,\Lambda_{\bot}^2\,}{\,m^2\,}
  \Lambda_{\bot}^2 \biggl ( \ds\frac{\,C_A\,}{2}-T_fN_f\biggr )
  \biggl ( 1-\ln \ds\frac{\,k_{\infty}^+\,}{\epsilon}\biggr )
  \biggr \} \, ,
\end{equation}
\def\theequation{5.55}
\begin{equation}
\begin{array}{rl}
  Z_3 = & 1 + \ds\frac{g^2}{\,8\pi^2\,}\biggl \{ \biggl (
  \ds\frac{\,11\,}{6}-2\ln \ds\frac{\,q^+\,}{\epsilon}\biggr ) C_A \ln
  \ds\frac{\,\Lambda_{\bot}^2\,}{\,\mu_G^2\,}-\ds\frac{2}{\,3\,}T_f
  N_f\ln \ds\frac{\,\Lambda_{\bot}^2\,}{\,m^2\,} \\[1.5\eqnskip]
  & -2C_A\ln^2 \ds\frac{\,q^+\,}{\epsilon} \biggr \} \, .
\end{array}
\end{equation}

\hspace{0.3cm} All severe divergences appear in the gluon sector:
quadratic and logarithmic UV divergences, linear and logarithmic
IR divergences, a gluon mass singularity and an unusual large longitudinal
momentum logarithmic divergence.   Only the linear infrared divergences
are cancelled with the principal value prescription.  The gluon
mass correction is not zero.  The first term in (5.54) is a fermion
loop contribution (Fig. 6b), which is the same as the photon mass
correction in QED.  In addition, the gluon mass correction
also contains a severe mixing of the quadratic UV divergences with
the logarithmic IR and UV divergences of the longitudinal momentum.  It is
caused by the instantaneous fermion and gluon interaction contributions
plus the tadpole effect of the normal four gluon interactions (Fig. 6c-6e).
This kind of divergence behaves
in the same way from both the fermion contribution and the gluon contribution.
The non-zero gluon mass correction of Eq. (5.54) is not surprising
because it has the same divergence feature as the photon mass correction
in light-front QED [Eq. (5.54) will be reduced to the photon
mass correction when
we set $T_f=1$, $C_A=0$ and $N_f=1$].  In a covariant calculation,
the zero gluon mass correction is true only for dimensional
regularization which ``removes'' or drops the mass correction.
In the present calculation, maintaining
zero gluon mass requires a mass counterterm, as is known in QED.
The difference between QED and QCD is only
manifest in the gauge boson wavefunction renormalization.
For wavefunction renormalization, again there is an additional
mixing of UV and IR divergences, which provides the correct sign
for the wavefunction renormalization constant.  Besides this feature,
there is a contribution from the gluon loop (Fig. 6a).  As we will
see in the following, after the cancellation of the mixing
divergences, it is this contribution that leads to asymptotic freedom
in QCD.

\hspace{0.3cm} However, the above calculations show that the wave function
renormalization contains several complicated pure infrared divergences
and a mass singularity from the massless gluon.  This complexity can be
avoided if we introduce a mass scale $\mu$ for the minimum
cut-off for the transverse momentum:
\def\theequation{5.56}
\begin{equation}
  \Lambda_{\bot}^2 \geq \kappa_{\bot}^2 \geq \mu^2 \, ,
\end{equation}
and assume that $\mu$ is much larger than all other masses in the
theory.  With this regulator, the quark and gluon mass and
wavefunction renormalizations become simple,
\def\theequation{5.57}
\begin{equation}
  \delta m^2 = \ds\frac{g^2}{\,4\pi^2\,}C_f m^2\ln
\ds\frac{\,\Lambda_{\bot}^2\,}
  {\mu^2} \, ,
\end{equation}
\def\theequation{5.58}
\begin{equation}
  \delta \mu_G^2 = - \ds\frac{g^2}{\,4\pi^2\,}\biggl \{ T_fN_f m^2\ln
  \ds\frac{\,\Lambda_{\bot}^2\,}{\,\mu^2\,}(\Lambda_{\bot}^2-\mu^2)
  \biggl ( \ds\frac{\,C_A\,}{2}-T_fN_f\biggr ) \biggl ( 1-\ln
  \ds\frac{\,k_{\infty}^+\,}{\epsilon}\biggr ) \biggr \} \, , \\[\eqnskip]
\end{equation}
\def\theequation{5.59}
\begin{equation}
  Z_{2} = 1+\ds\frac{g^2}{\,8\pi^2\,}C_f\biggl ( \ds\frac{\,3\,}{2}
  -2\ln \ds\frac{\,p^+\,}{\epsilon}\biggr ) \ln \ds\frac{\,\Lambda_{\bot}^2\,}
  {\mu^2} \, ,
\end{equation}
\def\theequation{5.60}
\begin{equation}
  Z_3 = 1 + \ds\frac{g^2}{\,8\pi^2\,}\biggl \{ C_A \biggl ( \ds\frac{\,11\,}{6}
  -2\ln \ds\frac{\,q^+\,}{\epsilon}\biggr ) -\ds\frac{2}{\,3\,}
  T_fN_f\biggr \} \ln \ds\frac{\,\Lambda_{\bot}^2\,}{\,\mu^2\,} \, .
\end{equation}
This shows that all the pure infrared divergences in the wavefunction
renormalizations are removed.  The remaining unfamiliar divergences
are the mixing of quadratic UV with logarithmic IR divergences in
the gluon mass correction, and the mixing of UV and IR logarithmic divergences
in the wavefunction renormalizations.  The mixing divergences in the gluon
mass correction could be removed by a gluon mass counterterm, while
the mixing divergences in the wavefunction renormalizations are cancelled
in physical quantities.  This can be seen from the coupling constant
renormalization.

\hspace{0.3cm} iii). {\em Coupling constant renormalization}.  For convenience,
we set the external gluon momentum $q(q^+,q_{\bot}^i)=0$. The quark-gluon
vertex is then reduced:
\def\theequation{5.61}
\begin{equation}
  {\cal V}_0 = 2gT_{\beta \alpha}^a\ds\frac{p^i}{\,p^+\,}
  \delta_{\lambda_1\lambda_2}\varepsilon_{\sigma}^{i*} \, .
\end{equation}
In $x^+$-ordered perturbation theory, the one-loop vertex correction
is given by
\def\theequation{5.62}
\begin{equation}
  \delta {\cal V}_{0} = \{ V_1+V_2+V_{3}+V_{4}+V_{5}+V_{6} \}
  {\cal V}_{0} \, ,
\end{equation}
where $V_{n}$, $n=1-6$ are represented by Fig. 7.

\begin{figure}[hptb] 
\vspace{20.5cm}
\figcaption{
The $x^+$-ordered graphs for the one-loop correction
to the quark-gluon vectex renormalization.
}
\end{figure}

\def\theequation{5.63}
\begin{equation}
  V_1 = \ds\frac{g^2}{\,2\pi^2\,}\biggl ( \ds\frac{\,3\,}{2}
  -2\ln \ds\frac{\,p^+\,}{\epsilon}\biggr ) C_f\ln
\ds\frac{\,\Lambda\,}{\,\mu\,} \, ,
\end{equation}
\def\theequation{5.64}
\begin{equation}
  V_2 = \ds\frac{g^2}{\,8\pi^2\,}\biggl ( \ds\frac{\,11\,}{3}C_A
  -\ds\frac{\,4\,}{3}N_fT_f\biggr ) \, ,
\end{equation}
\def\theequation{5.65}
\begin{equation}
  V_3 = - \ds\frac{g^2}{\,4\pi^2\,}\biggl ( \ds\frac{\,3\,}{2}
  - 2\ln \ds\frac{\,p^+\,}{\epsilon}\biggr ) \biggl ( -\ds\frac{1}{\,2\,}C_A
  +C_f\biggr ) \ln \ds\frac{\,\Lambda\,}{\,\mu\,} \, ,
\end{equation}
\def\theequation{5.66}
\begin{equation}
  V_4 = - \ds\frac{g^2}{\,8\pi^2\,}\biggl ( \ds\frac{\,3\,}{2}
  - 2\ln \ds\frac{\,p^+\,}{\epsilon}\biggr ) C_A\ln \ds\frac{\,\Lambda\,}{\mu}
\, ,
\end{equation}
\def\theequation{5.67}
\begin{equation}
  V_5 = 0\, , \quad \quad V_{6}=0 \, .
\end{equation}

\hspace{0.3cm} To evaluate the contributions to the coupling constant we
have to multiply $V_1$ and $V_2$ by $\frac{1}{2}$ in order to
 take into account the proper correction due to the
renormalization of initial and final states [153,154].
Thus adding the contributions together, we have,
\def\theequation{5.68}
\begin{equation}
\begin{array}{lll}
  \delta {\cal V}_{0} & = & \biggl (
\ds\frac{1}{\,2\,}V_1+\ds\frac{1}{\,2\,}V_2
  +V_2+V_4+V_5+V_6\biggr ){\cal V}_{0} \\[1.5\eqnskip]
  & = & {\cal V}_{0}\ds\frac{g^2}{\,8\pi^2\,}\biggl ( \ds\frac{\,11\,}{6} C_A
  - \ds\frac{2}{\,3\,}N_fT_f\biggr ) \ln \ds\frac{\,\Lambda\,}{\,\mu\,} \, .
\end{array}
\end{equation}
Note that all mixed divergences cancel now.
The correction to the coupling constant is given by
\def\theequation{5.69}
\begin{equation}
  g_R = g(1+\delta g) =g\biggl \{ 1+ \ds\frac{g^2}{\,8\pi^2\,}
  \biggl ( \ds\frac{\,11\,}{6}C_A - \ds\frac{2}{\,3\,} N_f
  T_f\biggr ) \ln \ds\frac{\,\Lambda\,}{\,\mu\,}\biggr \} \, .
\end{equation}
By redefining the bare coupling constant $g$ such that $g_R$ is finite.
Thus we have given all renormalization quantities in QCD up to one-loop
order based on the old-fashioned Hamiltonian perturbation theory.

\hspace{0.3cm} From these results, the anomalous dimensions for quarks and
gluons
and the $\beta$ function up to one-loop can be easily calculated.
The anomalous dimension of the quark field to order $g^2$ is
\def\theequation{5.70}
\begin{equation}
  \gamma_F \equiv - \ds\frac{1}{\,2Z_2\,} \ds\frac{\partial Z_2}{\,\partial \ln
\Lambda\,}
  = \ds\frac{g^2}{\,8\pi^2\,}C_f \biggl ( 2\ln \ds\frac{\,p^+\,}{\epsilon}
  - \ds\frac{\,3\,}{2} \biggr ) \, .
\end{equation}
The momentum-dependent term implies that the quark anomalous dimension
is gauge dependent.  The anomalous dimension for the gluon field is
\def\theequation{5.71}
\begin{equation}
  \gamma_G \equiv - \ds\frac{1}{\,2Z_3\,}\ds\frac{\partial Z_3}{\,\partial \ln
\Lambda\,}
  = \ds\frac{g^2}{\,8\pi^2\,}\biggl \{ C_f\biggl ( 2
  \ln \ds\frac{\,q^+\,}{\epsilon} - \ds\frac{\,11\,}{6} \biggr )
  +\ds\frac{2}{\,3\,}T_fN_f\biggr \} \, ,
\end{equation}
which is also gauge-dependent. In the case of $q^+=0$,
the gauge dependent term can be removed, and Eq. (5.71) is reduced
to Gross and Wilczek's result in their Feynman calculation with $A_a^+=0$
and $q^+=0$ [42].  The $\beta$ function is
\def\theequation{5.72}
\begin{equation}
  \beta (g) = - \ds\frac{\partial g_R}{\,\partial \ln {\Lambda}\,}
  = - \ds\frac{g^3}{\,16 \pi^2\,}\biggl ( \ds\frac{\,11\,}{3}C_A
  -\ds\frac{4}{\,3\,}N_fT_f\biggr ) \, ,
\end{equation}
which is the well-known result to one loop order and is infrared
divergence free, as we expected.

\hspace{0.3cm} From the above result, we see that there are severe light-front
divergences in light-front QCD.  Systematic control of these
divergences is required {\em a priori} before we perform any
practical numerical calculation in light-front coordinates for
QCD bound states.  From the basic one-loop calculations, one can
see that, in the old-fashioned perturbation theory, light-front QCD involves
various UV and IR divergences. Some of the divergences have not
even been encountered in covariant and noncovariant Feynman
calculations to the same order.
Among various light-front divergences, there are two severe divergences
one has to deal with in the old-fashioned theory for light-front QCD.
The first is the mixing of UV and IR logarithmic divergences
in wavefunction renormalization.  The occurrence
of the mixing divergences may not be a severe problem.
The mixing divergences should be cancelled completely
for physical quantities, as we have seen from the coupling
constant renormalization. We expect that the problem of
mixing divergences may not exist when we consider real
physical processes.
The second problem is the infinite gluon mass correction. In the
old-fashioned Hamiltonian theory dimensional regularization is not
available to avoid the nonzero gluon mass correction.  To have
a massless gluon in perturbation theory, we have to introduce
a gluon mass counterterm. In the leading order (one-loop) calculation,
there is no difficulty arising from a gluon mass counterterm.  However, when
we go to the next order, it has been found that the gluon mass
counterterm leads to a noncancellation of infrared divergences.
The non-vanishing infrared divergences could introduce
non-local counterterms in both the longitudinal and transverse
directions.  In equal-time quantization, such non-local counterterms
are forbidden for a renormalizable theory.  Here, these
non-local counterterms are allowed by the light-front power
counting. This is a special feature of light-front QCD.  One speculation
from this property is that the non-local counterterms for
infrared divergences may also provide a source for quark confinement
[85].

\hspace{0.3cm} In summary, renormalization in {\em light-front QCD Hamiltonian
theory}
is very different from conventional Feynman theory and it is an entirely
new subject where investigations are still in their preliminary stage.
In perturbative
calculations, careful treatment could remove all severe infrared
divergences for interesting physical quantities in light-front QCD.
For nonperturbative studies, the cancellation of severe infrared
divergences may not work because certain approximations (e.g.,
Fock space truncation) might be used.  These
approximations may also break many important symmetries such as
gauge invariance and rotational invariance. It is the hope of the
current investigation of light-front renormalization theory
that the counterterms for the light-front
infrared divergences may restore the broken symmetries and
also provide an effective confining light-front QCD Hamiltonian for
hadronic bound states.

\vspace{15.5pt}
\noindent
{\bf V-4. Infrared Divergences and Nontrivial QCD Properties}

\hspace{0.3cm} Before we go on to discuss a possible approach to the
nonperturbative
dynamics of light-front QCD, I would like to further address
the infrared singularity and its relation with the nontrivial
structure of QCD on the light-front [150].

\hspace{0.3cm} The infrared divergences arise from the elimination of the
unphysical
gauge degrees of freedom.  In coordinate space, infrared divergences
are associated with the singularity at longitudinal infinity.
For physical states, finite energy density requires that the
longitudinal color electric field strength must vanish at
the longitudinal boundary:
\def\theequation{5.73}
\begin{equation}
  E_a^- |_{x^- = \pm \infty} = 0 \, ,
\end{equation}
which is a condition to canonically remove the light-front infrared
divergences and which leads to a constraint on the $A_a^i$ at longitudinal
infinity:
\def\theequation{5.74}
\begin{equation}
  \partial^i A_a^i |_{x^- = \pm \infty} =
  \mp \ds\frac{\,g\,}{2} \ds\int_{-\infty}^{\infty} dx^-
  j^+_a(x^-,x) \, .
\end{equation}
Eq. (5.74) is consistent with the choice of antisymmetric boundary
condition from the definition of Eq. (2.16).
Clearly, Eq. (5.74) is satisfied
only for physical states.  In perturbation theory, we cannot
use this condition because in perturbative QCD we consider
not only physical states but also color non-singlet states for
which Eq. (5.74) may not be satisfied.  Therefore,
the main effect of Eq. (5.74) should be manifested in
nonperturbative dynamics, i.e., bound states.
To address hadronic bound states, the existence
of nontrivial structure, namely, confinement potentials
and the mechanism of dynamical chiral symmetry breaking,
is crucial.  An explicit derivation of such nontrivial properties
from QCD is still lacking. Here I will show how Eq. (5.74)
can really influence the nontrivial properties of QCD.

\hspace{0.3cm} i). {\em Topological winding number}.
Let us consider the axial current equation (for zero quark mass)
\def\theequation{5.75}
\begin{equation}
  \partial_{\mu} j_5^{\mu} = N_f \ds\frac{g^2}{\,8\pi^2\,}\,
  \mbox{Tr}\,(F_{\mu \nu} \widetilde{F}^{\mu \nu})\, ,
\end{equation}
where the axial current is $j_5^{\mu}=\bar{\psi}\gamma^{\mu}\gamma_5 \psi$,
and the dual field strength is
$\widetilde{F}^{\mu\nu}=\frac{1}{2}\epsilon^{\mu\nu\sigma\rho}F_{\sigma\rho}$.
The winding number in light-front QCD is defined as the net charge between
$x^+=-\infty$ and $x^+=\infty$,
\def\theequation{5.76}
\begin{equation}
  \Delta Q_5 = N_f \ds\frac{g^2}{\,8\pi^2\,}\ds\int_M d^4x\,\mbox{Tr}\,
  (F_{\mu \nu}\widetilde{F}^{\mu \nu})\, .
\end{equation}
The integration on the r.h.s. of the above equation is defined in
Minkowski space ($M$) and can be replaced by a surface integral.  It has been
found [150] that
\def\theequation{5.77}
\begin{equation}
  \Delta Q_5 = -N_f\ds\frac{\,g^2\,}{\,\pi^2\,}\ds\int dx^+ d^2x_{\bot}\,
  \mbox{Tr}\,\bigl ( A^-[A^1,A^2]\bigr ) \Bigl |_{x^-=-\infty}^{x^-=\infty} \,
,
\end{equation}
where $A_a^- |_{x=\pm \infty}$ is determined by Eq. (5.27).
We see that a non-vanishing $\Delta Q_5$ is generated
from the asymptotic fields of $A_a^i, A_a^-$ and their
antisymmetric boundary conditions at longitudinal infinity.

\hspace{0.3cm} For physical states, it is particularly interesting to see from
Eq. (5.74) that the asymptotic physical gauge fields are
generated by the color charge densities integrated over $x^-$.
Thus, the topological winding number in the $A_a^+=0$ gauge
can be explicitly explored from Eq. (5.74).

\hspace{0.3cm} ii). {\em Non-local interactions in the transverse direction}.
{}From Eq. (5.74) we see that the asymptotic $A_a^i$ fields
at the longitudinal boundary are proportional to the color
charge density in the transverse space and also that they
involve non-local behavior in the transverse direction (induced by the
transverse derivative).  Intuitively, we may separate the
transverse gauge potentials into a normal part plus a boundary part,
\def\theequation{5.78}
\begin{equation}
  A_a^i = A_{aN}^i + A_{aB}^i \, ,
\end{equation}
where
\def\theequation{5.79}
\begin{equation}
  A_{aN}^i|_{x^-=\pm \infty} =0\, , \quad \quad
  \partial^i A_{aN}^i|_{x^-=\pm \infty}=0 \, ,
\end{equation}
\def\theequation{5.80}
\begin{equation}
  \partial^i A_{aB}^i|_{x^-=\pm \infty} = \mp \ds\frac{\,g\,}{2}
  \bigl ( \rho_a^g(x_{\bot})+\rho_a^q(x_{\bot})\bigr ) \, .
\end{equation}
In Eq. (5.80), $\rho_a(x_{\bot})$ denotes the color charge densities
integrated over $x^-$.  The conditions of Eqs. (5.79) and
(5.80) do not uniquely determine the separation of Eq. (5.78).
Generally, there are two types
of separation for Eq. (5.78).  One, is to consider $A_{aB}^i$
the long-distance fields generated by the boundary integrals and
$A_{aN}^i$ the short-distance fields determined by free theory.
In this case, if we are only interested in the low-energy dynamics,
the effect of the $A_{aN}^i$ fields may be ignored. This separation
is physically very interesting but it is practically very difficult
to realise analytically.  Another possibility is to choose
a simple solution for the $A_{aB}^i$ that satisfies Eq. (5.80).
In this case, the $A_{aN}^i$
have the trivial boundary condition, Eq. (5.79), but are not
determined by the free theory.  The Hamiltonian is then expressed only
in terms of the $A_{aN}^i$ and the boundary behavior of transverse gauge
fields is replaced by the effective interactions. A convenient
choice for $A_{aB}^i$ which satisfies Eq. (5.80) is
\def\theequation{5.81}
\begin{equation}
  A_{aB}^i(x) = - \ds\frac{\,g\,}{\,8\,} \ds\int dx'^-dx'^i
  \varepsilon (x^--x'^-)\varepsilon (x^i-x'^i)
  (\rho_a^g(x')+\rho_a^q(x'))\, , \quad i=1,2 \, .
\end{equation}
With such a formal solution, we obtain a new Hamiltonian in
terms of $A_{aN}^i$ that contains
many effective interactions induced by Eq. (5.74).  All
these effective interactions involve the color charge densities and
involve non-local behavior in both the longitudinal and transverse
directions. One of the lowest order interactions, for example, is given by
\def\theequation{5.82}
\begin{equation}
\begin{array}{rl}
  H_{b1} \propto \!\! & \ds\sum_{ij} \ds\int_{-\infty}^{\infty}dx^i dx^- dx^j
  dx'^- dx'^j \eta^{ij} \\[1.5\eqnskip]
  & \times \bigl \{ \partial^i \rho_a^q(x^-,x^i,x^j) \}
  |x^- - x'^-||x^j - x'^j| \{ \partial^i
  \rho_a^q(x'^-,x^i,x'^j) \bigr \} \, ,
\end{array}
\end{equation}
where $\eta^{ij}\equiv 1 (0)$ for $i\neq (=) j$. This
is a linear interaction in both the longitudinal and the transverse
directions. Hence, Eq. (5.74) leads to numerous many-body non-local color
charge interactions which may lead to confinement.

\hspace{0.3cm} Still the Hamiltonian contains, in principle, an infinite number
of
many-body interactions generated by the boundary integrals
(or obtained from the counterterms of the infrared divergences).
This is a consequence of the boundary integrals in a non-abelian
gauge theory due to the existence of nonlinear gluon interactions.
It is also true in other gauge choices, such as the Coulomb gauge
[155] or the axial gauge [157].
Practically, we do not know how to solve for hadrons from such a complicated
QCD Hamiltonian.  Nevertheless, in the light-front QCD,
one can set the vacuum to be trivial, the nontrivial QCD
features for physical states are switched
to the field operators and are manifested in the asymptotic
behavior of physical gauge fields at longitudinal infinity.
The trivial vacuum with nontrivial field variables
in light-front QCD may provide a practical
framework for describing hadrons. In the next two sections,
we will explore these problems from a totally different point of view.

\vspace{6pt}
\begin{center}
{\normalsize \bf VI. LIGHT-FRONT RENORMALIZATION}
\end{center}
\vspace{3pt}

\hspace{0.3cm} As I mentioned in the very beginning, the theory of
nonperturbative
QCD (at the hadronic scale) has indeed not been defined at the present
time.  What I mean is that since we do not know how to
renormalize QCD nonperturbatively, all results we obtain
(if we can calculate) may have nothing to do with
the data observed in various experiments. In other words, the bare
structure of QCD (or a field theory) does not directly connect to the
real hadronic world although it is constructed from the beautiful
requirement of various symmetries.   Without knowing precisely
the renormalization scheme for a field theory, the theory itself
has not indeed been defined.

\hspace{0.3cm} One may argue that since we even do not know what is
nonperturbative
field theory, how can we talk about nonperturbative renormalization.
It is true that there is no clear description in most field
theory text books of nonperturbative field theory.  However,
quantum field theory is developed within the framework of quantum
mechanics which is defined nonperturbatively in terms of the
Hamiltonian formalism.  From the descriptions in the previous
sections, we have seen that
light-front field theory can be formulated as an ordinary quantum
mechanical system.  This may provide a possible way to establish
a nonperturbative field theory for relativistic dynamics on the
light-front, as we have emphasized throughout this paper.

\hspace{0.3cm} In this section, I will discuss a new renormalization scheme for
Hamiltonian systems that was developed recently [83-85].
It begins with the light-front power counting plus a cutoff
scheme on the constituents to construct the basic structure of
the QCD Hamiltonian,
then using a similarity renormalization approach to remove the
cutoff dependence.  The renormalized theory is given by an effective
Hamiltonian that is characterized by a scale parameter.  It demands
that when the scale is taken at the hadronic energy level, the theory
should produce all the physics of the hadrons.  When the scale
is far from the hadronic scale, the theory must reproduce all
the results of perturbative QCD.

\vspace{15.5pt}
\noindent
{\bf VI-1. Light-Front Power Counting: Canonical Structure}

\hspace{0.3cm} Before discussing renormalization, I would like to introduce the
light-front power counting rules which determine the possible
structure of operators for the canonical Hamiltonian [85].

\hspace{0.3cm} Light-front power counting is done in terms of the longitudinal
coordinate $x^-$ and the transverse coordinate $x_\perp$. The power
counting analysis is to deduce the most general structure of
divergences that arises from increasing the powers of the interaction
Hamiltonian in perturbation theory. But power counting based on the
kinematical symmetries of the light front is different from power
counting based on the kinematical symmetries in equal-time
coordinates. This is immediately transparent from the light-front
dispersion relation for free particles,  $k^-=\frac{k_{\perp}^2+m^2}{k^+}$.
Because the energy
separates the $k^+$ and $k_\perp$ dependencies, the
subtractions are not constants.  For example, when $k_\perp$ gets
very large the energy diverges no matter what $k^+$ is. Thus, in
general, we get a divergent constant {\em multiplied by a function of}
$k^+$. In position space this translates into divergences at small
$x_\perp$ being nonlocal in $x^-$ and spread out over the light front.
A similar result follows for the case when $k^+$ gets very small.
It does scale differently under $x^-$- and $x_\perp$-scaling
(strictly speaking, a unique transverse scaling behavior holds only in
the absence of masses).
This situation is to be contrasted with the equal-time case where
the relationship between energy and momentum in the equal-time theory is
$E=\sqrt{\vec{k}^2+m^2}$. In this equal-time form, if
$k_\perp\rightarrow \infty$
while $k_z$ is fixed, the $k_z$ dependence becomes negligible and
arbitrary functions of $k_z$ cannot arise.

\hspace{0.3cm} i). {\em Power counting analysis and canonical light-front QCD
Hamiltonian}.
The scaling properties of the canonical Hamiltonian is determined
as follows. From the canonical commutation relations (5.20)
and (5.21), under the scale transformation $x^- \rightarrow
{x'}^-=s x^-, x_\perp \rightarrow x'_\perp = t x_\perp$, one can find
the power assignments for the field variables
\def\theequation{6.1}
\begin{equation}
  A_\perp : \ds\frac{1}{\,x_\perp\,} \, , \quad \quad
  \xi :\ds\frac{1}{\,\sqrt{x^-\,}\,} \ds\frac{1}{\,x_\perp\,} \, .
\end{equation}
The power assignments for the derivatives is obvious:
\def\theequation{6.2}
\begin{equation}
  \partial_\perp : \ds\frac{1}{\,x_\perp\,} \, , \quad \quad
  \partial^+ : \ds\frac{1}{\,x^-\,}\, , \quad \quad
  \ds\frac{1}{\,\partial^+\,} : x^- \, .
\end{equation}
{}From the light-front dispersion relation, we see that the Hamiltonian
$H=P^-$ scales just like $x^-/x_{\bot}^2$,
\def\theequation{6.3}
\begin{equation}
  H: \ds\frac{\,x^-\,}{\,x_\perp^2\,}\longrightarrow
  {\cal H}:\ds\frac{1}{\,x_\perp^4\,}\, , \quad \quad
  m: \ds\frac{1}{\,x_\perp\,} \, ,
\end{equation}
where ${\cal H}$ and $m$ are the Hamiltonian density and the mass
parameter. Note that under scale transformations masses scale
as constants.  Thus the Hamiltonian does not have a unique scaling
behavior when masses are present. Here the mass is assigned a power
only based on dimensional analysis.

\hspace{0.3cm} The canonical light-front QCD Hamiltonian density is a
polynomial in the six components $m$, $A_\perp$, $\xi$, $\partial^+$,
$\partial_\perp$, and $\frac{1}{\partial^+}$. The
most general structure we can build for the canonical Hamiltonian
density, which has dimension $\frac{1}{(x_\perp)^4}$, is
\def\theequation{6.4}
\begin{equation}
  (\xi \xi^{\dagger})^p(A_\perp ,\partial_\perp ,m)^{4-2p}
  (\partial^+)^{-p} \, .
\end{equation}
Here the expression $(A_\perp ,\partial_\perp ,m)^{4-2p}$ stands for
monomials of order $4-2p$ in any combination of the three variables
$A_\perp ,\partial_\perp ,m$, which immediately shows that the allowed values
of $p$ are $0,1,2$. The resulting canonical QCD Hamiltonian
structure is
\def\theequation{6.5}
\begin{equation}
\begin{array}{ll}
  p=0: & \quad \quad A_\perp^4, A_\perp^3 \partial_\perp, A_\perp^2
  \partial_\perp^2, m^2 A_\perp^2 \, \\[\eqnskip]
  p=1: & \quad \quad \ds\frac{1}{\,\partial^+\,}(\xi \xi^{\dagger})
  \bigl ( A_\perp^2,A_\perp \partial_\perp,\partial_\perp^2,m A_\perp,
  m\partial_\perp,m^2\bigr ) \, , \\[\eqnskip]
  p=2: & \quad \quad \biggl ( \ds\frac{1}{\,\partial^+\,}\biggr )^2
  (\xi \xi^{\dagger})^2 \, .
\end{array}
\end{equation}
Physically, the terms that are proportional to the operator
$\frac{1}{\partial^+}$ correspond to the elimination of the
dependent variables $\psi_-$ and $A^-$. Specifying the precise
way in which $\frac{1}{\partial^+}$ acts, we can enumerate
those terms which obey the canonical rules:
\def\theequation{6.6}
\begin{equation}
\begin{array}{ll}
  p=0: & \quad \quad (A_\perp)^4, (A_\perp)^3\partial_\perp,
  (A_\perp)^2(\partial_\perp)^2, m^2(A_\perp)^2 \, , \\[\eqnskip]
  & \quad \quad \quad \quad (A_\perp \partial^+
A_\perp)\ds\frac{1}{\,\partial^+\,}
  (\partial_\perp A_\perp),(A_\perp \partial^+ A_\perp)\biggl (
  \ds\frac{1}{\,\partial^+\,}\biggr )^2(A_\perp \partial^+ A_\perp ) \, ,
\\[1.5\eqnskip]
  p=1: & \quad \quad (\xi \xi^{\dagger})\biggl (
\ds\frac{1}{\,\partial^+\,}\biggr )^2
  (A_\perp \partial^+ A_\perp ), (\xi \xi^{\dagger})
  \ds\frac{1}{\,\partial^+\,}(\partial_\perp A_\perp ), \\[\eqnskip]
  & \quad \quad \quad \quad (m,\partial_\perp ,A_\perp )\xi^{\dagger}
  \ds\frac{1}{\,\partial^+\,}\bigl \{ (m,\partial_\perp,A_\perp )\xi \bigr \}
\, , \\[\eqnskip]
  p=2: & \quad \quad (\xi \xi^{\dagger})\biggl (
\ds\frac{1}{\,\partial^+\,}\biggr )^2
  (\xi \xi^{\dagger}) \, .
\end{array}
\end{equation}
Comparing with the Hamiltonian derived in the last section, one sees
that the free terms $m^2 A_\perp^2$ and
$m\xi^{\dagger}{\partial}_\perp \frac{1}{\partial^+}\xi$ are absent. The
absence of the first term is because of the presumed gauge
invariance. A term like the latter does appear in the free part when
$\psi_-$ is eliminated, but it is cancelled by a similar term and
leads to chiral symmetry for the light-front free Hamiltonian.

\hspace{0.3cm} ii). {\em Structure of light-front divergences.}
There are mainly two types of divergences in light-front field theory:
ultraviolet and infrared divergences.  The infrared divergences in
light-front QCD arises from the elimination of the unphysical
degrees of freedom.  In the Hamiltonian formulation, as we have shown
in the previous section, these divergences are manifested from the products of
interaction Hamiltonians. Each term in the interaction Hamiltonian
involves an integral over the product of three or four field
operators, some of them carry high momentum and others low
momentum.  Power counting allows us to estimate the divergences
arising from the high momentum operators after they are integrated
out, and the remaining low momentum operators provide the operator
structure of the corresponding counterterms.

\hspace{0.3cm} Explicitly, let us
consider some simple examples: Consider the second-order shift in
the energy of a gluon coming from a two-gluon intermediate state.
As candidates for $H_I$, we take
\def\theequation{6.7}
\begin{equation}
  H_{I(1)} = g f^{abc}\ds\int dz^- d^2 z_\perp \partial^i
  A^j_a A^i_b A^j_c \, ,
\end{equation}
\def\theequation{6.8}
\begin{equation}
  H_{I(2)} = -g f^{abc} \ds\int dz^- d^2z_\perp \biggl (
  \ds\frac{1}{\,\partial^+\,}\partial^i A^i_a \biggr )
  A^j_b \partial^+ A^j_c \, .
\end{equation}
The field operators are separated into low-momentum parts
which contain wavelets of width $\delta x_\perp$, $\delta x^-$, and
high-momentum parts, which contain wavelets
of width $\delta y_\perp$, $\delta y^-$.

\hspace{0.3cm} First let us consider the ultraviolet divergences.
To avoid confusion with infrared divergences, we set
$\delta x^-=\delta y^-$. For a candidate $H_I$ we choose $H_{I(1)}$.
Remember that to produce an ultraviolet divergence at least two
operators have to belong to the high-momentum sector.
a). Consider $H_{I(1)}\approx \int dz^-d^2z_\perp
(\partial^iA^jA^i)_{(y)}A^j_{(x)}$
and the energy shift $\Delta E\approx H_{I(1)}\frac{1}{E.D}H_{I(1)}$, where
$E.D$ denotes
the energy denominator. $H_{I(1)}$ scales like $\frac{1}{(\delta y_{\bot})^2}$,
$E.D$ produces a factor $(\delta y_{\bot})^2$, while the
$x$ wavelet operators in $H_{I(1)}$ produce a factor
$\bigl ( \frac{\delta y_{\bot}}{\delta x_{\bot}}\bigr )^2$ and there is an
overall factor from the
integral $\bigl ( \frac{\delta x_{\bot}}{\delta y_{\bot}}\bigr )^2$.  Thus the
scaling behavior of the energy shift is
$\frac{1}{(\delta y_\perp)^2}\times (\delta y_\perp)^2\times \frac{1}{(\delta
y_\perp)^2}
\times \bigl ( \frac{\delta x_\perp}{\delta y_\perp}\bigr )^2 \times \bigl (
\frac{\delta y_\perp}{\delta x_\perp}\bigr )^2$,
that is, $\Delta E \approx \frac{1}{(\delta y_\perp)^2}$, a quadratic
ultraviolet divergence; b). Consider
$H_{I(1)}\approx \int dz^-d^2z_\perp \partial^iA^j_{(x)}(A^iA^j)_{(y)}$
and the energy
shift $\Delta E\approx H_{I(1)}\frac{1}{E.D}H_{I(1)}$.
The only difference from case (a) is that the $x$ wavelet
operators in $H_{I(1)}$ produce a factor
$\bigl ( \frac{\delta y_{\bot}}{\delta x_{\bot}}\bigr )^2$.  Hence the scaling
behavior of the energy shift is
$\frac{1}{(\delta y_\perp)^2}\times (\delta y_\perp)^2\times \frac{1}{(\delta
y_\perp)^2}
\times \bigl ( \frac{\delta x_\perp}{\delta y_\perp}\bigr )^2 \times \bigl (
\frac{\delta y_\perp}{\delta x_\perp}\bigr )^4$,
that is, $\Delta E\approx \frac{1}{(\delta x_\perp)^2}$, a logarithmic
ultraviolet divergence.

\hspace{0.3cm} A complete analysis of light-front divergences and the
corresponding
counterterms based on light-front power counting and phase space
cell approach is given in [85]. It has been found that
the power counting rule for ultraviolet counterterms is the same
(in $\delta x_\perp$) as for the canonical Hamiltonian that there
can be at most four operators scaling as $\frac{1}{\delta x_\perp}$.
Thus the ultraviolet counterterms are built out of products
up to fourth order in $\xi$, $\xi^{\dagger}$, $A_\perp$, and
$\partial_\perp$ but they can have a completely arbitrary
longitudinal structure, except for the restriction from the power
counting and the longitudinal boost invariance. As a result of the
constraint, the vertex counterterms can involve
{\it a priori} unknown functions of longitudinal variables.

\hspace{0.3cm} Now consider infrared divergences.  We set
$\delta x_\perp =\delta y_\perp$. Remember that to produce an infrared
divergence at least two operators have to belong to the low sector. Thus,
one may take
$H_{I(1)}\approx \int dz^-d^2z_\perp \partial^iA^j_{(y)}(A^iA^j)_{(x)}$,
$H_{I(2)}\approx \int dz^-d^2z_\perp \bigl (
\frac{1}{\partial^+}\partial^iA^i_{(y)}
\bigr ) (A^j\partial^+A^j)_{(x)}$.
a). Consider the energy shift $\Delta E\approx H_{I(2)}\frac{1}{E.D}H_{I(2)}$.
$H_{I(2)}$ scales like $\delta y^-$. The energy denominator produces a
factor $\frac{1}{\delta y^-}$. Another $H_{I(2)}$ produces a
factor $\delta y^-$. Thus $\Delta E\approx \delta y^-$, which
results in a linear infrared divergence.
b). Consider the energy shift $\Delta E\approx H_{I(2)}\frac{1}{E.D}H_{I(1)}$.
Then, $H_{I(1)}$
scales like $\delta y^- \times \bigl ( \frac{\delta x^-}{\delta y^-}\bigr )$.
The energy denominator produces a
factor $\frac{1}{\delta y^-}$. $H_{I(2)}$ produces a factor
$\delta y^-$. Thus $\Delta E\approx \delta x^-$, which results in a logarithmic
infrared divergence.

\hspace{0.3cm} In general, the counterterms for infrared divergences involve
arbitrary numbers of quark and gluon operators. They have a complex
nonlocality in the transverse variables. This is in contrast to the
divergent counterterm for the canonical instantaneous four-fermion
interaction which is local in the transverse direction. If we identify
the counterterms arising from infrared gluons (small longitudinal
momentum) as the source of transverse confinement, then the unknown
nonlocal transverse behavior would have to include confining effects
at large transverse separation.

\hspace{0.3cm} One might worry that the appearance of
functions of momenta in the counterterms could destroy the predictive
power of the theory and lead to non-renormalizability. However, these
functions are needed to restore Lorentz covariance and gauge
invariance as $g\rightarrow g_s$ in a weak coupling treatment
(see the discussion in the next section), and without such counterterms
physical quantities will not even approach finite limits as cutoffs
are removed for any coupling.

\vspace{15.5pt}
\noindent
{\bf VI-2. A New Cutoff Scheme for the Light-Front Hamiltonian}

\hspace{0.3cm} The light-front infrared and ultraviolet divergences discussed
above should be regularized properly in order to develop a
renormalization
scheme. To provide a renormalization process for a realization
of nonperturbative Hamiltonian computation, we need to develop
a cutoff procedure that is applicable to the Hamiltonian as a whole
rather than just to order by order perturbative calculations that
we have used in the previous section. One way to accomplish this
is to cut off the single particle momenta appearing in the field
variables themselves. Cutoffs on the constituent momenta
$k^+$ and $k_\perp$ are called ``constituent cutoffs'' [85].
The resulting cutoff boundary on each constituent's relative momenta
$(k^+_i,k_{i\perp})$ is given by
\def\theequation{6.9}
\begin{equation}
  k_{i\perp}^2=2\Lambda^2\ds\frac{k_i^+}{\,P_0^+\,}
  \biggl ( 1-\ds\frac{k_i^+}{\,P_0^+\,}\biggr ) -m^2 \quad \quad
  \mbox{for} \quad \quad \ds\frac{m^2}{\,2\Lambda^2\,}P_0^+<k_i^+
  <\ds\frac{\,P_0^+\,}{2} \, ,
\end{equation}
and
\def\theequation{6.10}
\begin{equation}
  k_{i\perp}^2 = \ds\frac{\,\Lambda^2\,}{2}-m^2 \quad \quad
  \mbox{for} \quad \quad \ds\frac{\,P_0^+\,}{2}<k_i^+ \, ,
\end{equation}
where $m$ is the lowest of the constituent masses.  In order to define
such a constituent cutoff which still limits the invariant mass of
a state, however, we are forced to introduce a longitudinal
momentum scale $P_0^+$.  Dependence on this scale must also
be eliminated as part of the renormalization process. Then, in a
considerable range of center of mass domain
$\bigl ( P_\perp=0, P^+= \frac{P^+_0}{2}$ to
$P_\perp =\frac{\Lambda}{2\sqrt{2}}$, $P^+\geq \frac{3P^+_0}{4}\bigr )$, when
the above cutoff constituent appears in a state, the
internal mass of the state is guaranteed to be at least $\Lambda^2$.

\hspace{0.3cm} For future practical numerical computations, we will also
provide a buffer zone outside this constituent momentum cutoff
boundary extending $k_{i\perp}^2$ roughly by a factor of 2 and $k^+_i$
by a factor of 1/2.  We can accomplish this by setting the
outside of the buffer zone at
\def\theequation{6.11}
\begin{equation}
  k_{i\perp}^2 = 4\Lambda^2 \ds\frac{k_i^+}{\,P^+_0\,}
  \biggl ( 1-\ds\frac{k^+_i}{\,P_0^+\,}\biggr ) - m^2 \quad \quad \mbox{for}
  \quad \quad \ds\frac{m^2}{\,4\Lambda^2\,}P_0^+<k^+_i<\ds\frac{\,P_0^+\,}{2}\,
,
\end{equation}
and
\def\theequation{6.12}
\begin{equation}
  k_{i\perp}^2=\Lambda^2-m^2 \quad \quad \mbox{for} \quad \quad
  \ds\frac{\,P_0^+\,}{2}<k^+_i \, .
\end{equation}
The buffer zone allows the use of a smooth cutoff in order to
let the interactions die gradually. The cutoff $\Lambda$ eliminates
both ultraviolet transverse and
infrared longitudinal degrees of freedom.

\hspace{0.3cm} Practically, the cutoff on constituents means that
these cutoff boundaries are directly employed in the integrals
over momenta in the momentum-space expansion of the
field operators.  Thus, the quark field operator is
\def\theequation{6.13}
\begin{equation}
  \varphi (x) = \ds\sum_{\lambda}\ch_{\lambda} \ds\int_c
  \ds\frac{\,dk^+d^2k_\perp\,}{16\pi^3}\bigl [ b_{\lambda}(k)e^{-ik\cdot x}
  +d^{\dagger}_s{\lambda}(k)e^{ik\cdot x}\bigr ] \, ,
\end{equation}
and the gluon field operator is
\def\theequation{6.14}
\begin{equation}
  A^i(x) = \ds\sum_\lambda \ds\int_c \ds\frac{\,dk^+ d^2k_\perp\,}{16\pi^3k^+}
  \bigl [ \epsilon_\lambda^i a_\lambda(k)
  e^{-ik\cdot x}+\epsilon_\lambda^{i*} a^{\dagger}_\lambda
  (k)e^{ik\cdot x}\bigr ] \, ,
\end{equation}
where color indices have been suppressed. If we choose sharp momentum
cutoffs, the momentum integrals may be explicitly written as
\def\theequation{6.15}
\begin{equation}
\begin{array}{rl}
  \ds\int_c dk^+ d^2k_\perp \equiv & \ds\int_{k^+_{min}}^{P^+_0/2} dk^+
  \ds\int d^2k_\perp \theta (2\Lambda^2x(1-x)-m^2-k_\perp^2) \\[1.5\eqnskip]
  & +\ds\int_{P^+_0/2}^\infty dk^+ \ds\int d^2k_\perp
  \theta \biggl ( \ds\frac{\,\Lambda^2\,}{2}-m^2-k_\perp^2\biggr ) \, ,
\end{array}
\end{equation}
where now $k^+_{min}=\frac{m^2}{2\Lambda^2}P_0^+$. It should be
noted that sometimes a sharp cutoff may cause a so-called non-analytic
divergence in a practical calculation [85]. To avoid such
non-analyticities in the structure of counterterms, a smooth
cutoff can be introduced on the momentum integrals so that
\def\theequation{6.16}
\begin{equation}
  \ds\int_c dk^+ d^2k_\perp \equiv \ds\int_0^{\infty}dk^+ \ds\int d^2k_\perp
  {\cal C}\biggl (
\ds\frac{k_\perp^2+m^2}{\,k_\perp^2+m^2+4\Lambda^2x/(1+4x)\,}
  \biggr ) \, ,
\end{equation}
where $x=k^+/P^+_0$. ${\cal C}(y)$ is a  cutoff function which
equals 1 for $y =0$ and decreases analytically to 0 for $y=1$. The integration
over longitudinal momentum is thus
cutoff at a minimum $k^+_{\min}=\frac{m^2}{2\Lambda^2}P_0^+$.

\vspace{15.5pt}
\noindent
{\bf VI-3. Sigma Model with and Without Zero Modes}

\hspace{0.3cm} Using the power counting argument and a cutoff scheme, we may
generate
a theory (even at tree level) that may have already manifested some
nontrivial properties with a trivial vacuum.  Here we consider again the
sigma model to show such a possibility.  Obviously, with a cutoff on the
constituents, the zero modes are removed and the vacuum is trivial.
To construct the effective Hamiltonian we can rely on power counting
and locality. The Hamiltonian can generally be written as
\def\theequation{6.17}
\begin{equation}
  P^- = \ds\int dx^- d^2 x_\perp \biggl [ \ds\frac{1}{\,2\,}
  (\partial_\perp \phi \cdot \partial_\perp \phi + \partial_\perp \pi \cdot
  \partial_\perp \pi ) + V \biggr ] \, .
\end{equation}
Since the symmetry is broken only in the $\sigma$ sector,
$V$ should be even in the $\pi$-field so that the symmetry
under $\pi\rightarrow -\pi$ remains. Also, since no inverse powers
of mass are allowed, we can write
\def\theequation{6.18}
\begin{equation}
  V = \ds\frac{1}{\,2\,}m_\phi^2 \phi^2 + \ds\frac{1}{\,2\,}m_\pi^2 \pi^2
  +\lambda_1 \phi^4 + \lambda_2 \pi^4 + \lambda_3
  \phi^2 \pi^2 + \lambda_4 \phi^3 + \lambda_5 \pi^2 \phi \, .
\end{equation}

\hspace{0.3cm} On the other hand, the current operator $J^\mu$ can also be
constructed by power counting. The charge $ Q$
$\bigl ( =\frac{1}{2}\int dx^-d^2 x_\perp J^+\bigr )$ is dimensionless, which
leads to the canonical
dimension of $J^+$ being $\sim \frac{1}{x^-}\frac{1}{(x_\perp)^2}$.
Meanwhile, each term in $\partial^\mu J_\mu$
$\bigl ( = \frac{1}{2}(\partial^-J^++\partial^+J^-)-\partial^\perp \cdot
J^\perp\bigr )$
should have the same dimensions, which determines the
dimensions of the other components: $J^\perp \sim \frac{1}{(x_\perp)^3}$,
$J^-\sim \frac{x^-}{(x_\perp)^4}$.  The components of
$J^\mu$ are to be constructed from the operators $\pi$, $\phi$,
$\partial^+$, $\partial^-$, and $\partial^\perp$ and the constants.
The constants are allowed to have the dimensions of a negative power
of $x_\perp$ (as masses do) but no power of $x^-$. Since the
operator $\frac{1}{\partial^+}$ does not appear in the canonical
scalar field theory, the operator $\frac{1}{\partial^+}$ is not allowed to
appear in the canonical current operator. One should
also implement the symmetry that $J^\mu$ has to change sign under
$\pi \rightarrow -\pi$. Then the allowed structure of the components
of $J^\mu$ is
\def\theequation{6.19}
\begin{equation}
\begin{array}{lll}
  J^+ & = & a_1 \partial^+ \pi + a_2 (\partial^+ \pi ) \phi + a_3
  (\partial^+ \phi ) \pi \, , \\[\eqnskip]
  J_\perp & = & b_1 \partial_\perp \pi + b_2 (\partial_\perp \phi )
  \pi + b_3 (\partial_\perp \pi ) \phi \, , \\[\eqnskip]
  J^- & = & c_1\partial^- \pi + c_2 (\partial^- \pi ) \phi + c_3
  (\partial^- \phi ) \pi \, .
\end{array}
\end{equation}
Now all of the coefficients in the current and the Hamiltonian can be
determined by the requirement of current conservation, $\partial_{\mu}
j^{\mu}=0$, and covariance.  The results are:
\def\theequation{6.20}
\begin{equation}
\begin{array}{l}
  a_1=b_1=c_1, a_2 = -b_2 = c_2 = -a_3 = b_3 = - c_3=1 \, , \\[\eqnskip]
  \lambda_1 = \lambda_2 = \ds\frac{1}{\,2\,}\lambda_3 \equiv
  \ds\frac{\,\lambda\,}{4}, \;\; \lambda_5 =\lambda a_1, \;\; \lambda_4
  =\lambda a_1, \;\; m_\phi^2 = m_{\pi}^2 + 2\lambda a_1^2\, ,
\end{array}
\end{equation}
and
\def\theequation{6.21}
\begin{equation}
  a_1 = 0 \quad \quad \mbox{ or } \quad \quad  m_{\pi}=0 \, .
\end{equation}
If we take $a_1=0$ then the potential is reduced to the canonical
form, and the current is also of the canonical form.  This corresponds to
the full canonical theory with a symmetry preserving vacuum and a doublet
in the spectrum. If we choose $m_{\pi} =0$, we have
\def\theequation{6.22}
\begin{equation}
\begin{array}{l}
  V = \ds\frac{1}{\,2\,}m_\phi^2\phi^2+\ds\frac{\,\lambda\,}{4}
  (\phi^2+\pi^2)^2+\sqrt{\ds\frac{\,\lambda m_\phi^2\,}{2}\,}\,
  (\phi^2+\pi^2)\phi \, , \\[1.5\eqnskip]
  J^\mu = \phi \partial^\mu \pi -\pi \partial^\mu \phi
  +\sqrt{\ds\frac{\,m_\phi^2\,}{2\lambda}\,}\, \partial^\mu \pi \, ,
\end{array}
\end{equation}
which is the theory with zero modes eliminated and the pion is massless; then
Hamiltonian explicitly breaks the symmetry, the vacuum is trivial, and
there is no longer the notion of spontaneous symmetry breaking.
Current conservation is preserved and vastly reduces the number of
free parameters present in the effective Hamiltonian constructed from
power counting and forces the pion to remain massless.  The results
are the same as we have obtained by the use of the shift approach in
Sec. IV. We have argued that the shift technique cannot be
applied to QCD.  But the above solution is purely based on
light-front power counting and some symmetry consideration, and it
is generic for various theories. Of course, it should be emphasized
that the sigma model
discussed above is only at the tree level.  For QCD, one must
consider radiative corrections and the subsequent renormalization
effects, which are much much more difficult.

\vspace{15.5pt}
\noindent
{\bf VI-4. Similarity Renormalization Scheme}

\hspace{0.3cm} Now, I am going to introduce a new renormalization
procedure for a Hamiltonian formulation that was recently developed
by G{\l}azek and Wilson [83,85], which will allow
us to construct an effective Hamiltonian for light-front QCD.
The previous discussion shows that we can
write down a light-front QCD Hamiltonian based on the power counting
argument and some requirement of symmetries, with a cutoff on
constituents. To begin with such a bare cutoff Hamiltonian and to
approach the hadronic solution we need to develop a
renormalization procedure.  The renormalization should allow
us to remove all cutoff dependence, yet this is only a minor
requirement. The highlight of the renormalization we must develop
is that it allows us to reduce the effects of relativistic
momenta in an effective Hamiltonian
which is dominated in weak coupling by nonrelativistic relative
momenta.  Then, the bound states of the effective Hamiltonian in weak
coupling are pure $q\bar{q}$, pure $qqq$, or pure $gg$ states, exactly
as predicted in the CQM.  All major effects of gluon emission and
absorption --- or more complex processes --- are absorbed into the
effective Hamiltonian for these states.  Once such an effective
Hamiltonian is constructed, the computation of bound states
involves readily executed numerical computations.

\hspace{0.3cm} The renormalization procedure discussed next to realize these
requirements is called the similarity
renormalization scheme for a Hamiltonian formulation.
The basic idea is to develop a sequence of infinitesimal unitary
transformations that transform the initial bare Hamiltonian
$H_B$ to an effective Hamiltonian $H_\sigma$, where $\sigma$
is an arbitrarily chosen energy scale,
\def\theequation{6.23}
\begin{equation}
  H_\sigma = S_\sigma H_B S_\sigma^{\dagger} \, .
\end{equation}
The basic goal for the transformation $S_\sigma$ is that
$H_\sigma$ should be in band-diagonal form relative to the
scale $\sigma$, namely, the matrix
elements of $H_\sigma$ involving energy jumps much larger
than $\sigma$ (other than jumps between two large but
nearby energies) will all be zero, while matrix elements
involving smaller jumps or two nearby energies remain in
$H_\sigma$. The similarity transformation should satisfy
the condition that for $\sigma\rightarrow \infty$,
$H_\sigma\rightarrow H_B$ and $S_\sigma \rightarrow 1$.  The effective
Hamiltonian we
seek involves $H_\sigma$ with $\sigma$ on the order of the quark
or gluon mass.

\hspace{0.3cm} Consider an infinitesimal transformation, then Eq. (6.23)
is reduced to
\def\theequation{6.24}
\begin{equation}
   \ds\frac{\,dH_\sigma\,}{d\sigma} = [H_\sigma ,T_\sigma ] \, ,
\end{equation}
which is subject to the boundary condition
$\lim_{\sigma\rightarrow \infty}H_\sigma =H_B$.
To meet our demand, we need to specify the action
of $T_\sigma$.  This can be done by introducing a scale
$\sigma$:
\def\theequation{6.25}
\begin{equation}
  x_{\sigma ij} = \ds\frac{|E_i-E_j|}{\,E_i+E_j+\sigma\,} \, .
\end{equation}
into a function $f_{\sigma ij}=f(x_{\sigma ij})$ such that
\def\theequation{6.26}
\begin{equation}
  \left \{
\begin{array}{lll}
  0 \leq x\leq \frac{1}{\,3\,}, & f(x) = 1 & \quad (near\; diagonal\; region);
\\[\eqnskip]
  \frac{1}{\,3\,}\leq x\leq \frac{2}{\,3\,}, & f(x)\mbox{drops from 1 to 0} &
\quad (transition\; region); \\[\eqnskip]
  \frac{2}{\,3\,}\leq x\leq 1, & f(x) = 0 & \quad (far\; off\; diagonal\;
region);
\end{array}
  \right.
\end{equation}
where $f(x)$ is to be infinitely differentiable throughout
$0\leq x\leq 1$, including the transition points $1/3$ and $2/3$.
Then we rewrite Eq. (6.23) as follows:
\def\theequation{6.27}
\begin{equation}
\begin{array}{l}
  \ds\frac{\,dH_{\sigma ij}\,}{d\sigma} = f_{\sigma ij}[H_{I\sigma},
  T_\sigma ]_{ij}+\ds\frac{d}{\,d\sigma\,}(\ln f_{\sigma ij})
  H_{\sigma ij}\, , \\[\eqnskip]
  T_{\sigma ij} = \ds\frac{1}{\,E_j-E_i\,}\biggl \{
  (1-f_{\sigma ij})[H_{I\sigma},T_\sigma]_{ij}
  -\ds\frac{d}{\,d\sigma\,}(\ln f_{\sigma ij})H_{\sigma ij}\biggr \} \, .
\end{array}
\end{equation}

\hspace{0.3cm} Since $f(x)$ vanishes when $|x|\geq 2/3$, one can see that
$H_{\sigma ij}$ does indeed vanish in the far off-diagonal
region which will help identify divergent terms
and determine the form of counterterms necessary to remove
these divergences. It also can be seen that $T_{\sigma ij}$
is zero in the near-diagonal region which means
that a perturbative solution to $H_{\sigma ij}$ in terms of
$H^B_{Iij}$ will never involve vanishing energy
denominators.{\footnote{As we know, a vanishing energy denominator
in the old-fashioned Hamiltonian perturbative calculations
is a potential source of large errors in other
perturbative renormalization schemes.}}

\hspace{0.3cm} The solution for $H_{I\sigma}$ and $T_\sigma$ are
\def\theequation{6.28}
\begin{equation}
  H_{I\sigma} = H^B_{I\sigma}+{\underbrace{[H_{I\sigma'},T_{\sigma'}]}}_R\, ,
  \quad \quad T_\sigma = H^B_{I\sigma
T}+{\underbrace{[H_{I\sigma'},T_{\sigma'}]}}_T\, ,
\end{equation}
where $H^B_{I\sigma ij}=f_{\sigma ij}H^B_{Iij}$ and the
linear operation $R$ is
\def\theequation{6.29}
\begin{equation}
  {\underbrace{X_{\sigma' ij}}}_R
  = - f_{\sigma ij}\ds\int_\sigma^\infty d\sigma'X_{\sigma' ij}\, ,
\end{equation}
\def\theequation{6.30}
\begin{equation}
\begin{array}{l}
  H^B_{I\sigma Tij} = -\ds\frac{1}{\,E_j-E_i\,}\biggl (
  \ds\frac{d}{\,d\sigma\,}f_{\sigma ij}\biggr ) H^B_{Iij} \, , \\[1.5\eqnskip]
  {\underbrace{\ X_{\sigma' ij}\ }}_T = -\ds\frac{1}{\,E_j-E_i\,}
  \biggl ( \ds\frac{d}{\,d\sigma\,}f_{\sigma ij}\biggr )
  \ds\int_\sigma^\infty d\sigma'X_{\sigma' ij}
  + \ds\frac{1}{\,E_j-E_i\,}(1-f_{\sigma ij})X_{\sigma ij} \, .
\end{array}
\end{equation}
Finally, one can obtain an iterated solution for $H_{I\sigma}$
where all higher order terms have the substitution
${\underbrace{\cdots}}_{R\rightarrow T}$,
\def\theequation{6.31}
\begin{equation}
\begin{array}{lll}
  H_{I\sigma} & = & H^B_{I\sigma} +
  {\underbrace{[H^B_{I\sigma'},H^B_{I\sigma'T}]}}_R \\[1.5\eqnskip]
  & & + {\underbrace{[{\underbrace{[H^B_{I\sigma''},H^B_{I\sigma''T}]}}_{R'},
  H^B_{I\sigma'T}]}}_R \\[2.5\eqnskip]
  & & + {\underbrace{[H^B_{I\sigma'},{\underbrace{[H^B_{I\sigma'},
  H^B_{I\sigma''T}]}}_{T'}]}}_R
  + \cdots
\end{array}
\end{equation}
The counterterms in $H^B_I$ must be chosen to cancel the divergences
which occur in integrals over intermediate states
at higher orders in $H^B_I$, and such counterterms must
also then be included in higher-order iterations.
If a limit to this process exists the Hamiltonian is
said to be renormalizable. Now we see that the effective hamiltonian
$H_{\sigma}$ causes transitions only between states staying ``close
to the diagonal'' due to the factor $f_{\sigma ij}$, and the effects
of transitions to and from intermediate states which are ``far
off-diagonal'' have explicitly appeared in the effective Hamiltonian
$H_\sigma$ as a perturbative expansion.

\hspace{0.3cm} Thus, using the similarity renormalization scheme, one
transforms the
Hamiltonian $H_B$ into a manageable, band-diagonal form $H_\sigma$.
$H_B$ is the bare cutoff Hamiltonian, forced to be finite by the
imposition of some cutoff $\Lambda$. After we determine the form
of the counterterms and include them in $H_B$, each of the matrix elements of
the transformed Hamiltonian $H_{\sigma ij}$ has no large dependence on
$\Lambda$.  Thus as we send
$\Lambda\rightarrow \infty$, we get a renormalized, scale-dependent effective
Hamiltonian $H^R_\sigma$.  This does not finish the renormalization of
the Hamiltonian, however, for the finite parts of the counterterms in
$H_B$ will produce in $H_\sigma^R$ unknown constants and functions of
momenta which must be adjusted to reproduce physical observables and
to restore the symmetries which were broken by the cutoff $\Lambda$.
These quantities are to be fixed by solving $H_\sigma^R$;
one should be able to do this with a combination of few-body
Hamiltonian methods and weak-coupling diagrams.

\hspace{0.3cm} One can use the similarity transformation to bring the effective
Hamiltonian to any scale $\sigma_n$ and obtain a set
of sequences $\{H^N_n\}$.  As we change $\sigma$ we change
the characteristic scale, but the physics is invariant with
respect to this change for large enough $\Lambda_N$.
The utility of the similarity renormalization scheme is that
the transformation is invertible.  Thus, if one finds a
Hamiltonian that is finite and $\Lambda$-independent for
any one scale $\sigma_n$, the differential similarity
framework guarantees that one can obtain a Hamiltonian that is
finite and cutoff independent for all $\sigma$.  Note that
we require that each matrix element of $H_\sigma$ be
cutoff independent for external momenta which are small in
comparison to the cutoff, which is more restrictive than just
requiring this of the eigenvalues of $H_\sigma$.

\vspace{6pt}
\begin{center}
{\normalsize \bf VII. A WEAK COUPLING TREATMENT OF NONPERTURBATIVE QCD}
\end{center}
\vspace{3pt}

\hspace{0.3cm} In this section, I will discuss a new approach to QCD that was
recently proposed based on the theoretically successful approach
to bound states of QED and the phenomenologically successful
approach of the CQM [85]. The basic question we try
to answer is whether we can set up QCD to be renormalized and
solved by the same techniques that solve QED: namely, a
combination of weak-coupling perturbation theory and many-body
quantum mechanics, from which we can, at least, reproduce the
phenomenological success of CQM for hadrons. The
renormalization scheme discussed in the last section may provide
us with a practical framework to apply these two approaches together
to QCD on the light-front.

\hspace{0.3cm} To develop such a framework, one needs to change from the
standard
approach in QED by: a) the use of
light-front dynamics; b) the use of nonzero masses for both quarks and
gluons; c) the need to handle relativistic effects which give rise to,
for example, asymptotic freedom in QCD, which in turn leads to a
fairly strong renormalized coupling constant and hence relativistic
binding energies; d) the presence of artificial stabilizing and
confining potentials which vanish at relativistic values of the
coupling constant but nowhere else; and e) the suitable treatment
of light-front longitudinal infrared divergences, which cancel
perturbatively in light-front QED because of gauge invariance,
and which is the path for seeking the source of artificial potentials
in light-front QCD.

\hspace{0.3cm} The formulation starts with nonzero masses for both quarks and
gluons from which one can consider an arbitrary coupling constant $g$
which is small even at the quark-gluon mass scale. In the beginning,
this will sacrifice manifest gauge invariance and Lorentz covariance.
These symmetries are only implicitly
restored (if at all) when the renormalized coupling is increased to
its relativistic value, which we call $g_s$. The value $g_s$ is a
fixed number measured at the hadron mass scale. For
smaller values of $g$ the formulation lacks full covariance and is not
expected to have the predictive power of QCD, but it allows
phenomenology to guide renormalization and is defined to maximize the
ease of perturbative computations and extrapolation to $g_s$.
The nontrivial QCD structures arise from non-cancelling divergences
in the new framework, which are
necessarily the sources of true confining potentials and chiral
symmetry breaking in QCD. With the cutoff Hamiltonian one can have a
trivial vacuum. We expect light-front infrared
divergences to be sources of confinement and chiral symmetry breaking
because these are both vacuum effects in QCD and vacuum
effects can enter the light-front theory through light-front
longitudinal infrared effects, as we have explored in the previous
sections. Because of the unconventional scaling
properties of the light-front Hamiltonian, these effects include
renormalization counterterms with whole functions to be determined by
the renormalization process.

\hspace{0.3cm} The basic motivation of the approach is physical rather than
mathematical.  Physically, a
Hamiltonian with nonzero (constituent) quark and gluon masses and
confining potentials is closer to the physics of strong interactions
than a Hamiltonian with zero mass constituents and no confining
potentials.  Then, in the spirit of QED, renormalization effects analyzed
with the confining potential itself can be treated perturbatively,
but only for generating an effective few-body Hamiltonian which can be
solved nonperturbatively.  In this section, I will discuss the scheme
of how to address these arguments.  For a detailed derivation, please look at
our recently published paper [85].

\vspace{15.5pt}
\noindent
{\bf VII-1. Bare Cutoff Hamiltonian with Artificial Interactions}

\hspace{0.3cm} We begin with a bare cutoff light-front QCD Hamiltonian in which
the quark and gluon are massive constituents,
\def\theequation{7.1}
\begin{equation}
  H'_B(\Lambda) = \ds\int dx^- d^2x_{\bot} \ds\frac{1}{\,2\,}\biggl \{
  \partial^iA_a^j\partial^iA_a^j+m_G^2A_a^{i2}
  +\varphi^{\dagger}\biggl ( \ds\frac{\,-\nabla^2+m_F^2\,}{i\partial^+}
  \biggr ) \varphi \biggr \} +H_{I1}^B \, ,
\end{equation}
where $H_{I1}^B$ is the interaction part of the canonical light-front
QCD Hamiltonian given in Sec. V, the only difference with
the superscript $B$ is that all constituent field operators are imposed
by a cutoff of Eq. (6.13) or (6.14) on their momentum expansion.

\hspace{0.3cm} The choice of light-front dynamics, massive quarks and gluons,
and a particular cutoff scheme eliminates the traditional barriers
to a weak-coupling treatment of QCD and allows us to begin with a trivial
vacuum. In the equal-time theory, the QCD vacuum is thought to be a
complicated medium which presumably provides both confinement and the
spontaneous breaking of chiral symmetry. Here the vacuum is trivial
so we have to find other sources for confinement and spontaneous chiral
symmetry breaking in the cutoff theory.  In the bare cutoff
canonical Hamiltonian, one particular term of interest is
the instantaneous interaction in the longitudinal direction between
color charge densities, which provides a potential which is linear in
the longitudinal separation between two constituents that have the
same transverse positions (see the canonical light-front QCD
Hamiltonian in Sec. V).  In the absence of a gluon mass term, this
interaction is precisely cancelled by the emission and absorption of
longitudinal gluons [82].

\hspace{0.3cm} The presence of a gluon mass term automatically prevents
unbounded growth of the running coupling constant below the gluon mass scale
and
provides kinematic barriers to unlimited gluon emission. It eliminates
any equal-time type infrared problems. The cutoff procedure on
constituents for the Hamiltonian developed before is valid also only
with nonzero quark and gluon masses in a specific frame, while
a large number of states (the upper limit of their invariant masses
is guaranteed to be above a large cutoff) are still available for
study even in the boosted frames (see the discussion on the cutoff scheme
in the last section). However, the nonzero gluon mass results
in the cancellation of the linear longitudinal interaction being
incomplete.  The canonical Hamiltonian combined with a
one-gluon exchange term (in momentum space) is (see Fig. 8):

\begin{figure}[hptb] 
\vspace{3.5cm}
\figcaption{
The $x^+$-ordered $q\overline{q}$ interaction via
one-gluon exchange.
}
\end{figure}

\def\theequation{7.2}
\begin{equation}
\begin{array}{rl}
  H_{I ij }^{(2)} = & -4g^2T_{\alpha_3\alpha_1}^a
  T_{\alpha_4 \alpha_2}^a\delta_{s_1 s_3}\delta_{s_2 s_4}
  \ds\frac{1}{\,(p_1^+-p_3^+)^2\,}
  -g^2T_{\alpha_3\alpha_1}^aT_{\alpha_4 \alpha_2}^a
  {\cal M}_{2ij}\ds\frac{1}{\,p_1^+-p_3^+\,} \\[1.5\eqnskip]
  & \times \ds\frac{1}{\,2\,}\Biggl [ \ds\frac{1}
  {\,\frac{m_F^2+p_{1\perp}^2\,}{p_1^+}-\frac{\,m_F^2+p_{3\perp}^2\,}{p_3^+}
  -\frac{\,(p_{1\perp}-p_{3\perp})^2+m_G^2\,}{p_1^+-p_3^+}\,} \\[2\eqnskip]
  & +
\ds\frac{1}{\,\frac{\,m_F^2+p_{4\perp}^2\,}{p_4^+}-\frac{\,m_F^2+p_{2\perp}^2\,}{p_2^+}
  -\frac{\,(p_{1\perp}-p_{3\perp})^2+m_G^2\,}{p_1^+-p_3^+}\,} \Biggr ] \, ,
\end{array}
\end{equation}
where
\def\theequation{7.3}
\begin{equation}
\begin{array}{rl}
  {\cal M}_{2ij} = & \ch_{s_{3}}^{\dagger}\biggl [
2\ds\frac{p_1^{i_1}-p_3^{i_1}\,}{p_1^+-p_3^+}
  -im_F\biggl ( \ds\frac{1}{\,p_1^+\,}-\ds\frac{1}{\,p_3^+\,}\biggr )
  \sigma^{i_1} \\[1.5\eqnskip]
  & -\biggl ( \sigma^{i_1}\ds\frac{\,\sigma_\perp \cdot p_{1\perp}\,}{p_1^+}
  +\ds\frac{\sigma_\perp \cdot p_{3\perp}\,}{p_3^+}
  \sigma^{i_1}\biggr ) \biggr ] \ch_{s_{1}} \, , \\[1.5\eqnskip]
  & \ch_{-s_{2}}^{\dagger} \biggl [
2\ds\frac{\,p_1^{i_1}-p_3^{i_1}\,}{p_1^+-p_3^+}
  -im_F\biggl ( \ds\frac{1}{\,p_4^+\,}-\ds\frac{1}{\,p_2^+\,}\biggr )
  \sigma^{i_1} \\[1.5\eqnskip]
  & - \biggr ( \sigma^{i_1}\ds\frac{\,\sigma_\perp \cdot p_{4\perp}\,}{p_4^+}
  +\ds\frac{\sigma_\perp \cdot p_{2\perp}\,}{p_2^+}
  \sigma^{i_1}\biggr ) \biggr ] \ch_{-s_{4}} \, ,
\end{array}
\end{equation}
from which one can find an additional linear interaction due to
the nonzero gluon mass, which is proportional to
\def\theequation{7.4}
\begin{equation}
  g^2T^aT^a\ds\frac{1}{\,(p_1^+-p_3^+)^2\,}
  \ds\frac{m_G^2}{\,m^2_G+(p_{1\bot}-p_{2\bot})^2\,} \, .
\end{equation}
Yet, this interaction falls off too rapidly
in the transverse direction, but is too strong in the
longitudinal direction, which leads to an instability even
at weak-coupling.
To remove the above instabilities in hadronic bound states,
we need to subtract this linear interaction,
\def\theequation{7.5}
\begin{equation}
  \ds\frac{\,g^2\,}{4}\ds\int d^2x_{\bot}dx^-dy^-j_a^{+}
  (x^-,x_{\bot})|x^--y^-|j_a^{+}(y^-,x_{\bot}) \, ,
\end{equation}
from the bare cutoff hamiltonian.

\hspace{0.3cm} The cutoffs will violate Lorentz and gauge symmetries, forcing
the bare Hamiltonian to contain a larger than normal set of
counterterms to ensure a finite limit as the cutoffs are removed.
Of special interest are counterterms that reflect consequences
of zero modes (namely, modes with $k^+=0$), and thereby the
effect of confinement and chiral symmetry breaking in the full theory.
According to power-counting arguments, the counterterms for
longitudinal light-front infrared divergences may contain functions of
transverse momenta; there exists the possibility that
the {\em a priori} unknown functions in the finite parts of these
counterterms will include confining interactions in the transverse
direction. The $g\rightarrow g_s$ limit may then be smooth if such confining
functions are actually required to restore full covariance to the
theory.

\hspace{0.3cm} The task of solving the light-front Hamiltonian at the
relativistic
value of $g$ is far too difficult to attempt at the present time.
Instead, it may be helpful to define a plausible sequence of
simpler computations that can build a knowledge base which enables
studies of the full light-front Hamiltonian to be fruitful at some
future date.  A crucial step in simplifying the computation is
the introduction of artificial potentials in the Hamiltonian.

\hspace{0.3cm} The basic need is to incorporate the qualitative phenomenology
of QCD
bound states into the artificial potential.  This qualitative
phenomenology comes from three sources: kinetic energy, Coulomb-like
potentials, and linear potentials.  Three terms should be present
in the weak-coupling Hamiltonian and all should have
the same overall scaling behavior with $g$ in bound state
computations, namely $g^4$,  just as QED bound state energies scale
as $e^4$. The Coulomb-like terms can be constructed directly from
Eq. (7.2). By taking a nonrelativistic limit, the above effective
interaction in the $q\bar{q}$ sector due to {\em massless}
one-gluon exchange is reduced to
\def\theequation{7.6}
\begin{equation}
\begin{array}{rl}
  {\cal H}_{Iij}^{(2)NR} = & -4g^2T^aT^a\ds\frac{1}{\,2\,}
  \biggl [ \biggl ( \ds\frac{\,m_F\,}{\,p_1^+\,}\biggr )^2
  \ds\frac{1}{\,(p_{1\perp}-p_{3\perp})^2+\bigl (
\frac{\,m_F\,}{\,p_1^+\,}\bigr )^2(p_1^+-p_3^+)^2\,} \\[1.5\eqnskip]
  & +\biggl ( \ds\frac{\,m_F\,}{\,p_2^+\,}\biggr )^2
  \ds\frac{1}{\,(p_{2\perp}-p_{4\perp})^2+\bigl (
\frac{\,m_F\,}{\,p_2^+\,}\bigr )^2(p_2^+-p_4^+)^2\,}\biggr ] \, .
\end{array}
\end{equation}
By using an interpolating Fourier transformation,
\def\theequation{7.7}
\begin{equation}
\begin{array}{l}
  \biggl ( \ds\frac{\,m_F\,}{\,p_1^+\,}\biggr )^2
  \ds\frac{1}{\,(p_{1\perp}-p_{3\perp})^2+\bigl (
\frac{\,m_F\,}{\,p_1^+\,}\bigr )^2(p_1^+-p_3^+)^2\,} \\[1.5\eqnskip]
  \hspace{0.5cm} \longrightarrow \ds\frac{1}{\,4\pi\,}\ds\int dy^-d^2y_{\bot}
  e^{i\{ (p_1^+-p_3^+)y^--(p_{1\bot}-p_{3\bot})\cdot y_{\bot}\}}
  \ds\frac{\,m_F\,}{\,p_1^+\,}{1}{\,\sqrt{y_\perp^2+\bigl (
\frac{\,p_1^+\,}{\,m_F\,}\bigr )^2(y^-)^2\,}\,} \, ,
\end{array}
\end{equation}
${\cal H}_2^{NR}$ can further be expressed in position space with longitudinal
boost invariance:
\def\theequation{7.8}
\begin{equation}
  {\cal V}_C = g^2 \ds\frac{1}{\,2\,}
  \biggl [ \ds\frac{m_{1c}}{\,p_1^+r_1\,} +
\ds\frac{m_{2c}}{\,p_2^+r_2\,}\biggr ] \, .
\end{equation}
This is the light-front Coulomb potential, where $r$ is defined as
\def\theequation{7.9}
\begin{equation}
  r = \sqrt{\ds\frac{\,(p^+\delta x^-)^2\,}{m_c^2} + \delta x_{\bot}^2\, } \, ,
\end{equation}
$\delta x^-$ is the light-front longitudinal separation of two
constituents, $\delta x_{\bot}$ the transverse separation, $m_c$ the
constituent mass, and $p^+$ the constituent longitudinal momentum.
The $p^+/m_c$ in the definition of $r$ ensures that the dimensions
match. Of course, the positive or negative $SU(3)$ charges must also
be inserted.  It is easy to check that the light-front radial coordinate
is invariant under longitudinal
boosts.  However, it is not invariant under transverse boosts.
Relativistically, the spinor matrix elements are different in different
helicity sectors. In the nonrelativistic limit this helicity dependence
vanishes and we get the same interaction in all helicity sectors.  In
practical calculation, these helicity dependent interactions in
Eq. (7.2) must also be included.

\hspace{0.3cm} Finally, we need a linear potential term that is proportional to
$r$ and $r'$.  In Coulomb bound states both are of order $1/g^2$.
Hence, to achieve an energy of order $g^4$, the linear potential
must have a coefficient of order $g^6$.  Thus the linear potential
term, in position space, would be proportional to $g^6 r$.  To be
precise and get the dimensions straight, its form is
\def\theequation{7.10}
\begin{equation}
  {\cal V}_L = g^6\beta \biggl [ \ds\frac{\,m_c^3r\,}{p^+}
  +\ds\frac{\,m_c^{\prime 3}r'\,}{p'^+}\biggr ] \, ,
\end{equation}
with $\beta$ a numerical constant.

\hspace{0.3cm} The linear potential needs to exist between all possible pairs
of
constituents: $qq$, $q\bar{q}$, $\bar{q}\bar{q}$, $qg$, $\bar{q}g$, and $gg$,
where $q$, $\bar{q}$ and $g$ stand for quark, antiquark and gluon
respectively.  The linear potential must always be positive
(confining) rather than negative (destabilizing). It cannot
involve products of $SU(3)$ charges as the Coulomb
term does.  The Coulomb term could be given a Yukawa structure rather
than the pure $g^2/r$ term due to the nonzero gluon mass. All
of the potential have to be Fourier transformed to momentum
space and then  be restricted to the allowed range of both
longitudinal and transverse momenta after all cutoffs have
been imposed.

\hspace{0.3cm} A primary rule for the artificial potential is that it should
vanish in the relativistic limit.  For example, the artificial potential
might have an overall factor $(1-g^2/g_s^2)$ to ensure its vanishing
at $g = g_s$.  This rule leaves total flexibility in the choice of the
artificial potential since no relativistic physics is affected by it.
To ensure that the artificial potential vanishes at $g_s$ without
destroying its weak-coupling features, the subtraction term,
Eq. (7.5), of the longitudinal linear potential has to be
treated with care.  It is suggested that the subtracted linear potential be
multiplied by $(1-g^6/g_s^6)$ so that the subtraction begins to be
negated only in order $g^8$, which is smaller for small $g$ than the
artificial $g^6$ linear potential that needs to be dominant. All other
terms in the artificial potential can be multiplied by
$(1-g^2/g_s^2)$ instead.  To ensure that the Coulomb term shows Coulomb
behavior, at least roughly, at typical bound state sizes, it is important that
the mass used in any Yukawa-type modification of the Coulomb term scale as
$g^2$ rather than being a constant mass.

\hspace{0.3cm} The artificial potential is to give the
weak-coupling theory a structure close to the CQM so that past
experience with the quark model can be used to determine the precise
form of this potential and to fit it to experimental data.  Thus,
an initial calculation involves QCD
complications only in a very minimal form.
The artificial potential is also to incorporate a linear
potential in both the longitudinal and transverse directions to ensure
quark confinement for any $g$.  This is important for phenomenology.

\hspace{0.3cm} It is necessary to study where the artificial linear potential
might originate
from the counterterms of infrared divergences, and, especially in
the relativistic limit, where the artificial potential vanishes.
The light-front infrared singularities give rise to both linear and
logarithmic divergences. The linear divergences, however, contain the
inverse of the longitudinal cutoff $\frac{1}{\epsilon}$, which
violates longitudinal boost invariance; hence,
only the logarithmic infrared divergences can be the source of
the linear confinement interaction.  We expect that the logarithmic
infrared divergences could occur in the $g^4$-order correction to
Fig. 8, from which the counterterms might determine the structure of
such linear potentials.  But so far this calculation has not been
completed.

\hspace{0.3cm} Now, we add the artificial potential to the cutoff canonical
Hamiltonian.
\def\theequation{7.11}
\begin{equation}
\begin{array}{rl}
  H_B(\Lambda) = & \ds\int dx^-d^2x_{\bot}\ds\frac{1}{\,2\,}\biggl \{
  \partial^iA_a^j\partial^iA_a^j + m_G^2A_a^{i2} + \varphi^{\dagger} \biggl (
  \ds\frac{\,-\nabla^2+m_F^2\,}{i\partial^+}\biggr ) \varphi \biggr \}
\\[1.5\eqnskip]
  & +H_{I1}^B + H_{I2}^B \, ,
\end{array}
\end{equation}
where
\def\theequation{7.12}
\begin{equation}
\begin{array}{rl}
  H_{I2}^B = \!\! & \biggl ( 1-\ds\frac{g^2}{\,g_s^2\,}\biggr ) \ds\int dx^-
  d^2x_{\bot}dy^-d^2y_{\bot} \\[1.5\eqnskip]
  & \cdot \biggl \{ -\ds\frac{1}{\,4\pi\,}j_a^{+}(x){\cal V}_C(x,y)j_a^{+}(y)
  +j^+(x){\cal V}_L(x,y)j^+(y)\biggr \} \\[1.5\eqnskip]
  & \cdot \ds\frac{\,g^2\,}{4}\biggl ( 1-\ds\frac{g^6}{\,g_s^6\,}\biggr )
  \ds\int
dx^-d^2x_{\bot}dy^-j_a^{+}(x^-,x_{\bot})|x^--y^-|j_a^{+}(y^-,x_{\bot})\, ,
\end{array}
\end{equation}
$j_a^{+}$ is the color vector charge density and $j^+$ the color
singlet charge density. This starting bare cutoff Hamiltonian
will allow us to determines an effective Hamiltonian via the
similarity renormalization scheme.

\vspace{15.5pt}
\noindent
{\bf VII-2. Effective Light-Front QCD Hamiltonian and Bound State computations}

\hspace{0.3cm} In this subsection we outline the construction of the
effective light-front QCD Hamiltonian for low-energy
hadrons, and discuss the increasing levels of complexity
in the light-front QCD bound state computations in this
formulation.

\hspace{0.3cm} In the similarity renormalization scheme, the effective
light-front QCD Hamiltonian that will be used to compute the hadronic bound
states is
\def\theequation{7.13}
\begin{equation}
  H_{\sigma ij} = \ds\lim_{\Lambda \rightarrow \infty}
  f_{\sigma ij}\Biggl [ H^B_{ij}+\ds\sum_k\ds\frac{1}{\,2\,}H^B_{Iik}H^B_{Ikj}
  \biggl ( \ds\frac{g_{\sigma ijk}}{\,P^-_j-P^-_k\,}
  +\ds\frac{g_{\sigma jik}}{\,P^-_i-P^-_k\,}\biggr ) + \cdots \Biggr ] \, .
\end{equation}
Here the bare cutoff Hamiltonian is divided as $H_B=H^B_0+H^B_I$.  For
the determination of the effective Hamiltonian $H_\sigma$, the
unperturbed part of the bare Hamiltonian $H^B_0$ is chosen to be that
of free massive quarks and gluons with the standard relativistic
dispersion relation in light-front kinematics. The interaction part
$H^B_I$ then contains the canonical interaction terms $H_{I1}^B$,
the artificial potential $H_{I2}^B$ plus a complete set of counterterms.
Recall that $H^B_{ij}$ has the cutoff $\Lambda$ which violates both
longitudinal and transverse boost invariance. First one needs to identify
the counterterms that must be included in $H^B$ so that the matrix
elements of the effective Hamiltonian $H_{\sigma ij}$ have  no
divergent dependence on $\Lambda$. Then one can send
$\Lambda \rightarrow \infty$ so that the cutoff dependence is removed from the
effective
Hamiltonian $H_\sigma$.  As in standard perturbative renormalization
theory, $\Lambda$ dependence will be removed order by order in
$g_\sigma$, where $g_\sigma$ is the running coupling constant at the
similarity scale in the Hamiltonian matrix elements. This avoids having to
solve the nonperturbative bound state problem in order to identify and remove
the $\Lambda$ dependence.

\hspace{0.3cm} Consider the effective Hamiltonian $H_\sigma$ generated to
second
order in the renormalized coupling constant. The counterterms
$H^{CT}_B$ up to this order contain i). the counterterms for the
canonical instantaneous interactions, which are
\def\theequation{7.14}
\begin{equation}
  H^{CT}_{B2} = - \ds\frac{g^2}{\,4\pi \epsilon\,} \ds\int d^2x_{\bot}
  \biggl ( \ds\int dx^-j^{+a}(x^-,x_{\bot})\biggr )^2 \, ,
\end{equation}
and ii). the counterterms from one-gluon exchange given in momentum
space, plus a similar term in the gluon-gluon sector.
There are also quark and gluon mass counterterms, some of which
have given explicitly in [85]. The effect of these counterterms
is to completely cancel the leading radiative corrections from
instantaneous gluon exchange.

\hspace{0.3cm} There is a question whether $H_{\sigma ij}$ satisfies boost
invariance after $\Lambda \rightarrow \infty$. There can be finite terms in
$H_\sigma$ which violate boost invariance yet cannot justifiably be
subtracted.  In this case violations
of boost invariance can only disappear at special values of $g$ where
the coefficient of the boost violating terms vanishes. Clearly, one such
special value has to be $g=g_s$, as part of the restoration of
covariance at $g_s$.
In the following it is assumed that the effective Hamiltonian
$H_\sigma$ has boost invariance simply as a result of taking the
$\Lambda \rightarrow \infty$ limit. $H_\sigma$ is generated as the first step
in solving the bare cutoff Hamiltonian $H_B$.

\hspace{0.3cm} The second step is to construct bound states from $H_\sigma$,
which are determined by the  bound state equation of Eq. (4.11) with
$H_{\sigma}$ as the Hamiltonian.
One can solve this field theoretic bound state problem in the standard
fashion, using the bound state approach of QED.  This requires us to
identify a part of $H_\sigma$, $H_{\sigma 0}$, which is treated
nonperturbatively to produce bound states.  The essential
simplification that makes further calculation possible is that
$H_{\sigma 0}$ does not contain any interactions that change particle
number, so that the methods of few-body quantum mechanics can be used
to solve this initial nonperturbative problem.  All field theoretic
corrections the interaction part of the effective Hamiltonian
that arise from particle creation and annihilation will
then be treated perturbatively, as in QED.

\hspace{0.3cm} Now, one might ask, what is the dependence of this effective
Hamiltonian
$H_\sigma$ on the scale $\sigma$ and what is the physical consequence
of the scale dependence? Suppose we restrict the states to only a $q\bar{q}$
pair. Then only number conserving (potential-like) terms in $H_\sigma$
contribute. Now, if we let the scale $\sigma $ approach a very large
number, only the original interactions in $H_B$ survive since the
factor $(1-f_{\sigma ik}f_{\sigma jk})$ $\rightarrow 0$. Thus the
canonical four-fermion interaction together with all counterterms
survive, whereas the transversely smeared four-fermion interaction
diminishes in strength. In the limit $\sigma \rightarrow \infty$ we
are recovering $H_B$, which is no surprise. In this limit the effects
of higher Fock sectors (for example $q\bar{q}g$) can be recovered
only by including them explicitly, and so it clearly becomes a poor
approximation to include only the $q\bar q$ sector when $\sigma$ becomes
too large.

\hspace{0.3cm} By lowering the similarity scale $\sigma$, we reduce the allowed
range of gluon momenta which can contribute, for example, to the
binding of a quark and antiquark in a meson. These effects must appear
elsewhere; we see that, through the similarity transformation, they
are added directly to the Hamiltonian via the second and higher order
terms in $H_B$.  Thus, by lowering $\sigma$, we put the bare
gluon exchange effects of $H_B$ into a $q\bar q$ potential in the
effective Hamiltonian $H_\sigma$ perturbatively. This clearly changes
the character of the bound state calculation.  It changes from a field
theoretic computation with arbitrary numbers of constituents to a
computation dominated by an effective $q\bar q$ potential. If we
choose the similarity scale $\sigma$ to be just above the hadronic
mass scale the major effects come from the $q\bar q$ sector.   The
resulting nonrelativistic calculation will not see the scale $\sigma$
in the first approximation, since only states close to the diagonal,
for which $f_{\sigma ij}=1$, will contribute.  This can be outlined
with the following schematic equation:
\def\theequation{7.15}
\begin{equation}
\begin{array}{lll}
  H_{QCD}^R & = & H_0^R(g_s)+H_I^R(g_s) \\[\eqnskip]
  & = & \bigl \{ H_0^R(g_s)+H_{artificial}^R (g_s)\bigr \}
  +\bigl \{ H_I^R(g_s)-H_{artificial}^R(g_s)\bigr \} \\[\eqnskip]
  & \equiv & H_{\sigma 0}(g)+H_{\sigma I}(g) \, .
\end{array}
\end{equation}
Equivalently, $H_{\sigma
0}(g)=H_{0}^R(\sigma,g_{\sigma})+(1-g_{\sigma}^2/g^2_s)H_{artificial}^R(\sigma,g_{\sigma})$
and $H_{\sigma I}(g)=H_I^R(\sigma,g_{\sigma})$.

\hspace{0.3cm} A basic goal of this approach is to construct a sequence of
computations in light-front QCD with growing levels of complexity. The major
sources of the complexity are analyzed in [85].  The further
computations are i). an initial nonperturbative calculation of the
hadronic masses to determine the parameters in $H_{\sigma 0}$ in the hadronic
scale with the binding energies up to the fourth order of the coupling
constant.  This is a straightforward calculation; ii). the next, the
leading radiative correction to these masses, which requires a calculation
of the effective Hamiltonian up to $g^4$ and the binding energies up
to $g^6$, in which logarithmic infrared divergences should occur and the
transverse linear potentials as the counterterms of these divergences
could be justified; iii) the extensions to the perturbative
theory to study the strong coupling structure of the bound states
which will be the real challenge to this approach.

\hspace{0.3cm} Practically, even if we are successful in finding the
light-front
hadronic bound states in this direction, the theory is still not completely
established. We have seen in Sec. III that the matrix elements
associated with various hadronic structures are defined on the {\em canonical}
light-front field operators in the light-front bound states. Since
the bound states obtained in the above approach come from the use of the
light-front effective QCD Hamiltonian that contains many nontrivial
counterterms for the severe light-front divergences that are allowed
by the light-front power counting.  Thus the matrix elements
in terms of the canonical light-front field operators may not
provide the correct physical information of hadrons due to the
fact that many light-front physical operators associated with the
observables are also dynamically dependent. Typical examples are
the transverse and longitudinal components of currents that have
``bad'' components $\psi_-$ of fermions, and the operators in
the hadronic structure function that belong to the
high-twist contributions [94].  Therefore
renormalization which introduces many noncanonical counterterms
to the light-front Hamiltonian must also change the canonical
structure of the operators used in the structure function
definition. Obviously, the renormalization of physical operators
besides the light-front Hamiltonian is another essential problem
in the investigations of the light-front QCD bound states which
has yet to be developed [156].

\vspace{6pt}
\begin{center}
{\normalsize \bf VIII. A BRIEF SUMMARY}
\end{center}
\vspace{3pt}

\hspace{0.3cm} In this article, I have reviewed the basic formulation of the
light-front field theory, the light-front expression of various
strong interaction phenomena measured or extracted from
various high energy experiments, and the possible understanding of
these phenomena from QCD on the light-front.  The essential problem
for solving QCD on the light-front is to find the light-front
hadronic bound states, since most of the interesting hadronic quantities,
such as the parton distribution functions, parton fragmentation
functions, hadronic form factors and coupling constants can
naturally be formulated on the light-front in terms of simple matrix
elements of light-front field operators in light-front hadronic
bound states.  Light-front field theory gives us a simple
Schr\"{o}dinger picture of bound state equations for relativistic
particles, in which the bound states are simply a Fock states
expansion on the trivial light-front vacuum.  With such a picture,
we may develop a reliable approach for addressing the relativistic bound
states,
as well as nonperturbative field theory for QCD.  The main problem
in this framework is reduced then, to constructing an effective QCD
Hamiltonian on the light-front with a trivial vacuum so that it can
reproduce the well-known phenomenology of hadrons on the low energy
scale and can deduce perturbative QCD with asymptotic freedom at
the high energy scale.  A schematic approach to solving this problem has been
discussed; further applications need to be explored.

\vspace{6pt}
\begin{center}
{\normalsize \bf ACKNOWLEDGMENTS}
\end{center}
\vspace{3pt}

\hspace{0.3cm} The author is grateful to H. Y. Cheng, C. Y. Cheung, S. C. Lee,
S. P. Li, G. L. Lin, H. L. Yu, M. C. Huang and C. L. Wu  for many useful
discussions.  The author also thanks H. Y. Cheng for giving him
the opportunity to present the lectures on this subject at ``the Second
Workshop on Particle Physics Phenomenology'' held in Kenting National
Park, and to S. C. Lee for support through
the National Science Council Grant: NSC83-0208-M-001-069.

\vspace{6pt}
\begin{center}
{\normalsize \bf APPENDIX A: TWO-COMPONENT FEYNMAN PERTURBATIVE THEORY}
\end{center}
\vspace{3pt}

\hspace{0.3cm} For most field theorists Feynman's formulation is more
convenient.  In this appendix, I present the two-component
light-front QCD Feynman theory.  The two-component light-front
QCD Lagrangian has a canonical form,
\def\theequation{A1}
\begin{equation}
  {\cal L} = {\cal L}_0 + {\cal L}_{int} \, ,
\end{equation}
where
\def\theequation{A2}
\begin{equation}
\begin{array}{l}
  {\cal L}_0 = \ds\frac{1}{\,2\,}\partial_{\mu}A_a^i
  \partial^{\mu}A_a^i - \xi^{\dagger} \biggl ( \ds\frac{1}{\,i\partial^+\,}
  \biggr ) (\Box^2+m^2)\xi \, , \\[1.5\eqnskip]
  {\cal L}_{int} = - {\cal H}_{int} \, .
\end{array}
\end{equation}
The Feynman rules can be derived from the path integral formalism [158].
Here we use the procedure of 't Hooft and Veltman [159].
The result follows.
The free quark and gluon propagators in two-component light-front
QCD are
\def\theequation{A3}
\begin{equation}
  S_{0\alpha \beta}(p) = \ds\frac{\,ip^+\delta_{\alpha \beta}\,}{\,p^2-m^2\,}\,
,
  \quad \quad D_{0ab}^{ij} = \delta_{ij}\delta_{ab}\ds\frac{i}{\,k^2\,} \, .
\end{equation}
The diagrammatic rules for various vertices are listed in Table III.
The rest of the Feynman rules are the same as in the instant-form
for boson and fermion theories.  With these basic ingredients,
the conventional Feynman approach to various physical processes
can be directly discussed within the above two-component
formulation of light-front QCD.

\hspace{0.3cm} It may be worth mentioning that
one advantage of the two-component Feynman perturbation theory is that
we can use the ML prescription and dimensional regulator to regulate
the light-front infrared divergences and ultraviolet divergences in
Feynman loop integrals, and may recover the multiplicative
renormalizability of the theory, at least in the one-loop approximation,
as shown in Ref. [60].

\vspace{70pt}

\begin{table}[hptb] 
\tblcaption{
Two-component Feynman Diagrammatic Rules
}

\vspace{20cm}
\end{table}

\pagebreak

$\mbox{ }$
\vspace{5cm}

\vspace{6pt}
\begin{center}
{\normalsize \bf REFERENCES}
\end{center}

\begin{enumerate}
\item [{[$\;\;$1$\;\;$]}] For an extensive list of the reference, on
light-front
dynamics, see ``Sources for Light-front Physics'' available
via anonymous FTP from pacific.mps.ohio-state.edu in the
subdirectory pub/infolight.
\vspace{-5pt}

\item [{[$\;\;$2$\;\;$]}] P. A. M. Dirac, Rev. Mod. Phys. {\bf 21}, 392
(1949).
\vspace{-5pt}

\item [{[$\;\;$3$\;\;$]}] S. Tomonoga, Prog. Theor. Phys. {\bf 1}, 27 (1946).
\vspace{-5pt}

\item [{[$\;\;$4$\;\;$]}] J. Schwinger, Phys. Rev. {\bf 74}, 1439 (1948);
{\bf 75}, 651; {\bf 76}, 790 (1949).
\vspace{-5pt}

\item [{[$\;\;$5$\;\;$]}] R. P. Feynman, Phys. Rev. {\bf 76}, 769 (1949).
\vspace{-5pt}

\item [{[$\;\;$6$\;\;$]}] F. J. Dyson, Phys. Rev. {\bf 75}, 486; 1736 (1949).
\vspace{-5pt}

\item [{[$\;\;$7$\;\;$]}] S. Weinberg, Phys. Rev. {\bf 150}, 1313 (1966).
\vspace{-5pt}


\item [{[$\;\;$8$\;\;$]}] S. Fubini and G, Furlan, Physics {\bf 1}, 229 (1965).
\vspace{-5pt}

\item [{[$\;\;$9$\;\;$]}] R. Dashen and M. Gell-Mann, Phys. Rev. Lett. {\bf
17},
340 (1966).
\vspace{-5pt}

\item [{[$\;$10$\;$]}] R. P. Feynman, Phys. Rev. Lett. {\bf 23}, 1415 (1969).
\vspace{-5pt}

\item [{[$\;$11$\;$]}] R.$\,$ P.$\,$ Feynman,$\,$ {\em Photon-hadron$\,$
interactions},$\,$ Benjamin,
$\,$ Reading,$\,$ Massachusetts (1972).
\vspace{-5pt}

\item [{[$\;$12$\;$]}] J. D. Bjorken and E. A. Paschos, Phys. Rev. {\bf 185},
1975 (1969).
\vspace{-5pt}

\item [{[$\;$13$\;$]}] D. J. Gross and S. Treiman, Phys. Rev.  {\bf D4},
1059 (1971).
\vspace{-5pt}

\item [{[$\;$14$\;$]}] S. D. Drell, D. J. Levy, and T. M. Yan, Phys. Rev.
{\bf 181}, 2159 (1969).
\vspace{-5pt}

\item [{[$\;$15$\;$]}] S. D. Drell, D. J. Levy, and T. M. Yan, Phys. Rev.
{\bf D1}, 1035 (1970).
\vspace{-5pt}

\item [{[$\;$16$\;$]}] S. D. Drell, D. J. Levy, and T. M. Yan, Phys. Rev.
{\bf D1}, 1617 (1970).
\vspace{-5pt}

\item [{[$\;$17$\;$]}] J. D. Bjorken, J. B. Kogut, and D. E. Soper, Phys. Rev.
{\bf D3}, 1382 (1971).
\vspace{-5pt}

\item [{[$\;$18$\;$]}] see an early review by J. B. Kogut and L. Susskind,
Phys. Rep. {\bf C8}, 75 (1973).
\vspace{-5pt}

\item [{[$\;$19$\;$]}] S.-J. Chang and S.-K. Ma, Phys. Rev. {\bf 180}, 1506
(1969).
\vspace{-5pt}

\item [{[$\;$20$\;$]}] H. Leutwyler, J. R. Klauder and L. Streit, Nuo. Cim.
Vol. {\bf LXVIA}, No. 3 (1970).
\vspace{-5pt}

\item [{[$\;$21$\;$]}] J. B. Kogut and D. E. Soper, Phys. Rev. {\bf D1}, 2901
(1970).
\vspace{-5pt}

\item [{[$\;$22$\;$]}] F. Rohrlich, Acta. Phys. Austr. {\bf 32}, 87 (1970).
\vspace{-5pt}

\item [{[$\;$23$\;$]}] R. A. Neville and F. Rohrlich, Nov. Cim. {\bf A1},
625 (1971).
\vspace{-5pt}

\item [{[$\;$24$\;$]}] S.-J. Chang, R. G. Root, and T.-M. Yan, Phys. Rev.
{\bf D7}, 1133 (1973).
\vspace{-5pt}

\item [{[$\;$25$\;$]}] S.-J. Chang and T.-M. Yan, Phys. Rev. {\bf D7}, 1147
(1973).
\vspace{-5pt}

\item [{[$\;$26$\;$]}] T.-M. Yan, Phys. Rev. {\bf D7}, 1760 (1973).
\vspace{-5pt}

\item [{[$\;$27$\;$]}] T.-M. Yan, Phys. Rev. {\bf D7}, 1780 (1973).
\vspace{-5pt}

\item [{[$\;$28$\;$]}] D. Flory, Phys. Rev. {\bf D1}, 2795 (1970).
\vspace{-5pt}

\item [{[$\;$29$\;$]}] L. Susskind, Phys. Rev., {\bf 165}, 1535 (1968);
and in {\em Lectures in Theoretical Physics}, Vol. Xi-D,
edited by K. T. Mahanthappa and E. E. Brittin,
(Gordon and Breach Sci. Pub., New York, 1969) p. 135.
\vspace{-5pt}

\item [{[$\;$30$\;$]}] J. M. Cornwell and R. Jackiw, Phys. Rev. {\bf D4},
367 (1971).
\vspace{-5pt}

\item [{[$\;$31$\;$]}] D. A. Dicus, R. Jackiw, and V. L. Teplitz, Phys. Rev.
{\bf D4}, 1733 (1971).
\vspace{-5pt}

\item [{[$\;$32$\;$]}] R. Jackiw, {\em Springer Tracts in Mod. Phys.} {\bf 62},
1972.
\vspace{-5pt}

\item [{[$\;$33$\;$]}] R. F. Dashen and M. Gell-Mann, in
{\em Proc. of the 3rd Coral Gables Conf. on Symmetry Principles at High
Energy},
ed. by A. Perlmutter et at. (1966).
\vspace{-5pt}

\item [{[$\;$34$\;$]}] S. Fubini, Nuovo Cimento, {\bf 34A}, 475 (1966).
\vspace{-5pt}

\item [{[$\;$35$\;$]}] H. Fritzsch and M. Gell-Mann, in
{\em Proc. XVI Int. Conf. on High Energy Physics}, edited by J. D. Jackson
and A. Roberts. (Fermilab, 1972) Vol. 2, p. 135.
\vspace{-5pt}

\item [{[$\;$36$\;$]}] D. J. Gross and F. Wilczek, Phys. Rev. Lett. {\bf 30},
1343 (1973).
\vspace{-5pt}

\item [{[$\;$37$\;$]}] H. D. Politzer, Phys. Rev. Lett. {\bf 30}, 1346 (1973).
\vspace{-5pt}

\item [{[$\;$38$\;$]}] For an example, see R. L. Jaffe, in
{\em Relativistic Dynamics and Quark-Nuclear Physics}, edited by M. B. Johnson
and A. Picklesimer, (Wiley, New York, 1986), p. 537.
\vspace{-5pt}

\item [{[$\;$39$\;$]}] E. Tomboulis, Phys. Rev. {\bf D8}, 2736 (1973).
\vspace{-5pt}

\item [{[$\;$40$\;$]}] J. M. Cornwell, Phys. Rev. {\bf D10}, 1067 (1974).
\vspace{-5pt}

\item [{[$\;$41$\;$]}] R. J. Crewther, {\em Weak and Electromagnetic
interactions
at high energy}, ed M. Levy et al., (Plenum, New York, 1976) p. 345.
\vspace{-5pt}

\item [{[$\;$42$\;$]}] D. Gross and  F. Wilczek, Phys. Rev. {\bf D9}, 980
(1974).
\vspace{-5pt}

\item [{[$\;$43$\;$]}] A. Casher, Phys. Rev. {\bf D14}, 452 (1976).
\vspace{-5pt}

\item [{[$\;$44$\;$]}] W. A. Bardeen and R. B. Pearson, Phys. Rev.
{\bf D13}, 547 (1976).
\vspace{-5pt}

\item [{[$\;$45$\;$]}] W. A. Bardeen, R. B. Pearson, and E. Rabinovici, Phys.
Rev. {\bf D21}, 1037 (1980).
\vspace{-5pt}

\item [{[$\;$46$\;$]}] C. B. Thorn, Phys. Rev. {\bf D19}, 639; {\bf D20},
1934 (1979).
\vspace{-5pt}

\item [{[$\;$47$\;$]}] G. P. Lepage and S. J. Brodsky, Phys. Rev. {\bf D22},
2157 (1980).
\vspace{-5pt}

\item [{[$\;$48$\;$]}] G. P. Lepage, S. J. Brodsky, T. Huang, and P. B.
Mackenzie,
in {\em Particles and Fields 2}, edited by
A. Z. Capri and A. N. Kamal (Plenum, New York, 1983).
\vspace{-5pt}

\item [{[$\;$49$\;$]}] S. J. Brodsky and G. P. Lepage, in {\em Perturbative
Quantum Chromodynamics} ed. A. H. Mueller, (World Scientific,
Singapore, 1989).
\vspace{-5pt}

\item [{[$\;$50$\;$]}] V. A. Franke, Y. V. Novozhilov, and E. V. Prokhvatilov,
Lett. Math. Phys. {\bf 5}, 239; 437 (1981).
\vspace{-5pt}

\item [{[$\;$51$\;$]}] St. D. G{\l}azek, Phys. Rev. {\bf D38}, 3277 (1988).
\vspace{-5pt}

\item [{[$\;$52$\;$]}] G. Altarelli and G. Parisi, Nucl. Phys. {\bf B126},
298 (1977).
\vspace{-5pt}

\item [{[$\;$53$\;$]}] G. Curci, W. Furmanski, and R. Petronzio, Nucl. Phys.
{\bf B175}, 27 (1980).
\vspace{-5pt}

\item [{[$\;$54$\;$]}] R. K. Ellis, W. Furmanski and R. Petronzio, Nucl. Phys.
{\bf B207}, 1 (1982).
\vspace{-5pt}

\item [{[$\;$55$\;$]}] J. C. Collins, D. E. Soper, and G. Sterman, in
{\em Perturbative Quantum Chromodynamics} ed. A. H. Mueller, (World
Scientific, Singapore, 1989).
\vspace{-5pt}

\item [{[$\;$56$\;$]}] K. G. Wilson, Phys. Rev. {\bf 179}, 1499 (1969).
\vspace{-5pt}

\item [{[$\;$57$\;$]}] S. Mandelstam, Nucl. Phys. {\bf B213}, 149 (1983).
\vspace{-5pt}

\item [{[$\;$58$\;$]}] G. Leibbrandt, Phys. Rev. {\bf D29}, 1699 (1984).
\vspace{-5pt}

\item [{[$\;$59$\;$]}] D. M. Capper, J. J. Dulwich, and M. J. Litvak, Nucl.
Phys. {\bf B241}, (1984) 463.
\vspace{-5pt}

\item [{[$\;$60$\;$]}] H. C. Lee and M. S. Milgram, Nucl. Phys. {\bf B268},
543 (1986).
\vspace{-5pt}

\item [{[$\;$61$\;$]}] L. D. Faddeev and A. A. Slavnov, {\em Gauge Firlds},
(Benjamin/Cummings, reading, MAssachusetts, 1980).
\vspace{-5pt}

\item [{[$\;$62$\;$]}] J. A. Dixon and J. C. Taylor, Nucl. Phys.
{\bf B78}, (1974) 552.
\vspace{-5pt}

\item [{[$\;$63$\;$]}] S. D. Joglekar and B. W. Lee,  Ann. of Phys. {\bf 97},
160 (1976).
\vspace{-5pt}

\item [{[$\;$64$\;$]}] R. Hamberg and W. L. van Neerven, Nucl. Phys.
{\bf B379}, (1992) 143, and references therein.
\vspace{-5pt}

\item [{[$\;$65$\;$]}] J. C. Collins and R. J. Scalise, Preprint PSU/TH/141,
hep-ph/9403231 (1994).
\vspace{-5pt}

\item [{[$\;$66$\;$]}] G. 't Hooft, Phys. Rev. {\bf D14}, 3432 (1976);
Phys. Rep. {\bf 142}, 357 (1986).
\vspace{-5pt}

\item [{[$\;$67$\;$]}] K. G. Wilson, Phys. Rev. {\bf D14}, 2455 (1974).
\vspace{-5pt}

\item [{[$\;$68$\;$]}] K. G. Wilson, Phys. Rev. {\bf 140}, B445 (1965).
\vspace{-5pt}

\item [{[$\;$69$\;$]}] K. G. Wilson, Phys. Rev. {\bf B4}, 3184 (1971).
\vspace{-5pt}

\item [{[$\;$70$\;$]}] C. Rebbi, ed., {\em Lattice Gauge Theories and
Monte Carlo Simulations}, (World Scientific, Singapore, 1983).
\vspace{-5pt}

\item [{[$\;$71$\;$]}] M. A. Shifman,$\,$ A. I. Vainshtein,$\,$ and$\,$ V. I.
Zakharov,
$\,$ Nucl.$\,$ Phys.$\,$ {\bf B147},$\,$ 385 (1979).
\vspace{-5pt}

\item [{[$\;$72$\;$]}] See, for example, G. T. Bodwin and D. R. Yennie, Phys.
Rep. {\bf 43C}, 267 (1978).
\vspace{-5pt}

\item [{[$\;$73$\;$]}] W. E. Caswell and G. P. Lepage, Phys. Rev.
{\bf A18}, 810 (1978).
\vspace{-5pt}

\item [{[$\;$74$\;$]}] J. R. Sapirstein, E. A. Terray, and D. R. Yennie,
Phys. Rev. {\bf D29}, 2990 (1984).
\vspace{-5pt}

\item [{[$\;$75$\;$]}] G. T. Bodwin, D. R. Yennie, and M. A. Gregorio,
Rev. Mod. Phy. {\bf 57}, 723 (1985).
\vspace{-5pt}

\item [{[$\;$76$\;$]}] M. Gell-Mann, Phys. Lett. {\bf 8}, 214 (1964).
\vspace{-5pt}

\item [{[$\;$77$\;$]}] G. Zweig, CERN Reports Th. 401 and 412 (1964),
and in Proc. Int. School  of Phys. ``Ettore Majorana'', Erice,
Italy (1964), ed.  A. Zichichi, p. 192 (Academic, New York, 1965).
\vspace{-5pt}

\item [{[$\;$78$\;$]}] H. J. Lipkin, Phys. Rep. {\bf 8C}, 173 (1973).
\vspace{-5pt}

\item [{[$\;$79$\;$]}] O. W. Greenberg, Ann. Rev. Nucl. Sci. {\bf 28}, 327
(1978).
\vspace{-5pt}

\item [{[$\;$80$\;$]}] for reviews, see:  F. E. Close, {\em Quarks and
Partons},
Academic Press, London (1979); also
W. Lucha, F. F. Sch\"{o}berl, and D. Gromes,
Phys. Rep. {\bf 200}, 127 (1991).
\vspace{-5pt}

\item [{[$\;$81$\;$]}] K. G. Wilson, Nucl. Phys.  (Proc. Suppl.), {\bf B17},
82 (1989).
\vspace{-5pt}

\item [{[$\;$82$\;$]}] W. M. Zhang and A. Harindranath, Phys. Rev. {\bf D48},
4881 (1993).
\vspace{-5pt}

\item [{[$\;$83$\;$]}] St. D. G{\l}azek and K. G. Wilson, Phys. Rev. {\bf D48},
5863 (1993).
\vspace{-5pt}

\item [{[$\;$84$\;$]}] St. D. G{\l}azek and K. G. Wilson, Phys. Rev. {\bf D49},
4214 (1994).
\vspace{-5pt}

\item [{[$\;$85$\;$]}] K. G. Wilson, T. S. Walhout, A. Harindranath, W. M.
Zhang,
R. J. Perry, and St. D. G{\l}azek, Phys. Rev. {\bf D49}, 6720 (1994).
\vspace{-5pt}

\item [{[$\;$86$\;$]}] P. A. M. Dirac, {\em Lectures on Quantum Mechanics},
(Yeshiva Univ., New York, 1964).
\vspace{-5pt}

\item [{[$\;$87$\;$]}] L. D. Faddeev and R. Jackiw, Phys. Rev. Lett. {\bf 60},
1692 (1988).
\vspace{-5pt}

\item [{[$\;$88$\;$]}] W. M. Zhang and A. Harindranath, Phys. Rev. {\bf D48},
4868 (1993).
\vspace{-5pt}

\item [{[$\;$89$\;$]}] S. L. Adler and R. F. Dashen, {\em Current Algebra},
(Benjamin, New York, 1968).
\vspace{-5pt}

\item [{[$\;$90$\;$]}] K. Baedakci and G. Segre, Phys. Rev. {\bf 153},
1263 (1967).
\vspace{-5pt}

\item [{[$\;$91$\;$]}] J. Jersak and J. Stern, Nuovo Cimento, {\bf 59}, 315
(1969).
\vspace{-5pt}

\item [{[$\;$92$\;$]}] F. Feinberg, Phys. Rev. {\bf D7}, 540 (1973).
\vspace{-5pt}

\item [{[$\;$93$\;$]}] R. Carlitz, D. Heckathorn, J. Kaur, and W. K. Tang,
Phys. Rev. {\bf D11}, 1234 (1975).
\vspace{-5pt}

\item [{[$\;$94$\;$]}] R. Jaffe and X. Ji, Nucl. Phys. {\bf B375}, 527 (1992).
\vspace{-5pt}

\item [{[$\;$95$\;$]}] J. C. Collins and D. E. Soper,  Nucl. Phys. {\bf B194},
445 (1982).
\vspace{-5pt}

\item [{[$\;$96$\;$]}] R. L. Jaffe and L. Randall, Nucl. Phys. {\bf B412}, 79
(1994).
\vspace{-5pt}

\item [{[$\;$97$\;$]}] M. Neubert, CERN-TH.7113/93, Dec. 1993.
\vspace{-5pt}


\item [{[$\;$98$\;$]}] T. Maskawa and K. Yamawaki, Prog. Theor. Phys. {\bf 56},
1649 (1976).
\vspace{-5pt}

\item [{[$\;$99$\;$]}] A. Harindranath and J. P. Vary, Phys. Rev. {\bf D37},
421 (1988).
\vspace{-5pt}

\item [{[100]}] G. McCartor, Z. Phys. {\bf C41}, 271 (1988).
\vspace{-5pt}

\item [{[101]}] Th. Heinzl, St. Krusche, and E. Werner, Z. Phys.
{\bf A334}, 443 (1989).
\vspace{-5pt}

\item [{[102]}] E. V. Prokhvatilov and V. A. Franke, Sov. J.
Nucl. Phys. {\bf 49}, 688 (1989).
\vspace{-5pt}

\item [{[103]}] G. McCartor, Zeit. Phys. {\bf C52}, 611 (1991).
\vspace{-5pt}

\item [{[104]}] F. Lenz, M. Thies, S. Levit, and K. Yazaki, Ann. Phys. (NY)
{\bf 208}, 1 (1991).
\vspace{-5pt}

\item [{[105]}] G. Robertson, Phys. Rev. {\bf D47}, 2549 (1993).
\vspace{-5pt}

\item [{[106]}] M. Burkardt, Phys. Rev. {\bf D47}, 4628 (1993).
\vspace{-5pt}

\item [{[107]}] S. Huang and W. Lin, UW preprint, (submitted to Ann. Phys.).
\vspace{-5pt}

\item [{[108]}] H. C. Pauli and S. J. Brodsky, Phys. Rev. {\bf D32}, 1993;
2001 (1985).
\vspace{-5pt}

\item [{[109]}] A. C. Tang, S. J. Brodsky, and H. C. Pauli, Phys. Rev.
{\bf D44}, 1842 (1991).
\vspace{-5pt}

\item [{[110]}] M. Kaluza and H. C. Pauli, Phys. Rev. {\bf D45},
2968 (1992).
\vspace{-5pt}

\item [{[111]}] M. Kaluza and H.-J. Pirner, Phys. Rev. {\bf D47},
1620 (1993).
\vspace{-5pt}

\item [{[112]}] R. J. Perry, A. Harindranath, and K. G. Wilson,
Phys. Rev. Lett. {\bf 65}, 2959 (1990).
\vspace{-5pt}

\item [{[113]}] R. J. Perry and A. Harindranath, Phys. Rev. {\bf D43},
4051 (1991).
\vspace{-5pt}

\item [{[114]}] S. G{\l}azek, A. Harindranath, S. Pinsky, J. Shigemitsu,
and K. G. Wilson, Phys. Rev. {\bf D47}, 1599 (1993).
\vspace{-5pt}

\item [{[115]}] A. Harindranath, J. Shigemitsu, and W. M. Zhang,
``Study of Utilizing Basis Functions to$\,$ Solve$\,$ Simple
$\,$ Light-Front$\,$ Equations$\,$ in$\,$ 3+1 Dimensions'',$\,$ unpublished
(1993).
\vspace{-5pt}

\item [{[116]}] I. Tamm. J. Phys. (USSR) {\bf 9}, 449 (1949).
\vspace{-5pt}

\item [{[117]}] S. M. Dancoff, Phys. Rev. {\bf 78}, 382 (1950).
\vspace{-5pt}

\item [{[118]}] G. t'Hooft, Nucl. Phys. {\bf B75}, 461 (1974).
\vspace{-5pt}

\item [{[119]}] H. Bergknoff, Nucl. Phys. {\bf B122}, 215 (1977).
\vspace{-5pt}

\item [{[120]}] T. Eller, H. C. Pauli, and S. J. Brodsky, Phys. Rev.
{\bf D35}, 1493 (1987).
\vspace{-5pt}

\item [{[121]}] A.$\,$ Harindranath$\,$ and$\,$ J.$\,$ Vary,$\,$ Phys.$\,$ Rev.
$\,$ {\bf D36},$\,$ 1141$\,$ (1987);$\,$ {\bf D37},$\,$ 1064;$\,$ 3010 (1988).
\vspace{-5pt}

\item [{[122]}] M. Burkardt, Nucl. Phys. {\bf A504}, 762 (1989).
\vspace{-5pt}

\item [{[123]}] K. Hornbostel, S. J. Brodsky, and H. C. Pauli, Phys.
Rev. {\bf D41}, 3814 (1990).
\vspace{-5pt}

\item [{[124]}] St. D. G{\l}azek and R. J. Perry, Phys. Rev. {\bf D45},
3740 (1992).
\vspace{-5pt}

\item [{[125]}] A. Harindranath, R. J. Perry, and J. Shigemitsu,
Phys. Rev. {\bf D46}, 4580 (1992).
\vspace{-5pt}

\item [{[126]}] M. Burkardt, Nucl. Phys. {\bf B373}, 613 (1992).
\vspace{-5pt}

\item [{[127]}] Y. Mo and R. J. Perry, J. Comp. Phys. {\bf 108}, 159 (1993).
\vspace{-5pt}

\item [{[128]}] S. Dalley and I. R. Klebanov, ``String Spectrum
of 1+1 Dimensional Large N QCD with Adjoint Matter'', Princeton
Univ. preprint PUPT-1342 (1993).
\vspace{-5pt}

\item [{[129]}] M. V. Terent'ev, Sov. J. Nucl. Phys. {\bf 24},
106 (1976).
\vspace{-5pt}

\item [{[130]}] V. B. Berestetsky and M. V. Terent'ev, Sov. J.
Nucl. Phys. {\bf 24}, 547 (1976); {\em ibid}, {\bf 25}, 347 (1977).
\vspace{-5pt}

\item [{[131]}] H. J. Melosh, Phys. Rev. {\bf D9}, 1095 (1974).
\vspace{-5pt}

\item [{[132]}] E. Eichten, F. Feinberg, and J. F. Willemen, Phys. Rev.
{\bf D8}, 1204 (1973).
\vspace{-5pt}

\item [{[133]}] F. Schlumpf, Phys. Rev. {\bf D47}, 4114 (1993).
\vspace{-5pt}

\item [{[134]}] P. L. Chung, F. Coester, and W. N. Polyzou,
Phys. Lett. {\bf B205}, 545 (1988).
\vspace{-5pt}

\item [{[135]}] P. L. Chung, F. Coester, B. D. Keister, and W. N. Polyzou,
Phys. Rev. {\bf C37}, 2000 (1988).
\vspace{-5pt}

\item [{[136]}] Z. Dziembowski and J. Franklin, Phys. Rev.
{\bf D42}, 905 (1990).
\vspace{-5pt}

\item [{[137]}] W. Jaus, Phys. Rev. {\bf D44}, 2851 (1991).
\vspace{-5pt}

\item [{[138]}] V. L. Chernyak and A. R. Zhitnitsky, Nucl. Phys.
{\bf B201}, 492 (1982); Phys. Rep. {\bf 112}, 173 (1984).
\vspace{-5pt}

\item [{[139]}] T. Huang, B. Q. Ma, and Q. X. Shen, Phys. Rev. {\bf D49},
1490 (1994).
\vspace{-5pt}

\item [{[140]}] V. Braun and I. Halperin, preprint, hep-ph/9402270.
\vspace{-5pt}

\item [{[141]}] R. L. Jaffe and M. Soldate,  Phys. Rev. {\bf D26},
49 (1982).
\vspace{-5pt}

\item [{[142]}] D. J. Pritchard and W. J. Stirling, Nucl. Phys.
{\bf B165}, 237 (1980).
\vspace{-5pt}

\item [{[143]}] For a review, see G. Leibbrandt, Rev. Mod. Phys.
{\bf 59}, 1067 (1987).
\vspace{-5pt}

\item [{[144]}] D.$\,$ Mustaki,$\,$ S.$\,$ Pinsky,$\,$ J.$\,$ Shigemitsu,$\,$
and$\,$ K.$\,$ Wilson,
$\,$ Phys.$\,$ Rev.$\,$ {\bf D43},$\,$ 3411 (1991).
\vspace{-5pt}

\item [{[145]}] M. Burkardt and A. Langnau, Phys. Rev. {\bf D44},
1187, 3857 (1991).
\vspace{-5pt}

\item [{[146]}] A. Langnau and M. Burkardt, Phys. Rev. {\bf D47},
3452 (1993).
\vspace{-5pt}

\item [{[147]}] R. J. Perry, Phys. Lett. {\bf 300B}, 8 (1993).
\vspace{-5pt}

\item [{[148]}] A. Harindranath and W. M. Zhang, Phys. Rev. {\bf D48},
4903 (1993).
\vspace{-5pt}

\item [{[149]}] J. Schwinger, {\em Phys. Rev.} {\bf 130}, 402 (1963).
\vspace{-5pt}

\item [{[150]}] W. M. Zhang and A. Harindranath, Phys. Lett.
{\bf 314B}, 223 (1993).
\vspace{-5pt}

\item [{[151]}] A. C. Tang, Phys. Rev. {\bf D37}, 3014 (1988).
\vspace{-5pt}

\item [{[152]}] K. G. Wilson, OSU internal reprot (unpublished, 1990);
also see [85].
\vspace{-5pt}

\item [{[153]}] J. D. Bjorken and S. D. Drell, {\em Relativistic
Quantum Mechanics}, (McGraw-Hill, New York, 1964).
\vspace{-5pt}

\item [{[154]}] W.$\,$ Heitler,$\,$ {\em The$\,$ Quantum$\,$ Theory$\,$ of$\,$
radiation},
$\,$ (Oxford,$\,$ Univ.$\,$ Press,$\,$ London,$\,$ 1936).
\vspace{-5pt}

\item [{[155]}] N. Christ and T. D. Lee, Phys. Rev. {\bf D22}, 939 (1980).
\vspace{-5pt}

\item [{[156]}] W. M. Zhang, Phys. Lett. {\bf B333}, 158 (1994).
\vspace{-5pt}

\item [{[157]}] I. Bars and F. Green, Nucl. Phys. {\bf B142}, 157 (1978).
\vspace{-5pt}

\item [{[158]}] E. S. Abers and B. W. Lee, Phys. Rep. {\bf 9C}, 1 (1973).
\vspace{-5pt}

\item [{[159]}] G. 't Hooft and M. Veltman, {\em Diagrammar}, CERN
preprint (1973).
\end{enumerate}

\vspace{340pt}

\pagebreak

\footnotesize

\noindent
Of An It Strictly If Since The As

\end{document}